\documentclass[showpacs,amsmath,amsfonts,amssymb,aps,superscriptaddress]{revtex4}

\usepackage{makecell, tabularx}

\usepackage{booktabs}
\usepackage[dvipsnames, x11names]{xcolor}

\usepackage{comment}
\usepackage{enumitem}

\usepackage{graphicx}
\graphicspath{{figs/}}

\usepackage{float}
\usepackage{dcolumn}
\usepackage{bm}
\usepackage{mathrsfs}
\usepackage{amsmath}

\usepackage{amsfonts}
\usepackage{dsfont}
\makeatletter
\DeclareFontShape{OT1}{cmr}{m}{scit}{<->ssub * cmr/m/sc}{}
\makeatother
\begin{document}

\title{Exact solutions for slowly rotating wormholes in the presence of an  anisotropic fluid}

\author{Davide Batic}
\email{davide.batic@ku.ac.ae (corresponding author)}
\affiliation{Department of Mathematics, Khalifa University of Science and Technology, PO Box 127788, Abu Dhabi, United Arab Emirates}

\author{Denys Dutykh}
\email{denys.dutykh@ku.ac.ae}
\affiliation{Department of Mathematics, Khalifa University of Science and Technology, PO Box 127788, Abu Dhabi, United Arab Emirates}

\author{Mark Essa Sukaiti}
\email{100064482@ku.ac.ae}
\affiliation{Department of Mathematics, Khalifa University of Science and Technology, PO Box 127788, Abu Dhabi, United Arab Emirates}
\date{\today}

\begin{abstract}
We construct slowly rotating traversable wormholes in the presence of an anisotropic fluid. Starting from a Teo-type stationary, axisymmetric extension of the Morris–Thorne metric, we perform a slow-rotation expansion, fix a gauge that preserves the geometric meaning of the radial coordinate, and introduce two complementary prescriptions for treating the throat (fixed and free). Within this framework, the Einstein equations and conservation laws form a closed system, from which we obtain analytic expressions for the leading frame dragging and for the second-order rotational backreaction. We apply the construction to the spatial–Schwarzschild and Morris–Thorne wormholes, derive the induced corrections to the stress–energy tensor, analyse the redistribution of null energy condition (NEC) violations, and characterise quadrupolar deformations, curvature diagnostics, and possible ergoregions.
\end{abstract}
\maketitle

\section{Introduction}

Rotation is ubiquitous in compact object astrophysics. If traversable wormholes exist, they would almost certainly rotate, and their angular momentum would characterise the geometry through frame dragging, possible ergoregions, and characteristic deformations of the throat and its neighbourhood. By contrast to the static sector, where the geometry, matter content, and energy‑condition requirements are well understood, rotating traversable wormholes remain comparatively underdeveloped. From an astrophysical perspective, the familiar static, spherically symmetric wormhole solutions should therefore be viewed as idealised non-rotating limits. Realistic compact objects and their environments almost invariably carry angular momentum, so any traversable wormhole embedded in an astrophysical setting would generically be expected to spin. Extending well-understood static wormhole geometries to stationary, axisymmetric configurations is thus essential because it allows one to quantify how rotation modifies the location and shape of the throat, redistributes the exotic matter required by the flare-out condition, and introduces frame dragging and possible ergoregions. In this way, rotating wormhole models are brought closer to realistic astrophysical scenarios and become suitable for confronting potential observational signatures with data. Static, spherically symmetric wormholes are commonly expressed in terms of a redshift (lapse) function and a shape function, with a minimal‑area throat and a flare‑out condition that enforces the presence of exotic matter content near the throat (\emph{e.g.}, violations of the NEC). The canonical pedagogical treatment is attributed to Morris and Thorne \cite{Morris1988AJP}, while a thorough systematisation is offered in Visser’s monograph \cite{Visser1996}.  Prototypical examples include the massless Ellis–Bronnikov spacetime, independently obtained by Ellis \cite{Ellis1973JMP, Ellis1974JMP, Ellis1979GRG} and Bronnikov \cite{Bronnikov1973APP} as solutions of General Relativity minimally coupled to a scalar field with a reversed sign kinetic term.

Carrying these constructions into the stationary, axisymmetric regime raises technical and conceptual difficulties. One needs a metric ansatz rich enough to encode frame dragging and angular deformations, boundary conditions that ensure regularity at the throat and asymptotic flatness at both ends, and a matter model whose conservation law closes consistently with the geometry. Early analyses also derived general constraints on the stress–energy tensor capable of sourcing stationary, axisymmetric wormhole metrics  \cite{Bergliaffa2000arxiv}. A natural starting point is Teo's stationary wormhole metric \cite{Teo1998PRD}, which is characterised by the presence of a generalised shape function, an angular distortion factor, and a frame‑dragging function. The latter has also been used to study astrophysical signatures such as thin accretion disks, offering potential observational contrast with Kerr black holes \cite{Harko2008PRD}. Linear scalar perturbations in rotating wormhole backgrounds have likewise been investigated by \cite{Kim2005NCB}. Furthermore, electromagnetic magnetospheres have been analysed in rotating wormhole backgrounds. Solving the force‑free stream equation to second order in spin yields the magnetic‑field geometry within the ergoregion and the associated Poynting flux, revealing quantitative differences relative to Kerr \cite{Urtubey2025EPJC}.

Beyond kinematical ans\"{a}tze, rotating traversable wormholes have been pursued along two complementary routes. First, perturbative approaches in the slow-rotation scenario expand the geometry and matter fields in powers of the angular‑momentum parameter $J$. For wormholes supported by a phantom scalar field as in the Ellis--Bronnikov case, Kashargin and Sushkov carried out the first‑order construction and then the second‑order expansion, finding that rotation increases the asymptotic mass and can slightly reduce the magnitude of NEC violation relative to the static case \cite{Kashargin2008GC, Kashargin2008PRD}. In addition, axially symmetric rotating solutions with a less severe weak energy condition (WEC)  violation than in the static case have been reported by  \cite{Kuhfittig2003PRD}. More recently, slow‑rotation truncations to $\mathcal{O}(J^2)$ have been used to analyse radial stability, with indications that rotation can mitigate the unstable mode present in the static Ellis–Bronnikov background \cite{Azad2023, Azad2024PLB}. Second, fully nonlinear numerical families have been constructed. For example,  \cite{Kleihaus2014PRD} obtained globally regular rotating Ellis wormholes supported by a phantom scalar that acquire finite mass and quadrupole, develop ergoregions, and approach extremal Kerr at a maximal rotation. Furthermore, related studies show that rotation can be triggered by co‑rotating bosonic matter and by a complex phantom scalar field \cite{Hoffmann2018PRD, Chew2019PRD}. These perturbative and numerical families reveal the characteristic signatures of rotation in wormhole geometries and supply benchmarks for validating analytical approaches. Complementing perturbative and numerical approaches, exact rotating wormholes have also been constructed via Ehlers/Harrison transformations applied to a static wormhole seed in Einstein–Maxwell–scalar theory, yielding stationary solutions with controlled ergoregions \cite{Cisterna2023PRD}. In Einstein–Maxwell theory, the overcharged Kerr–Newman–NUT spacetime admits an exact rotating, geodesically complete wormhole. The exoticity is localised on counter‑rotating Misner–Dirac strings, and the spacetime exhibits an ergoregion without superradiance \cite{Clement2023PLB}. Moreover, rotating wormholes supported by Casimir stresses have also been explored. In that context, rotation largely preserves the static structure, while constant angular velocity configurations require an exponential cut-off and an additional scale to prevent rotations at spatial infinity \cite{Garattini2025PRD}. Finally, rotating wormholes supported by three‑form fields have also been constructed. For suitable parameter choices, the matter threading the geometry can satisfy the NEC and WEC, with the three‑form sector carrying the exoticity \cite{Tanghpati2024NPB}. In many of these constructions, the supporting exotic matter exhibits anisotropic stresses. The Ellis–Bronnikov wormhole, for instance, arises from General Relativity minimally coupled to a phantom scalar field, and its stress–energy tensor can be written in terms of an energy density and distinct radial and tangential pressures (see Sec.~IV~B). Likewise, Casimir-type vacuum stresses and semiclassical vacuum polarisation generically produce negative energy densities with direction-dependent principal pressures, while higher-form fields and multi-field sectors also yield stress–energy tensors with nontrivial radial and angular structure. In what follows, we therefore model the exotic matter as an effective anisotropic fluid whose principal pressures encode these direction-dependent stresses. Our slow-rotation framework should be viewed as extending these physically motivated sources (phantom fields, Casimir vacua, semiclassical backreaction, and higher-form sectors) into the stationary, axisymmetric regime in a unified, analytically tractable way.

To address this gap, we develop a general slow rotation framework for traversable wormholes in the presence of an anisotropic fluid that co-rotates with the geometry. The matter threading the wormhole is modelled as an effective anisotropic fluid, providing a macroscopic description of the exotic stress–energy required by the flare-out condition. In the stationary, axisymmetric regime considered below, we follow the same spirit for rotation, i.e. rather than postulating a microphysical equation of state, we treat the angular-velocity profile $\Omega(r,\chi)$ as part of an effective constitutive relation for the anisotropic fluid, in close analogy with the rotation laws prescribed in slow-rotation studies of compact stars \cite{Hartle1967AJ, Hartle1968AJ}. The corresponding stress–energy tensor is then reconstructed from the Einstein equations and the conservation law. In the static limit, many familiar wormhole solutions admit precisely such a representation. For instance, the Ellis–Bronnikov wormhole, originally obtained as a solution of General Relativity minimally coupled to a phantom scalar field, can be rewritten in terms of a diagonal anisotropic stress–energy tensor with distinct radial and tangential pressures (see Sec. IV B). Our slow-rotation framework thus extends standard exotic matter sources into the stationary, axisymmetric regime, tracking how rotation redistributes the energy density and principal pressures while preserving regularity at the throat and asymptotic flatness. The construction is based on a stationary, axisymmetric extension of the familiar Morris–Thorne wormhole, organised as a perturbative expansion in a small rotation parameter. Rotational effects are incorporated into the metric in a controlled manner. Those associated with frame dragging appear at first order in the expansion, while deformations of the spatial geometry arise only at second order due to equatorial symmetry. A key ingredient of our approach is a gauge choice that preserves the geometric meaning of the radial coordinate, ensuring that constant radius surfaces retain their correct area throughout the expansion. We also formulate two complementary prescriptions for treating the throat. One approach keeps the throat fixed at its static radial location, while the other determines its rotationally deformed position directly from the minimal-area condition. This distinction allows us to separate physical throat deformations from coordinate artefacts. Within this structure, the Einstein equations, together with the conservation law for the anisotropic fluid, form a closed system that admits a systematic analytic treatment, allowing us to derive explicit expressions for the leading frame-dragging profile and the second-order rotational backreaction on the geometry. The resulting solutions place rotating wormholes on the same conceptual footing as their static counterparts, while providing explicit spacetimes suitable for applications ranging from geodesic analysis and lensing/shadow modelling to quasinormal-mode and wave-propagation studies, and for benchmarking observational signatures against Kerr and other compact-object backgrounds. In the present work, we derive exact leading-order rotational solutions and the full second-order backreaction for two canonical zero-redshift wormhole backgrounds: the spatial–Schwarzschild wormhole with a constant shape function and the Morris–Thorne wormhole. In each case, we impose regularity at the throat and enforce asymptotic flatness on both ends, and we systematically track how rotation redistributes violations of the NEC among the principal null directions, both near and away from the throat. We further quantify the associated quadrupolar deformations of the throat, evaluate curvature diagnostics, and analyse the emergence and geometry of possible ergoregions through the zero set of the temporal metric component.

In comparison with existing rotating wormhole constructions ranging from Teo’s kinematical stationary metric \cite{Teo1998PRD}, through slow-rotation Ellis-Bronnikov solutions supported by phantom scalars \cite{Kashargin2008GC, Kashargin2008PRD, Azad2024PLB} and axisymmetric models with milder energy-condition violation \cite{Kuhfittig2003PRD}, to fully nonlinear rotating Ellis and related wormholes \cite{Kleihaus2014PRD, Hoffmann2018PRD, Chew2019PRD} and exact stationary solutions obtained via solution generating techniques or Casimir stresses \cite{Cisterna2023PRD, Clement2023PLB, Garattini2025PRD}, the present work provides a complementary, fully analytic slow rotation framework based on an effective anisotropic fluid. The main new ingredients are: (i) a general Teo-type construction in which a co-rotating anisotropic fluid with a prescribed rotation law sources the geometry; (ii) an area gauge combined with fixed- and free-throat prescriptions that disentangle physical deformations of the throat from coordinate artefacts; and (iii) closed form expressions for the leading Lense-Thirring frame dragging and the full second order backreaction for spatial Schwarzschild and Morris-Thorne seeds, together with a systematic analysis of NEC violations, curvature diagnostics, quadrupolar throat deformations, and ergoregions. These results complement perturbative, numerical, and exact rotating wormhole models by supplying a unified, analytically tractable family of solutions that can serve both as benchmarks for more detailed matter models and as backgrounds for phenomenological applications.

The paper is organised as follows. Section II reviews the static framework and fixes notation for the anisotropic fluid. Section III develops the slow‑rotation expansion, fixes gauge and throat schemes, and derives the order‑by‑order field equations. Section IV applies the framework to the spatial Schwarzschild and Morris-Thorne seeds, deriving closed-form expressions for frame dragging, metric backreaction, induced stresses, energy-condition diagnostics, and ergoregions. Finally, Section V summarises our findings and outlines possible directions for future work.

\section{STATIC WORMHOLES}

The line element of a static,  spherically symmetric wormhole, expressed in natural units where $c = G_N = 1$, is given by  \cite{Morris1988AJP, Ellis1973JMP, Ellis1979GRG, Ellis1974JMP, Bronnikov1973APP}
\begin{equation}\label{metric}
  ds^2 = -e^{2\Phi(r)}dt^2+\frac{dr^2}{1-\frac{b(r)}{r}} + \frac{r^2}{1-\chi^2}d\chi^2 + r^2(1-\chi^2)d\varphi^2, \quad \chi\in[-1,1], \quad \varphi\in[0,2\pi),
\end{equation}
For later use, it is convenient to work with the angular coordinate $\chi=\cos{\vartheta}$, so that $\chi\in[-1,1]$ corresponds to the usual polar angle $\vartheta\in[0,\pi]$. A short calculation shows that
\begin{equation}
  \frac{d\chi^2}{1-\chi^2} + (1-\chi^2)d\varphi^2= d\vartheta^2 + \sin^2{\vartheta}d\varphi^2,
\end{equation}
and hence the angular sector in \eqref{metric} is exactly the standard metric on the unit 2–sphere. In particular, the line element \eqref{metric} coincides with the usual Morris–Thorne static wormhole metric, written in terms of $\chi$ for convenience in later Legendre expansions. Furthermore, $\Phi$ and $b$ are the redshift (or lapse) and shape functions, respectively, as defined in \cite{Garattini2009PLB, Nicolini2010CQG}. We assume that the radial coordinate $r$ increases monotonically from its minimum value $r_0$, representing the throat of the wormhole, to spatial infinity. The matter content acting as a source of  the geometry described by \eqref{metric} is modelled as an anisotropic fluid, with energy-momentum tensor given by
\begin{equation}\label{emt}
 T^\alpha{}_\beta=(\rho+p_t)u^\alpha u_\beta+p_t\delta^\alpha{}_\beta+(p_r-p_t)\ell^\alpha\ell_\beta,    
\end{equation}
where $\ell^\alpha$ is a unit space-like vector orthogonal to the fluid four-velocity $u^\alpha$, that is $\ell^\alpha \ell_\alpha=1$ and $\ell^\alpha u_\alpha=0$. Moreover, $u^\alpha$ must satisfy the condition $g_{\alpha\beta}u^\alpha u^\beta=-1$. These constraints require that
\begin{equation}\label{utnorot}
    u^\alpha=e^{-\Phi(r)}\delta^\alpha{}_t,\quad
    \ell^\alpha=\sqrt{1-\frac{b(r)}{r}}\delta^\alpha{}_r.
\end{equation}
Hence, the mixed energy-momentum tensor can be represented in terms of the diagonal matrix
\begin{equation}
    T^\alpha{}_\beta=\text{diag}(-\rho(r),p_r(r),p_t(r),p_t(r)),
\end{equation}
where $\rho$ is the energy density, $p_r$ the radial pressure, and $p_t$ the tangential pressure measured orthogonally to the radial direction. Let us recall that
\begin{eqnarray}
T_{\alpha\beta}=\text{diag}\left(e^{2\Phi(r)}\rho(r),\frac{p_r(r)}{1-\frac{b(r)}{r}},\frac{r^2 p_t(r)}{1-\chi^2},r^2(1-\chi^2)~p_t(r)\right),\\
    T^{\alpha\beta}=\text{diag}\left(e^{-2\Phi(r)}\rho(r),\left[1-\frac{b(r)}{r}\right]p_r(r),\frac{1-\chi^2}{r^2}p_t(r),\frac{p_t(r)}{r^2(1-\chi^2)}\right).
\end{eqnarray}
We emphasise that the anisotropic fluid form \eqref{emt} is generic for static, spherically symmetric geometries of the form \eqref{metric}. Given a choice of redshift $\Phi(r)$ and shape function $b(r)$, the Einstein equations and the conservation law $\nabla_\alpha T^\alpha{}_{\beta}=0$ determine the functions $\rho(r)$, $p_r(r)$, and $p_t(r)$ via equations \eqref{eq1}–\eqref{eq4} or, equivalently, via the explicit expressions \eqref{ed}-\eqref{pt}, as derived below. In this sense, the anisotropic fluid description does not introduce a new constraint at the static level but simply provides a convenient decomposition of the stress–energy tensor into an energy density and three principal pressures. As a concrete example, in Sec. IV B we show that the Morris–Thorne wormhole can be written as an anisotropic fluid with $\rho(r)=p_r(r)=-r_0^2/(8\pi r^4)$ and $p_t(r)=+r_0^2/(8\pi r^4)$, reproducing the well-known NEC-violating behaviour $\rho+p_r<0$ and $\rho+p_t=0$ near the throat. In what follows, we therefore interpret \eqref{emt} as an effective description of the exotic matter supporting the static wormhole geometry. The same anisotropic-fluid framework is also appropriate for effective stress–energy tensors arising from Casimir-type vacuum effects, semiclassical vacuum polarisation, and higher-form fields, which likewise generically exhibit negative energy densities and direction-dependent principal pressures. By applying the Einstein field equations $G_{\alpha\beta}=8\pi T_{\alpha\beta}$ alongside the conservation equation $\nabla_\alpha T^{\alpha\beta}=0$, we obtain the following system of equations, where an  overdot denotes differentiation with respect to the radial coordinate
\begin{eqnarray}
&&\dot{b}-8\pi r^2\rho=0,\label{eq1}\\
&&2r(r-b)\dot{\Phi}-b-8\pi r^3 p_r=0,\label{eq2}\\
&&r^2(r-b)\ddot{\Phi}+\left[r(r-b)\dot{\Phi}+\frac{b-r\dot{b}}{2}\right](1+r\dot{\Phi})-8\pi r^3 p_t=0,\label{eq3}\\
&&r(\rho+p_r)\dot{\Phi}+r\dot{p}_r+2(p_r-p_t)=0.\label{eq4}
\end{eqnarray}
Using equations \eqref{eq1}–\eqref{eq3}, we can express the energy density, the radial and tangential pressures in terms of the redshift and shape functions as follows
\begin{eqnarray}
\rho&=&\frac{\dot{b}}{8\pi r^2},\label{ed}\\
p_r&=&\frac{1}{8\pi}\left[\frac{2}{r}\left(1-\frac{b}{r}\right)\dot{\Phi}-\frac{b}{r^3}\right],\label{pr}\\
p_t&=&\frac{1}{8\pi}\left(1-\frac{b}{r}\right)\left[\ddot{\Phi}+\dot{\Phi}^2-\frac{r\dot{b}-b}{2r(r-b)}\dot{\Phi}+\frac{\dot{\Phi}}{r}-\frac{r\dot{b}-b}{2r^2(r-b)}\right].\label{pt}
\end{eqnarray}
It is straightforward to verify that substituting equations \eqref{ed}–\eqref{pt} into the conservation equation \eqref{eq4} shows that it is identically satisfied, as expected from the consistency of the Einstein field equations. Moreover, in the case $\Phi=0$, from \eqref{ed}–\eqref{pt}, we find that $p_t$ is determined by $\rho$ and $p_r$ according to the relation
\begin{equation}\label{pteos}
p_t=-\frac{\rho+p_r}{2}. 
\end{equation}
For completeness, we also provide the expressions for the Ricci scalar and the Kretschmann invariant, which will be instrumental for the subsequent analysis. They are given by
\begin{eqnarray}
R&=&-2\left(1-\frac{b}{r}\right)\left(\ddot{\Phi}+\dot{\Phi}^2\right)+\frac{2\dot{b}}{r^2}+\frac{r\dot{b}+3b-4}{r^2}\dot{\Phi},\label{Ricci}\\
R^{\alpha\beta\gamma\delta}R_{\alpha\beta\gamma\delta}&=&\frac{4}{r^4}\left\{r(r-b)\ddot{\Phi}+\left[r(r-b)\dot{\Phi}-\frac{r\dot{b}-b}{2}\right]\dot{\Phi}\right\}^2+\frac{8(r-b)^2}{r^4}\dot{\Phi}^2+\frac{2(r\dot{b}-b)^2}{r^6}+\frac{4b^2}{r^6},
\end{eqnarray}
and they reduce to the following compact form when $\Phi=0$
\begin{equation}
R=16\pi \rho,\quad
\mathcal{K}=R^{\alpha\beta\gamma\delta}R_{\alpha\beta\gamma\delta}=256\pi^2\left[p_r^2+\frac{(\rho+p_r)^2}{2}\right].
\end{equation}

\section{Wormhole solutions in the slow rotation limit and zero redshift}
In the static sector discussed in Sec. II, once a redshift function and a shape function are specified, the system \eqref{eq1}–\eqref{eq4} determines the corresponding energy density and principal pressures, thereby defining the background wormhole solution \eqref{metric} supported by an anisotropic fluid. This static configuration serves as the seed geometry for our slow-rotation construction, organised as a perturbative expansion in the small rotation parameter $J$ about this background. As a rotating extension of the static wormhole metric \eqref{metric}, we therefore adopt a Teo-like ansatz \cite{Teo1998PRD} written in Boyer–Lindquist–type coordinates $(t,r,\chi,\varphi)$ with $\chi=\cos{\theta}$.
\begin{equation}\label{rotmet}
  ds^2=-[N^2-r^2(1-\chi^2) K^2\omega^2]dt^2+\frac{dr^2}{1-\frac{B}{r}}-2\omega r^2(1-\chi^2) K^2 dtd\varphi+r^2 K^2\left[\frac{d\chi^2}{1-\chi^2}+(1-\chi^2)d\varphi^2\right].
\end{equation}
Here $N(r,\chi)$ is the lapse, $\omega(r,\chi)$ is the angular velocity of inertial frames, $B(r,\chi)$ generalises the shape function, and $K(r,\chi)$ encodes the departure of the $r=$const two-geometry from sphericity. The metric reduces to the static configuration \eqref{metric} in the non-rotating limit $J\to0$, i.e.
\begin{equation}
N(r,\chi)\to e^{\Phi(r)},\qquad
B(r,\chi)\to b(r),\qquad
K(r,\chi)\to 1,\qquad
\omega(r,\chi)\to 0.
\end{equation}
Furthermore, as already observed by \cite{Teo1998PRD}, $J$ can be interpreted as the total angular momentum per unit mass of the wormhole if we impose the following asymptotic behaviour on $\omega$, that is
\begin{equation}
\omega(r,\chi)=\frac{2J}{r^3}+\mathcal{O}\left(\frac{J^3}{r^7}\right).    
\end{equation}
In order to prevent misreadings, notice that Teo’s $a$ appearing in \cite{Teo1998PRD} is the total angular momentum $J$, not the Kerr parameter $a =J/M$. In the following, we work in the slow-rotation regime $(J\ll 1)$ and expand the metric functions in powers of $J$, keeping $\omega$ to $O(J^{3})$ and $(N,B,K)$ to $O(J^{2})$. More precisely, we have
\begin{eqnarray}
N(r,\chi)&=&N_0(r)+N_2(r,\chi)J^2+\mathcal{O}(J^4),\quad N_0(r)=e^{\Phi(r)},\quad
N_2(r,\chi)=N_0^{(2)}(r)+N_2^{(2)}(r)P_2(\chi),\label{N21}\\    
K(r,\chi)&=&1+K_2(r,\chi)J^2+\mathcal{O}(J^4),\quad
K_2(r,\chi)=K_0^{(2)}(r)+K_2^{(2)}(r)P_2(\chi),\\
B(r,\chi)&=&b(r)+B_2(r,\chi)J^2+\mathcal{O}(J^4),\quad
B_2(r,\chi)=B_0^{(2)}(r)+B_2^{(2)}(r)P_2(\chi),\label{expB}\\
\omega(r,\chi)&=&\omega_1(r)J+\omega_3(r,\chi)J^3+\mathcal{O}(J^5),\quad 
\omega_3(r,\chi)=\omega_0^{(3)}(r)+\omega_2^{(3)}(r)P_2(\chi).\label{N24}
\end{eqnarray}
For later convenience, we introduce the dimensionless spin parameter $j=J/r_0^2$, where $r_0$ denotes the throat radius of the static seed. In terms of $j$, the
functions $N$, $B$, $K$, and $\omega$ are analytic and split into even and odd
sectors according to \eqref{N21}–\eqref{N24}: $\omega$ contains only odd powers of $j$,
while $N$, $B$, and $K$ contain only even powers. Our slow rotation scheme is therefore
a regular Taylor expansion around $j=0$, truncated at $\mathcal{O}(j^3)$ in
$\omega$ and $\mathcal{O}(j^2)$ in $N$, $B$, and $K$. The corresponding domain of validity can be estimated from the explicit solutions in Sec. IV. As a representative example, the frame-dragging function for Model I can be written as
\begin{equation}
  \omega(x,\chi) = \frac{2j}{x^3} + F_I(x,\chi)j^3 + \mathcal{O}(j^5),
\end{equation}
with $x=r/r_0$, $\chi=\cos{\theta}$, and a smooth bounded coefficient $F_I(x,\chi)$ given in \eqref{102}. From its explicit expression one sees that $F_I(x,\chi)$ is a finite sum of terms $c_n x^{-n}$ and $d_n x^{-n}P_2(\chi)$ with $n\geq 3$ and rational coefficients $c_n$, $d_n$, so that $F_I(x,\chi)\,x^3$ is uniformly bounded for $x\geq 1$ and $|\chi|\leq 1$. In particular, one finds
\begin{equation}
  \frac{|\omega^{(3)} J^3|}{|\omega_1 J|}
  = \frac{1}{2}\,\bigl|x^3 F_I(x,\chi)\bigr|j^2
  \leq C_{I}j^2,
\end{equation}
with a finite constant $C_{I}$ of order $10$. Thus, for $|j|\leq 0.1$ the cubic correction modifies the leading frame dragging by at most $\sim 10\%$ even at the throat, and for $|j|\lesssim 0.15$ the maximal change remains at the $\sim 20$–$25\%$ level in this model. Analogous estimates based on \eqref{110}, \eqref{shapeB}, \eqref{141}, and \eqref{145} show that, in all the rotation models considered, the functions multiplying $j^2$ and $j^3$ are smooth and uniformly bounded on $x \geq 1$ and decay at least as $x^{-3}$ at large radius. Consequently, for each model, the ratio between the cubic and linear frame-dragging terms is bounded by a model-dependent constant times $j^2$. Moreover, equatorial reflection $(\chi\to -\chi)$ enforces that $N, B, K$ are even in $\chi$, hence their $J^{2}$ corrections are expanded in even Legendre modes (here truncated to $\ell=0,2$ for simplicity). Reversal of the sense of rotation $(J\to -J)$ requires $N, B, K$ to be even in $J$ and $\omega$ to be odd, which is why only odd powers appear in $\omega$. With equatorial symmetry, $\omega$ must also be even in $\chi$, so its $O(J^{3})$ term makes use of even $\ell$ (here $\ell=0,2$). Here, $\ell$ is the usual angular multipole index. Furthermore, at first order in $J$ we write $\omega(r,\chi)=\omega_{1}(r)J+O(J^{3})$ and determine $\omega_{1}$ from the $(t\varphi)$ Einstein equation, subject to regularity at the throat and the asymptotic normalization $\omega_{1}(r)\to 2/r^{3}$ as $r\to\infty$, which fixes the integration constants and identifies $J$ with the total angular momentum. In what follows, we fix the area gauge, which eliminates the residual radial reparametrization freedom $r\!\to\!\bar r(r)$ and keeps $r$ radial in the geometric sense. To achieve that, we first observe that on each $t=$const slice, the 2–surface at fixed $r$ has induced metric $r^{2}K^{2}\big(d\chi^{2}/(1-\chi^2)+(1-\chi^2)\,d\varphi^{2}\big)$, and then, we impose that its area equal that of a sphere of radius, i.e. $A(r)=4\pi r^{2}$. This  implies that 
\begin{equation}\label{gauge}
\langle K^{2}\rangle(r):=\frac{1}{2}\!\int_{-1}^{1}\!K^{2}(r,\chi)\,d\chi=1
\end{equation}
order by order in the slow–rotation expansion, in particular $\langle K_{2}\rangle=0$ at $O(J^{2})$. Geometrically, this area gauge ensures that the coordinate $r$ keeps its usual areal meaning, i.e. the area of a $t=const$, $r=const$ two–surface remains $A(r)=4\pi r^2$ as in the static seed. At the level of the multipolar expansion in $\chi$, the condition $\langle K^2\rangle=1$ precisely removes the $\ell=0$ (monopole) part of $K$, so that no artificial spherically symmetric distortion is absorbed into $K$ and only genuine angular deformations (quadrupole and higher) are kept. Furthermore, such a gauge choice ensures that asymptotic quantities are read off in the standard way, decouples angular multipoles when projecting the field equations onto Legendre modes, and simplifies the characterisation of the throat as the minimal area surface. A straightforward computation shows that \eqref{gauge} requires that for all $r$ in the coordinate domain
\begin{equation}
K_0^{(2)}(r)=0.  
\end{equation}
Following Teo’s canonical form, we allow $B(r,\chi)$ to depend on both $r$ and $\chi$. Regularity at the throat requires $\partial_\chi B|_{r_0}=0$, ensuring that the throat lies at a constant coordinate radius $r=r_0$ even though $B$ may have angular dependence away from the throat. Within our slow-rotation expansion for $B$, to keep the throat at fixed $r = r_0$, we impose the following boundary condition
\begin{equation}\label{B2cond}
  B_2(r_0,\chi)=0~~\forall\chi\in[-1,1],
\end{equation}
which selects a subclass of physical solutions, i.e. those whose throat stays at the same coordinate radius in our gauge. If this condition is not enforced, the throat generally shifts by an amount $\delta r(\chi)$, so that the throat location is $r_t=r_0+\delta r(\chi)$. To quantify $\delta r$, we define the function
\begin{equation}\label{F}
F(r,\chi;J):=1-\dfrac{B(r,\chi;J)}{r},
\end{equation}
so the throat satisfies $F(r_t(\chi;J),\chi;J)=0$ with $r_t(\chi;0)=r_0$ and $b(r_0)=r_0$. Since the metric functions $N$, $B$, and $K$ are even in $J$, also $F$ is even in $J$. Notice that $\partial_{r}F(r_{0},\chi;0)=\dfrac{1-b'(r_{0})}{r_{0}}\neq 0$ because of the flare-out condition in the static case. Then, by the implicit function theorem, $r_{t}$ admits a Taylor series in $J$
\begin{equation}\label{expa}
r_{t}(\chi;J)=r_{0}+c_{1}(\chi)J+c_{2}(\chi)J^{2}+\cdots .
\end{equation}
Differentiating $F=0$ with respect to $J$ at $J=0$ yields
\begin{equation}
0=\Big(\partial_{r_t}F\Big)_{J=0}\,c_{1}(\chi)+\Big(\partial_{J}F\Big)_{J=0}.
\end{equation}
But $\partial_{J}F|_{J=0}=0$ because $F$ depends on $J$ only through even powers. Since $\partial_{r_t}F|_{J=0}\neq 0$, it follows that $c_{1}(\chi)=0$. Therefore, the first term in $\delta r$ is even in $J$. More precisely, we have
\begin{equation}
\delta r(\chi):=r_{t}-r_{0}=c_{2}(\chi)J^{2}+O(J^{4}).
\end{equation}
In order to determine the unknown function $c_2(\chi)$, we replace the first expansion in \eqref{expB} in \eqref{F} and we set $F$ to first order in $\delta r$. This procedure gives
\begin{equation}
  0=\underbrace{\Big(1-\tfrac{b(r_{0})+J^{2}B_{2}(r_{0},\chi)}{r_{0}}\Big)}_{(I)} + \underbrace{\frac{1-b'(r_{0})}{r_{0}}}_{(II)}\delta r(\chi)\ +\ O(J^{4}).
\end{equation}
Let us recall that $b(r_0)=r_0$, then (I) reduces to $-\tfrac{J^{2}}{r_{0}}B_{2}(r_{0},\chi)$ while (II) comes from $(\partial F /\partial (r_0+\delta r)|_{\delta r=0}$. 
Hence, we find that
\begin{equation}\label{deltarchi}
\delta r(\chi)=\frac{J^{2}\,B_{2}(r_{0},\chi)}{\,1-b'(r_{0})\,}+O(J^{4}).
\end{equation}
Terms arising from the expansion $B_2(r_0+\delta r,\chi)=B_2(r_0,\chi)+(\partial_{r_0+\delta r} B_2)|_{\delta r=0}\,\delta r+\cdots$ appear multiplied by the overall $J^2$ and by $\delta r=O(J^2)$, hence they contribute only at $O(J^4)$. They are therefore neglected at this order. In summary, the discrete reversal symmetry $J \to -J$ implies that the even sector of the solution, and hence $F$, is an even function of $J$. Consequently, any linear term $c_1(\chi)J$ in the expansion \eqref{expa} must vanish. Equivalently, a linear order throat displacement $\delta r \propto J$ is forbidden by symmetry, and rotational deformations of the throat start at quadratic order, $\delta r = O(J^2)$, as explicitly shown in \eqref{deltarchi}.  The only way to obtain $\delta r=O(J)$ would be to introduce an $O(J)$ term in $B$, thereby breaking the $J\to -J$ symmetry, which we do not consider.  For later use, we adopt two complementary implementations when solving the Einstein field equations:
\begin{itemize}
\item {\bf{A fixed-throat scheme}}: Impose $B_{2}(r_{0},\chi)=0$ for all $\chi$ and with area gauge already in place. Then the throat remains at $r=r_{0}$ through $O(J^2)$, and the quadrupolar deformation is carried by $K_2$ (and $N_2$).
\item {\bf{ A free-throat scheme}}: Do not impose \eqref{B2cond}. Solve the Einstein Field Equations and locate the throat a posteriori from $F=0$, which gives
\begin{equation}
r_{t}(\chi)=r_{0}+\frac{J^{2}B_{2}(r_{0},\chi)}{1-b'(r_{0})}+O(J^{4}). 
\end{equation}
\end{itemize}

Moreover, we model the matter content of the rotating wormhole as an anisotropic fluid. Let $\{U^{\alpha},n_{(r)}^{\alpha},n_{(\chi)}^{\alpha},n_{(\varphi)}^{\alpha}\}$ be an orthonormal tetrad comoving with the fluid, with $U_{\alpha}U^{\alpha}=-1$, $n_{(i)\,\alpha}n_{(j)}^{\alpha}=\delta_{ij}$, and $U_{\alpha}n_{(i)}^{\alpha}=0$. In its mixed-index form, the energy–momentum tensor is
\begin{equation}\label{AEMT}
\widetilde{T}^\alpha{}_\beta=\widetilde{\rho}U^\alpha U_\beta
+P_r n^{\alpha}_{(r)}n_{(r)\beta}
+P_\chi n^{\alpha}_{(\chi)}n_{(\chi)\beta}
+P_\varphi n^{\alpha}_{(\varphi)}n_{(\varphi)\beta},    
\end{equation}
where $\widetilde{\rho}(r,\chi)$ is the energy density measured by comoving observers and $P_{r}(r,\chi), P_{\chi}(r,\chi), P_{\varphi}(r,\chi)$ are the principal pressures along the radial, polar, and azimuthal directions, respectively. The four-velocity is taken co-rotating with the geometry, $U^{\alpha}\propto \partial_{t}+\Omega(r,\chi)\,\partial_{\varphi}$, with the normalization fixed by $U_{\alpha}U^{\alpha}=-1$. A straightforward computation gives
\begin{equation}\label{norm}
U^\alpha=U^t(1,0,0,\Omega),\quad
U^t=\frac{1}{\sqrt{N^2-r^2(1-\chi^2) K^2(\Omega-\omega)^2}}.
\end{equation}
Equation \eqref{norm} implies that $U^\alpha$ is timelike if and only if $N^2 - r^2(1-\chi^2)K^2(\Omega-\omega)^2>0$. In other words, the physical azimuthal 3-velocity of the fluid as measured by zero-angular-momentum observers remains subluminal provided $|\Omega-\omega|$ is sufficiently small compared to $N/(rK\sqrt{1-\chi^2})$. This causality requirement will be used as a basic admissibility criterion for the rotation laws considered below. In our slow rotation ansatz one has $N=1+\mathcal{O}(j^2)$, $K=1+\mathcal{O}(j^2)$, and $\Omega-\omega=\mathcal{O}(j/r)$ once the rotation laws of Table~I are imposed. The condition $N^2-r^2(1-\chi^2)K^2(\Omega-\omega)^2>0$ therefore reduces schematically to $1-\mathcal{O}(j^2)>0$, so that for sufficiently small $|j|$ the comoving four-velocity is automatically timelike throughout the domain. In particular, for the range $0\le j\le 0.15$ used in Sec.~IV, our explicit solutions satisfy this inequality everywhere, and the comoving azimuthal 3-velocity remains strictly subluminal. Furthermore, notice that $U_\varphi=g_{\varphi\varphi}(\Omega-\omega)U^t$. Hence, if $\Omega=\omega$, the fluid has zero azimuthal velocity, i.e. $U_\varphi=0$, and corotates with the local inertial frames (a ZAMO flow). Since the fluid’s EOS is left unspecified, Einstein’s equations do not determine its kinematic angular velocity. We must therefore prescribe a physically admissible rotation law $\Omega$ from the outset. A convenient choice that satisfies regularity, causality, and asymptotic flatness requirements will be introduced later on. Moreover, the vanishing of the $U^r$ and $U^\chi$ components ensures that the fluid motion is confined entirely to the $(t, \varphi)$-plane, indicating purely azimuthal circulation. Furthermore, if we impose $n_{(i)\,\alpha}n_{(j)}^{\alpha}=\delta_{ij}$, and $U_{\alpha}n_{(i)}^{\alpha}=0$, a straightforward but tedious computation shows that
\begin{equation}
n^{\alpha}_{(r)}=\left(0,\sqrt{1-\frac{B}{r}},0,0\right),\quad
n^{\alpha}_{(\chi)}=\left(0,0,\frac{\sqrt{1-\chi^2}}{rK},0\right),\quad
n^{\alpha}_{(\varphi)}=\left(n^{t}_{(\varphi)},0,0,n^{\varphi}_{(\varphi)}\right)
\end{equation}
with
\begin{equation}
n^{t}_{(\varphi)}=\frac{rK(\Omega-\omega)\sqrt{1-\chi^2}}{N\sqrt{N^2-r^2(1-\chi^2) K^2(\Omega-\omega)^2}},\quad
n^{\varphi}_{(\varphi)}=\frac{N^2+r^2(1-\chi^2)K^2\omega(\Omega-\omega)}{r\sqrt{1-\chi^2}NK\sqrt{N^2-r^2(1-\chi^2) K^2(\Omega-\omega)^2}}.
\end{equation}
Finally, the non-trivial Einstein field equations are
\begin{eqnarray}
G_{tt}&=&8\pi T_{tt},\label{Gtt}\\
G_{t\varphi}&=&8\pi T_{t\varphi},\label{Gtphi}\\
G_{\varphi\varphi}&=&8\pi T_{\varphi\varphi},\label{Gphiphi}\\
G_{rr}&=&\frac{8\pi P_r}{1-\frac{B}{r}},\label{Grr}\\
G_{\chi\chi}&=&\frac{8\pi r^2 K^2 P_\chi}{1-\chi^2},\label{Gchichi}\\
G_{r\chi}&=&0.
\end{eqnarray}
These are supplemented by the conservation laws $\nabla_\alpha T^{r\alpha}=0$ and $\nabla_\alpha T^{\chi\alpha}=0$. With the rotation law $\Omega$ prescribed, the geometry encoded in the metric unknowns $\{N, K, \omega, B\}$ is determined as follows. The off–diagonal relation $G_{r\chi}=0$, projected onto $\ell=2$ at $\mathcal{O}(J^2)$, fixes the quadrupole $B_2^{(2)}$ of the shape function according to \eqref{B22} thereby eliminating $B_2^{(2)}$ from the remaining equations. The algebraic system formed by the $(tt)$, $(t\varphi)$, and $(\varphi\varphi)$ Einstein equations solves directly for the matter pair $(\widetilde{\rho},P_\varphi)$. Demanding consistency via \eqref{cons1} yields the metric integrability conditions, i.e. the $\mathcal{O}(J)$ frame-dragging equation for $\omega_1$ (see \eqref{omega1eq}) and the coupled $\mathcal{O}(J^3)$ relations \eqref{L30}–\eqref{L32} for $\omega^{(3)}_{0}$, and $\omega^{(3)}_{2}$. The $(rr)$- and $(\chi\chi)$-sector then define the principal stresses $P_r,P_\chi$ once the geometry is fixed. In particular, expanding \eqref{Grr}–\eqref{Gchichi} to $\mathcal{O}(J^2)$ gives the coefficients appearing in the expansions \eqref{expPR} and \eqref{expPchi}. The conservation laws are identically satisfied at $\mathcal{O}(1)$ while at $\mathcal{O}(J^2)$ they reduce to consistency/regularity conditions that force the angular deformation $K^{(2)}_2$ of $K$ to vanish while the area gauge \eqref{gauge} already sets $K^{(2)}_0=0$. We consider two classes of prescriptions for $\Omega$. 
\begin{itemize}
    \item 
    {\bf{Rigid fluids $\Omega=\Omega(r)$}}:
    \begin{eqnarray}
    \Omega_{I}(r)&=&\frac{\kappa J}{r^3},\quad
    \Omega_{II}(r)=\frac{\kappa Jr}{r^4+\kappa^2 J^2},\quad
    \Omega_{III}(r)=\frac{\kappa J}{r^3}e^{-\kappa^2 J^2/r^4}.\label{rl1}
    \end{eqnarray}
\item 
{\bf{Differential fluids $\Omega=\Omega(r,\chi)$}}: We will consider the following model
    \begin{equation}\label{DFluid}
        \Omega_{IV}(r,\chi)=\frac{\kappa J r}{r^4+\kappa^2 J^2\chi^2}.
    \end{equation}
\end{itemize}
We treat $\Omega$ as a constitutive choice for the fluid. More precisely, we regard $\Omega=\Omega(r,\chi)$ as part of a phenomenological constitutive relation for the effective anisotropic fluid, in the same spirit as prescribing a rotation profile for barotropic stellar models in the Hartle–Thorne framework \cite{Hartle1968AJ}. For each static seed $(\Phi,b)$ and each admissible $\Omega$, the metric functions $N$, $K$, $\omega$, and $B$, and the fluid variables $\widetilde{\rho}$, $P_r$, $P_\chi$, and $P_\varphi$ are then fixed by the Einstein equations and the conservation laws, subject to regularity at the throat and asymptotic flatness. In particular, the condition that $U^\alpha$ remain timelike translates into $N^2 - r^2(1-\chi^2)K^2(\Omega-\omega)^2>0$. In the slow-rotation regime considered here, the profiles in Table~\ref{coeff} satisfy $\Omega-\omega = \mathcal{O}(J)/r^3$ at large radii and remain $\mathcal{O}(J)$ throughout the domain, so this inequality is automatically fulfilled for sufficiently small $J$. The Teo-type ansatz \eqref{rotmet} is therefore not employed purely kinematically: it is tied to an effective co-rotating anisotropic fluid whose stress–energy tensor is fully determined by the field equations.  Among admissible, causal profiles we introduce an adjustable scale $\kappa>0$ so that the geometry-determined frame dragging obeys asymptotically the standard Lense–Thirring tail. Moreover, all models can be encompassed by the slow-rotation ansatz
\begin{equation}\label{OmegaAnsatz}
  \Omega(r,\chi) = \Omega_1(r)J + \left[\Omega^{(3)}_0(r) + \Omega^{(3)}_2(r)P_2(\chi)\right]J^3 + \mathcal{O}(J^5).    
\end{equation}
Within this class, the rotation laws \eqref{rl1}–\eqref{DFluid} are chosen as simple analytic representatives that satisfy a common set of physical requirements such as regularity at the throat and on the axis, asymptotic flatness with $\Omega-\omega = \mathcal{O}(J/r^3)$, and timelikeness of the comoving four-velocity $U^\alpha$ in the slow-rotation regime $0\leq j\leq 0.15$ used in Sec. IV. Models I–III describe rigid rotation, with a common leading profile $\Omega_1(r)=\kappa/r^3$ and different saturation mechanisms in the inner region, while Model IV introduces a mild angular dependence through $P_2(\chi)$, mimicking a configuration that is slightly more rapidly rotating away from the equatorial belt. This setup is directly analogous to the uniform and mildly differential rotation laws adopted in Hartle-Thorne descriptions of neutron stars \cite{Hartle1967AJ, Hartle1968AJ}. Our goal here is not to single out a unique microphysical rotation law, but to sample qualitatively distinct causal profiles that share the same asymptotic behaviour and to test the robustness of the geometric and energy-condition diagnostics against these choices. At the level of the slow rotation expansion \eqref{OmegaAnsatz}, the specific choices \eqref{rl1}–\eqref{DFluid} enter only through the three radial profiles
$\Omega_{1}(r)$, $\Omega^{(3)}_{0}(r)$, and $\Omega^{(3)}_{2}(r)$. For ease of
reference, these functions are listed for Models I–IV in Table~\ref{coeff}.

\begin{table}[ht]
\caption{Expansion coefficients in \eqref{OmegaAnsatz} for all rotation models considered (rigid and differential), listed to $O(J^{3})$. Here, $D$ stands for 'differential'. Models II and III coincide up to order $J^3$.}
\label{coeff}
\vspace*{1em}
\begin{tabular}{||c|c|c|c|c|c||}
\hline\hline
\text{Model} & $\Omega_1(r)$   & $\Omega^{(3)}_0(r)$ & $\Omega^{(3)}_2(r)$\\ [0.5ex]
\hline\hline
I            &$\kappa/r^3$                   & $0$               & $0$\\
II           &$\kappa/r^3$                    & $-\kappa^3/r^7$  & $0$\\
III          &$\kappa/r^3$                    & $-\kappa^3/r^7$  & $0$\\
IV           &$\kappa/r^3$ & $-\kappa^3/(3r^7)$& $-2\kappa^3/(3r^7)$\\[1ex]
\hline\hline 
\end{tabular}
\end{table}
Projecting the Einstein equation $G_{r\chi}=0$ onto the $\ell=2$ sector and retaining terms up to $O(J^{2})$, one solves algebraically for the quadrupolar piece of the shape function. This yields
\begin{equation}\label{B22}
B_2^{(2)}(r)=2rF(r\dot{\Sigma}-N_2^{(2)}),\quad
F=1+8\pi r^2 p_r,\quad\Sigma=K_2^{(2)}+N_2^{(2)}
\end{equation}
where a dot denotes $d/dr$. Thus $B_{2}^{(2)}$ is completely fixed once $p_r(r)$, $K_{2}^{(2)}(r)$, and $N_{2}^{(2)}(r)$ are known. In deriving the remaining equations, we henceforth implement \eqref{B22}. First, we observe that \eqref{Gtt}-\eqref{Gphiphi} is equivalent to the following algebraic system of equations for the energy density $\widetilde{\rho}$ and the azimuthal pressure $P_\varphi$
\begin{eqnarray}
  \mathfrak{A}\widetilde{\rho}+\mathfrak{B}P_\varphi&=&G_{tt},\label{EQ1}\\
  \mathfrak{C}\widetilde{\rho}+\mathfrak{D}P_\varphi&=&G_{t\varphi},\label{EQ2}\\
  \mathfrak{E}\widetilde{\rho}+\mathfrak{F}P_\varphi&=&G_{\varphi\varphi},\label{EQ3}   
\end{eqnarray}
where the coefficients are defined as
\begin{eqnarray}
\mathfrak{A}&=&\frac{8\pi\left[N^2+r^2(1-\chi^2)K^2\omega(\Omega-\omega))\right]^2}
{N^2-r^2(1-\chi^2)K^2(\Omega-\omega)^2},\quad
\mathfrak{B}=\frac{8\pi r^2(1-\chi^2)K^2 N^2\Omega^2}{N^2-r^2(1-\chi^2)K^2(\Omega-\omega)^2},\\
\mathfrak{C}&=&-\frac{8\pi r^2(1-\chi^2)K^2(\Omega-\omega)\left[N^2+r^2(1-\chi^2)K^2\omega(\Omega-\omega)\right]}{N^2-r^2(1-\chi^2)K^2(\Omega-\omega)^2},\quad
\mathfrak{D}=-\frac{8\pi r^2(1-\chi^2)\Omega K^2 N^2}{N^2-r^2(1-\chi^2)K^2(\Omega-\omega)^2},\\
\mathfrak{E}&=&\frac{8\pi r^4(1-\chi^2 )^2 K^4(\Omega-\omega)^2}{N^2-r^2(1-\chi^2)K^2(\Omega-\omega)^2},\quad
\mathfrak{F}=\frac{8\pi r^2(1-\chi^2)K^2 N^2}{N^2-r^2(1-\chi^2)K^2(\Omega-\omega)^2}.
\end{eqnarray}
The system represented by \eqref{EQ1}-\eqref{EQ2} is consistent if, for example, we can find functions $\alpha = \alpha(r,\chi)$ and $\beta = \beta(r,\chi)$ such that $\alpha\eqref{EQ1} + \beta\eqref{EQ2} = \eqref{EQ3}$. This requirement leads to the following conditions
\begin{eqnarray}
\mathfrak{A}\alpha+\mathfrak{C}\beta&=&\mathfrak{E},\label{sys1}\\
\mathfrak{B}\alpha+\mathfrak{D}\beta&=&\mathfrak{F}\label{sys2}
\end{eqnarray}
along with
\begin{equation}\label{cons1}
\alpha G_{tt}+\beta G_{t\varphi}=G_{\varphi\varphi}.    
\end{equation}
For the system \eqref{sys1}-\eqref{sys2} to admit a unique solution, the determinant must be non-zero, that is
\begin{equation}
\mathfrak{A}\mathfrak{D}-\mathfrak{B}\mathfrak{C}=-\frac{64\pi^2 r^2(1-\chi^2)\Omega K^2 N^2\left[N^2+r^2(1-\chi^2)K^2\omega(\Omega-\omega)\right]}{N^2-r^2(1-\chi^2)K^2(\Omega-\omega)^2}\neq 0.
\end{equation}
When this condition is met, the system yields a unique solution for $\alpha$ and $\beta$, given by 
\begin{equation}\label{alphabeta}
\alpha=-\frac{r^2(1-\chi^2)K^2(\Omega-\omega)}{\Omega\left[N^2+r^2(1-\chi^2)K^2\omega(\Omega-\omega)\right]},\quad
\beta=-\frac{N^2+r^2(1-\chi^2)K^2(\Omega^2-\omega^2)}{\Omega\left[N^2+r^2(1-\chi^2)K^2\omega(\Omega-\omega)\right]}.
\end{equation}
By substituting the expressions for $\alpha$ and $\beta$ from \eqref{alphabeta} into \eqref{cons1}, and expanding in powers of $J$ with the aid of \textsc{Maple}, we obtain the following equations
\begin{eqnarray}
&&rF\ddot{\omega}_1+4\left[1+\pi r^2(7p_r-\rho)\right]\dot{\omega}_1-8\pi r\Delta(p_r-\rho)=0,\quad 
\Delta=\Omega_1-\omega_1,\label{omega1eq}\\
&&5r^2\mathcal{L}(\omega^{(3)}_0)-r^2\mathcal{L}(\omega^{(3)}_2)+H(r)=0,\label{L30}\\
&&7r^2\mathcal{L}(\omega^{(3)}_0)-5r^2\mathcal{L}(\omega^{(3)}_2)+G(r)=0\label{L32}
\end{eqnarray}
with the operator $\mathcal{L}$ defined as
\begin{equation}
\mathcal{L}=2rF\frac{d^2}{dr^2}+8\left[1+\pi r^2(7p_r-\rho)\right]\frac{d}{dr}+16\pi r(p_r-\rho)
\end{equation}
and
\begin{eqnarray}
H(r)&=&2r\left\{r^3 F\dot{\omega}_1\ddot{K}^{(2)}_2-r\left[3rF\dot{\omega}_1-8\Delta(1+\pi r^2(5p_r-\rho))\right]\dot{K}^{(2)}_2-8\Delta K^{(2)}_2\right\}\nonumber\\
&&+2r\left\{r^2F\left(r\dot{\omega}_1-4\Delta\right)\ddot{N}^{(2)}_2+r^2\left[F\dot{\omega}_1-8\pi r\Delta(p_r-\rho)\right]\dot{N}^{(2)}_2 
+16\Delta N^{(2)}_2\right\}\nonumber\\ 
&&+10r^2\left\{2\Delta\left[r F\ddot{N}^{(2)}_0+(1+4\pi r^2(p_r-\rho))\dot{N}^{(2)}_0\right] 
-rF\dot{\omega}_1\dot{N}^{(2)}_0\right\}\nonumber\\
&&-5\left[r(r\dot{\omega}_1-2\Delta)\dot{B}^{(2)}_0+(2r^2\ddot{\omega}_1+7r\dot{\omega}_1-2\Delta)B^{(2)}_0\right]-16\pi r^3(p_r-\rho)(5\Omega^{(3)}_0-\Omega^{(3)}_2)\nonumber\\
&&-16r^5\Delta\left[F\dot{\omega}_1^2-4\pi\Delta^2(p_r-\rho)\right],\\   
G(r)&=&+72r\omega^{(3)}_2+ 2r\left\{5r^3 F\dot{\omega}_1\ddot{K}^{(2)}_2
-r\left[15rF\dot{\omega}_1-4\Delta(1+10\pi r^2(5p_r-\rho))\right]\dot{K}^{(2)}_2-40\Delta K^{(2)}_2\right\}\nonumber\\ 
&&+2r\left\{5r^2F(r\dot{\omega}_1-4\Delta)\ddot{N}^{(2)}_2
+r\left[5rF\dot{\omega}_1-4\Delta(9+10\pi r^2(p_r-\rho))\right]\dot{N}^{(2)}_2+80\Delta N^{(2)}_2\right\}\nonumber\\
&&+14r^2\left\{2\Delta\left[rF\ddot{N}^{(2)}_0+(1+4\pi r^2(p_r-\rho))\dot{N}^{(2)}_0\right]-rF\dot{\omega}_1\dot{N}^{(2)}_0\right\}+16\pi r^3(p_r-\rho)(5\Omega^{(3)}_2-7\Omega^{(3)}_0)\nonumber\\
&&-7\left[r(r\dot{\omega}_1-2\Delta)\dot{B}^{(2)}_0+(2r^2\ddot{\omega}_1+7r\dot{\omega}_1-2\Delta)B^{(2)}_0\right]
-32r^5\Delta\left[F\dot{\omega}_1^2-4\pi\Delta^2(p_r-\rho)\right].   
\end{eqnarray}
For the matter sector, we expand the energy density and the azimuthal pressure in even powers of $J$, retaining the monopole and quadrupole contributions
\begin{eqnarray}
\widetilde{\rho}(r,\chi)&=&\rho(r)+\left[\rho_0^{(2)}(r)+\rho_2^{(2)}(r)P_2(\chi)\right]J^2+\mathcal{O}(J^4),\\
P_\varphi(r,\chi)&=&p_t(r)+\left[P_{0\varphi}^{(2)}(r)+P_{2\varphi}^{(2)}(r)P_2(\chi)\right]J^2+\mathcal{O}(J^4),\\
\end{eqnarray}
where $\rho(r)$ and $p_t(r)$ are the static (spherically symmetric) density and tangential pressure, respectively. From \eqref{EQ1} and \eqref{EQ2}, we find that
\begin{equation}
\widetilde{\rho}=\frac{\mathfrak{D}G_{tt}-\mathfrak{B}G_{t\varphi}}{\mathfrak{A}\mathfrak{D}-\mathfrak{B}\mathfrak{C}},\quad
P_\varphi=\frac{\mathfrak{A}G_{t\varphi}-\mathfrak{C}G_{tt}}{\mathfrak{A}\mathfrak{D}-\mathfrak{B}\mathfrak{C}}.
\end{equation}
After expanding around $J=0$, we find that
\begin{eqnarray}
\rho_0^{(2)}(r)&=&\frac{\dot{B}^{(2)}_0}{8\pi r^2}-\frac{r^2}{48\pi}F\dot{\omega}_1^2 +\frac{r^2}{3}\Delta^2(p_r-\rho),\label{r20}\\    
\rho_2^{(2)}(r)&=&-\frac{r^2\Delta}{3}\left[p_r\Delta-\rho(\Omega_1-2\omega_1)\right]+\frac{F}{4\pi}\ddot{N}^{(2)}_2+
\frac{1-2\pi r^2\rho}{\pi r^2}(r\dot{N}^{(2)}_2-N^{(2)}_2)\nonumber\\
&&+\frac{1}{2\pi r}\left[1-2\pi r^2(rp_r+\rho)\right]\dot{K}^{(2)}_2+\frac{K^{(2)}_2}{2\pi r^2}+\frac{r^2 F}{48\pi}\dot{\omega}_1^2,\label{r22}\\
P^{(2)}_{0\varphi}(r)&=&-\frac{\mathcal{L}(\omega^{(3)}_0)}{32\pi r\Omega_1}
+\frac{\omega^{(3)}_2}{8\pi r^2\Omega_1}+\frac{\omega_1\dot{\Sigma}}{8\pi r\Omega_1}+\frac{\omega_1 F}{8\pi\Omega_1}\ddot{N}^{(2)}_0+ \frac{1}{16\pi r\Omega_1}\left\{rF\dot{\omega}_1+2\left[1+4\pi r^2(p_r-\rho)\right]\omega_1\right\}\dot{N}^{(2)}_0\nonumber\\
&&+\frac{r^2\Delta^2}{3\Omega_1}(p_r-\rho)\omega_1+\frac{r^2 F}{48\pi\Omega_1}(\Omega_1-4\omega_1)\dot{\omega}_1^2+\frac{1}{2\Omega_1}(p_r-\rho)\Omega^{(3)}_0+\frac{1}{32\pi r^2\Omega_1}(r\dot{\omega}_1+2\omega_1-4\Omega_1)\dot{B}^{(2)}_0\nonumber\\
&&
+\frac{1}{32\pi r^3\Omega_1}(2r^2\ddot{\omega}_1+7r\dot{\omega}_1+2\omega_1)B^{(2)}_0,
\end{eqnarray}
\begin{eqnarray}
P^{(2)}_{2\varphi}(r)&=&-\frac{\mathcal{L}(\omega^{(3)}_2)}{32\pi r\Omega_1}+\frac{5\omega^{(3)}_2}{8\pi r^2\Omega_1}+\frac{rF\dot{\omega}_1\ddot{\Sigma}}{16\pi\Omega_1}+\frac{p_r-\rho}{2\Omega_1}\Omega^{(3)}_2-\frac{r^2}{48\pi\Omega_1}\left[F(\Omega_1-4\omega_1)\dot{\omega}_1^2+16\pi\Delta^2(p_r-\rho)\omega_1\right]\\\nonumber
&&-\frac{1}{16\pi r^2\Omega_1}\left\{r\left[3rF\dot{\omega}_1+8\Omega_1(1-8\pi r^2 p_r)-2(1-4\pi r^2(5p_r-\rho))\omega_1\right]\dot{K}^{(2)}_2+8\Delta K^{(2)}_2\right\},\\\nonumber
&&-\frac{1}{16\pi r^2\Omega_1}\left\{4r^2 F\Delta\ddot{N}^{(2)}_2-r^2 F\dot{\omega}_1\dot{N}^{(2)}_2-2r\left[(5+4\pi r^2(p_r-\rho))\omega_1-8\Omega_1(1-\pi r^2(p_r+\rho))\right]\dot{N}^{(2)}_2\right.\\\nonumber
&&\left.+ 16\left[\pi r^2\Omega_1(p_r+\rho)-\Delta\right]N^{(2)}_2\right\}.
\end{eqnarray}
Since \eqref{L30} and \eqref{L32} yield
\begin{eqnarray}
\mathcal{L}(\omega^{(3)}_0)&=&\frac{G(r)-5H(r)}{18r^2}=\frac{4\omega^{(3)}_2}{r}-4\Delta\dot{\Sigma}-2rF\left(2\Delta\ddot{N}^{(2)}_0+\dot{\omega}_1\dot{N}^{(2)}_0\right)-4\Delta\left[1+4\pi r^2(p_r-\rho)\right]\dot{N}^{(2)}_0,\nonumber\\
&&+\left(\dot{\omega}_1-\frac{2\Delta}{r}\right)\dot{B}^{(2)}_0+\frac{1}{r}\left(2r\ddot{\omega}_1+7\dot{\omega}_1-\frac{2\Delta}{r}\right)B^{(2)}_0
+16\pi r(p_r-\rho)\Omega^{(3)}_0\nonumber\\
&&+\frac{8r^2}{3}\Delta\left[F\dot{\omega}_1^2-4\pi\Delta^2(p_r-\rho)\right],\label{L30ode}\\
\mathcal{L}(\omega^{(3)}_2)&=&\frac{5G(r)-7H(r)}{18r^2}=\frac{20\omega^{(3)}_2}{r}+2r^2 F\dot{\omega}_1\ddot{\Sigma}
-6rF\dot{\omega}_1\dot{K}^{(2)}_2-4\Delta\left[1-4\pi r^2(5p_r-\rho)\right]\dot{K}^{(2)}_2-\frac{16\Delta}{r}K^{(2)}_2,\nonumber\\
&&-8rF\Delta\ddot{N}^{(2)}_2+2rF\dot{\omega}_1\dot{N}^{(2)}_2-4\Delta\left[5+4\pi r^2(p_r-\rho)\right]\dot{N}^{(2)}_2+\frac{32\Delta}{r}N^{(2)}_2
+16\pi r(p_r-\rho)\Omega^{(3)}_2\nonumber\\
&&-\frac{8r^3}{3}\Delta\left[F\dot{\omega}_1^2-4\pi\Delta^2(p_r-\rho)\right],\label{L32ode}
\end{eqnarray}
it follows that $P^{(2)}_{0\varphi}(r)$ and $P^{(2)}_{2\varphi}(r)$ can be written as
\begin{eqnarray}
P^{(2)}_{0\varphi}(r)&=&\frac{\dot{\Sigma}}{8\pi r}+\frac{F}{8\pi}\ddot{N}^{(2)}_0+\frac{1+4\pi r^2(p_r-\rho)}{8\pi r}\dot{N}^{(2)}_0-\frac{1}{16\pi r^3}\left(r\dot{B}^{(2)}_0-B^{(2)}_0\right)-\frac{r^2 F}{16\pi}\dot{\omega}_1^2+\frac{r^2\Delta^2}{3}(p_r-\rho),\\
P^{(2)}_{2\varphi}(r)&=&-\frac{3-4\pi r^2(3p_r+\rho)}{8\pi r}\dot{\Sigma}-(p_r+\rho)N^{(2)}_2+\frac{r^2 F}{16\pi}\dot{\omega}_1^2-\frac{r^2\Delta^2}{3}(p_r-\rho).
\end{eqnarray}
Finally, by introducing the following expansions for the remaining pressures,
\begin{eqnarray}
P_r(r,\chi)&=&p_r(r)+\left[P_{0r}^{(2)}(r)+P_{2r}^{(2)}(r)P_2(\chi)\right]J^2+\mathcal{O}(J^4),\label{expPR}\\
P_\chi(r,\chi)&=&p_t(r)+\left[P_{0\chi}^{(2)}(r)+P_{2\chi}^{(2)}(r)P_2(\chi)\right]J^2+\mathcal{O}(J^4),\label{expPchi}
\end{eqnarray}
the unknown radial functions can be determined by first rewriting \eqref{Grr} and \eqref{Gchichi} as
\begin{equation}
P_r=\frac{1}{8\pi}\left(1-\frac{B}{r}\right)G_{rr},\quad
P_\chi=\frac{1-\chi^2}{8\pi r^2 K^2}G_{\chi\chi},
\end{equation}
and then expanding the right-hand sides in powers of $J$. Carrying out this procedure yields
\begin{eqnarray}
P_{0r}^{(2)}(r)&=&\frac{F}{4\pi}\left(\frac{r^2\dot{\omega}_1^2}{12}+\frac{\dot{N}^{(2)}_0}{r}\right)-\frac{B^{(2)}_0}{8\pi r^3},\\
P_{2r}^{(2)}(r)&=&-\frac{r^2 F}{48\pi}\dot{\omega}_1^2-\frac{1}{2\pi r^2}\left[K^{(2)}_2+(1-4\pi r^2 p_r)N^{(2)}_2\right],\\
P_{0\chi}^{(2)}(r)&=&-\frac{r^2 F}{48\pi}\dot{\omega}_1^2-\frac{\dot{\Sigma}}{8\pi r}+\frac{F}{8\pi}\ddot{N}^{(2)}_0+\frac{1+4\pi r^2(p_r-\rho)}{8\pi r}\dot{N}^{(2)}_0-\frac{1}{16\pi r^3}\left(r\dot{B}^{(2)}_0-B^{(2)}_0\right),\\
P_{2\chi}^{(2)}(r)&=&\frac{r^2 F}{48\pi}\dot{\omega}_1^2-\frac{1-4\pi r^2(3p_r+\rho)}{8\pi r}\dot{\Sigma}-\frac{p_r+\rho}{2}(K^{(2)}_2+2N^{(2)}_2).
\end{eqnarray}
Finally, expanding the conservation laws $\nabla_\alpha T^{r\alpha}=0$ and $\nabla_\alpha T^{\chi\alpha}=0$ up to order $J^2$,
\begin{equation}
\nabla_\alpha T^{r\alpha}=\mathcal{T}_0+\mathcal{T}_2 J^2+\mathcal{O}(J^4),\qquad
\nabla_\alpha T^{\chi\alpha}=\mathcal{S}_0+\mathcal{S}_2 J^2+\mathcal{O}(J^4),
\end{equation}
we obtain
\begin{equation}
\mathcal{T}_0=\frac{F}{r}\left[r\dot p_r+2(p_r-p_t)\right]=0,
\end{equation}
by virtue of \eqref{eq4} with zero redshift. At $\mathcal{O}(J^2)$ one finds $\mathcal{T}_2=\mathcal{T}_{20}(r)+\mathcal{T}_{22}(r)P_2(\chi)$ with 
$\mathcal{T}_{20}=0$ identically and
\begin{equation}
\mathcal{T}_{22}(r)=\frac{F}{2r}(p_r+\rho)K^{(2)}_{2},
\end{equation}
hence $\nabla_\alpha T^{r\alpha}=0$ implies $K^{(2)}_{2}=0$. Moreover, we have $\mathcal{S}_0=0$ identically, while
\begin{equation}
\mathcal{S}_2=\left[9P_3(\chi)-4P_1(\chi)\right](p_r+\rho)\frac{K^{(2)}_{2}}{10r^2},
\end{equation}
which vanishes since $K^{(2)}_{2}=0$. At this point, a comment is in order. Within the truncated slow-rotation expansion, the conservation equations $\nabla_\beta T^{\alpha\beta}=0$ serve as integrability/regularity conditions at $\mathcal{O}(J^{2})$. In particular, the $\ell=2$ projection of the radial conservation law is proportional to $(p_{r}+\rho)K^{(2)}_{2}$, implying $K^{(2)}_{2}=0$. In the exact, i.e. untruncated, system, the same constraint is enforced by the remaining Einstein equations together with the contracted Bianchi identity. Enforcing conservation at each order ensures that our truncated solution is consistent with the exact Einstein–matter system to the same accuracy. From a geometric standpoint, $K$ controls the proper area of the $t=const$, $r=const$ two–surfaces through the factor $r^2 K^2$ in the angular metric. With the area gauge $\langle K^2\rangle=1$ already removing the monopolar contribution $K^{(2)}_0$, the conservation law constraint $K^{(2)}_2=0$ shows that the $\ell=2$ (quadrupolar) piece of $K$ is likewise absent at $O(J^2)$. Quadrupolar deformations of the areal radius are therefore suppressed at this perturbative order: constant–$r$ two surfaces remain round spheres of area $4\pi r^2$, and rotation affects the geometry mainly through frame dragging via $\omega$ and through corrections to the shape function $B$ at this order. We conclude by expanding the curvature scalars in powers of $J$. For the Ricci scalar, we obtain
\begin{equation}
R=R_0+\left[R^{(2)}_0+R^{(2)}_2 P_2(\chi)\right]J^2+\mathcal{O}(J^4)    
\end{equation}
with  $R_0=16\pi\rho$ as in the static case, and 
\begin{eqnarray}
R^{(2)}_0&=&-2F\ddot{N}^{(2)}_0-\frac{4}{r}\left[1+2\pi r^2(3p_r-\rho)\right]\dot{N}^{(2)}_0+\frac{2}{r^2}\dot{B}^{(2)}_0+\frac{r^2 F}{3}\dot{\omega}_1^2,\label{R20}\\
R^{(2)}_2&=&-2F\ddot{N}^{(2)}_2-\frac{4}{r}\left[1+2\pi r^2(3p_r-\rho)\right]\dot{N}^{(2)}_2+\frac{12}{r^2}N^{(2)}_2+\frac{2}{r^2}\dot{B}^{(2)}_2+\frac{6B^{(2)}_2}{r^3 F}-\frac{r^2 F}{3}\dot{\omega}_1^2.\label{R22}
\end{eqnarray}
Since $F=1-b/r$ vanishes linearly at the throat, the term $6B^{(2)}_2/r^3 F$ can jeopardize regularity unless $B^{(2)}_2=\mathcal{O}(F)$ near $r=r_0$ or vanishes identically. We therefore adopt $B^{(2)}_2(r) = \mathcal{O}(F(r))$ as an explicit regularity condition at the throat, which guarantees that curvature invariants such as the Ricci scalar and the Kretschmann scalar remain finite at the wormhole throat. Concerning the Kretschmann scalar, we find
\begin{equation}
\mathcal{K}=\mathcal{K}_0+\left[\mathcal{K}^{(2)}_0+\mathcal{K}^{(2)}_2 P_2(\chi)\right]J^2+\mathcal{O}(J^4)    
\end{equation}
with $\mathcal{K}_0$ as in the static case, and
\begin{eqnarray}
\mathcal{K}^{(2)}_0&=&-\frac{4F}{3}\left[5+4\pi r^2(5p_r+\rho)\right]\dot{\omega}_1^2-\frac{256\pi r^2}{3}\Delta^2(p_r-\rho)^2+\frac{64\pi r}{3}F\Delta(p_r-\rho)\dot{\omega}_1\nonumber\\
&+&\frac{32\pi}{r^2}(p_r+\rho)\dot{B}^{(2)}_0-\frac{32\pi}{r^3}(3p_r+\rho)B^{(2)}_0,\label{K20}\\
\mathcal{K}^{(2)}_2&=&-\frac{16F}{3}\left[1-\pi r^2(5p_r+\rho)\right]\dot{\omega}_1^2+\frac{256\pi r^2}{3}\Delta^2(p_r-\rho)^2-\frac{64\pi r}{3}F\Delta(p_r-\rho)\dot{\omega}_1\nonumber\\
&+&\frac{32\pi}{r^2}(p_r+\rho)\dot{B}^{(2)}_2-\frac{64\pi}{r^3}\left[\rho-4\pi r^2(3p_r+\rho)\right]\frac{B^{(2)}_2}{F}.\label{K22}
\end{eqnarray}
Finally, the null energy inequalities take the form \cite{Visser1996, Lobo2005PRD}
\begin{equation}
    \widetilde{\rho}+P_r\geq 0,\quad
    \widetilde{\rho}+P_\chi\geq 0,\quad
    \widetilde{\rho}+P_\varphi\geq 0.
\end{equation}
Their behaviour will be examined and discussed in Sec .~IV. In the explicit solutions presented therein. we find that the $\mathcal{O}(J^2)$ corrections to $\widetilde{\rho}$ and $P_{r,\chi,\varphi}$ never remove the need for exotic matter. At least one of the combinations $\widetilde{\rho}+P_r$, $\widetilde{\rho}+P_\chi$, $\widetilde{\rho}+P_\varphi$ remains negative in a neighbourhood of the throat for all admissible spins. However, these corrections do control how the NEC violation is distributed. For the spatial-Schwarzschild seed, slow rotation generates an $\mathcal{O}(J^2)$ comoving density and anisotropic shifts of the principal pressures, so that the additional NEC violation or mitigation is concentrated mainly near the equatorial plane, while the polar regions remain close to their static values. For the Morris-Thorne seed, the magnitude of the radial NEC violation at the throat decreases with increasing $J$ (most strongly on the equatorial plane), whereas small meridional and azimuthal violations are induced in regions where the static configuration saturates the NEC. In this sense, the rotational backreaction primarily serves as a geometric mechanism that redistributes and mildly softens the exotic stress-energy, rather than as a means of eliminating it.

\section{APPLICATIONS}

We now apply the slow rotation framework to explicit wormhole geometries. Unless otherwise stated, all plots and diagnostics in this section are shown for $0 \le j \le 0.15$, i.e. for spins well inside the conservative regime where the slow rotation expansion remains perturbatively reliable. As a first step, we examine the spatial–Schwarzschild wormhole, and, in the following subsection, turn to the Morris--Thorne configuration.

\subsection{Schwarzschild wormhole: $b(r)=2M$ and $\Phi=0$}

We consider a Schwarzschild wormhole in the sense of \cite{Morris1988AJP}, i.e. a Morris-Thorne traversable wormhole with Schwarzschild spatial geometry and total ADM mass $M$. Such a wormhole is the classical counterpart of the noncommutative geometry-inspired wormhole in the limit of large mass parameter \cite{Nicolini2010CQG, Garattini2009PLB, Batic2025CQG}. Notice that in this geometry we have $\rho=0$ identically, while
\begin{equation}
  p_r(r) = -\frac{r_0}{8\pi r^3}, \quad p_t(r) = \frac{r_0}{16\pi r^3},\quad r_0 = 2M\,.
\end{equation}
So, the stress–energy is purely anisotropic pressure with vanishing density. The ADM mass $M$ is nevertheless fixed by the asymptotic/areal geometry rather than by a localised energy density. Thus, in this model, the mass is geometric, supported by stresses, not a Dirac delta layer. Note also $\rho+p_r=-M/(4\pi r^3)<0$, displaying the expected NEC violation that supports the throat. Let us start with the case $\Omega_{I}(r)=\kappa/r^3$. The general solution of the non-homogeneous equation \eqref{omega1eq} is
\begin{equation}
  \omega_1(r) = c_1\sqrt{1-\frac{r_0}{r}}+c_2\left(8-\frac{4r_0}{r}-\frac{r_0^2}{r^2}\right)+\frac{2\kappa}{5r^3}.  
\end{equation}
Because $F(r_0)=0$, regularity of \eqref{omega1eq} at the throat requires that $\dot{\omega}_1$ remains finite there. This imposes the condition
\begin{equation}
\dot{\omega}_1(r_0)=-\frac{2}{r_0}\left[\frac{2}{r_0^3}-\omega_1(r_0)\right],
\end{equation}
which automatically suppresses the square–root branch of the general solution, yielding $c_1=0$. Enforcing the asymptotic behaviour $\omega_1\to 2/r^3$ as $r\to\infty$ further eliminates the second homogeneous solution ($c_2=0$), and fixes the value of the model parameter to $\kappa=5$, ensuring that the geometry–determined frame dragging reproduces the standard Lense–Thirring falloff. Thus, in this background, the regular solution is
\begin{equation}
\omega_1(r)=\frac{2}{r^3}.    
\end{equation}
This choice of $\kappa$ fixes the overall normalisation of the frame–dragging tail so that the far field metric matches the standard Lense–Thirring form. In particular, it identifies the parameter $J$ in our slow–rotation expansion with the total angular momentum of the wormhole as read off from the asymptotic behaviour of $g_{t\varphi}$.

\begin{itemize}
\item 
{\bf{Case 1}}: $K^{(2)}_2=K^{(2)}_0=N^{(2)}_2=N^{(2)}_0=B^{(2)}_2=B^{(2)}_0=0$.\\ 
Imposing $K^{(2)}_2=K^{(2)}_0=0$ to make the conservation law satisfied up to order $J^2$, adopting $N^{(2)}_2=B^{(2)}_2=0$ (i.e. spherical lapse/shape at $\ell=2$) together with $N^{(2)}_0=B^{(2)}_0=0$ to preserve zero–tidal force and keep the throat fixed so that $B(r,\chi)=b(r)$ at $\mathcal O(J^2)$, we further require that $\omega^{(3)}_0$ and $\omega^{(3)}_2$ together with their derivatives to be finite at the throat. These conditions eliminate the square-root branch and any asymptotically growing homogeneous mode, yielding the regular solutions
\begin{equation}
  \omega^{(3)}_2(r)=\frac{4288}{2145 r_0^2 r^5}+\frac{2412}{715r_0 r^6}-\frac{252}{65r^7},\quad
  \omega^{(3)}_0(r)=-\frac{147456}{25025 r_0^4 r^3}-\frac{18432}{5005 r_0^3 r^4}-\frac{4672}{2145 r_0^2 r^5}-\frac{180}{143 r_0 r^6}+\frac{252}{65 r^7}.
\end{equation}
Note that if we impose $N=K=1$ and $B(r,\chi)=b(r)$ exactly, the off-diagonal equation $G_{r\chi}=0$ forces $\partial_\chi\omega=0$, so $\omega=\omega(r)$. Our construction, however, is perturbative. It enforces spherical geometry through $\mathcal O(J^{2})$, while at $\mathcal O(J^{3})$ nonlinear sources in the $(t\varphi)$ equation generate a small, regular quadrupolar correction $\omega^{(3)}_2(r)P_2(\chi)$. Thus, it would be conceptually incorrect to infer that one may set $N=K=1$ and $B(r,\chi)=b(r)$ at all orders because beyond $\mathcal O(J^{2})$ a mild $\chi$-dependence of $\omega$ is generically induced. Finally, if we introduce the rescalings $j=J/r_0^2$, and $x=r/r_0$, we find
\begin{eqnarray}
\omega_I(x,\chi)&=&\frac{2j}{x^3}+F_I(x,\chi)j^3+\mathcal{O}(j^5),\\
F_I(x,\chi)&=&-\frac{147456}{25025 x^3} - \frac{18432}{5005 x^4} - \frac{4672}{2145 x^5} - \frac{180}{14x^6} + \frac{252}{65 x^7}+\left(\frac{4288}{2145 x^5} + \frac{2412}{715 x^6} - \frac{252}{65 x^7}\right)P_2(\chi).\label{102}
\end{eqnarray}
The corresponding energy density becomes
\begin{equation}\label{SCHM1C1}
\widetilde{\rho}_I(x,\chi)=-\frac{3}{8\pi}\left(-\frac{2}{x^6}+\frac{1}{x^7}\right)\left[1-P_2(\chi)\right]j^2+\mathcal{O}(j^4).
\end{equation}
Although the static spatial-Schwarzschild wormhole already violates the NEC, slow rotation introduces additional NEC-violating contributions at $\mathcal{O}(j^2)$, as can be immediately seen from the result above. Rotation, therefore, acts as a source of exoticity, i.e. for a given geometry, it reduces the amount of exotic matter required in the static sector. In this background, rotation induces a negative density while leaving the ADM mass unchanged. The density attains its maximum magnitude on the equatorial plane and vanishes at the poles. To our knowledge, this explicit result has not been previously reported for the zero-tidal-force Morris–Thorne wormhole. The corresponding principal pressures read
\begin{eqnarray}
P_{r,I}(x,\chi)&=&-\frac{1}{8\pi x^3}+\frac{3}{4\pi}\left(\frac{1}{x^6}-\frac{1}{x^7}\right)\left[1-P_2(\chi)\right]j^2+\mathcal{O}(j^4),\\
P_{\chi,I}(x,\chi)&=&\frac{1}{16\pi x^3}-\frac{3}{4\pi}\left(\frac{1}{x^6}-\frac{1}{x^7}\right)\left[1-P_2(\chi)\right]j^2+\mathcal{O}(j^4),\\
P_{\varphi,I}(x,\chi)&=&\frac{1}{16\pi x^3}-\frac{3}{8\pi}\left(\frac{6}{x^6}-\frac{5}{x^7}\right)\left[1-P_2(\chi)\right]j^2+\mathcal{O}(j^4).  
\end{eqnarray}
The $1-P_2(\chi)$ factor enhances the pressure difference between the equatorial plane and the poles, reflecting the quadrupolar distortion of the rotating geometry. The signs of the corrections imply that rotation partially compensates the tension required to support the throat—consistent with the interpretation that rotation itself acts as a geometric source of exoticity. Together with the negative density shift derived above, these results provide the full leading-order stress–energy structure of the slowly rotating, zero–tidal-force Morris–Thorne wormhole. For Models II and III, recall that they coincide up to $j^3$. The only difference relative to Model I lies in the monopole correction $\omega^{(3)}_0$. All other physical quantities remain unchanged at this order. For this reason, we report only the final expression for the frame–dragging function
\begin{eqnarray}
\omega_{II}(x,\chi)&=&\frac{2j}{x^3}+F_{II}(x,\chi)j^3+\mathcal{O}(j^5),\\
F_{II}(x,\chi)&=&-\frac{1562368}{75075 x^3}-\frac{195296}{15015 x^4} - \frac{6224}{715 x^5} - \frac{80}{13 x^6} 
+\frac{2}{65 x^7}+\left(\frac{4288}{2145 x^5} + \frac{2412}{715 x^6} - \frac{252}{65 x^7}\right)P_2(\chi)\label{110}.
\end{eqnarray}
This expression preserves the same quadrupolar structure as Model I while incorporating the distinct monopole contribution characteristic of the corresponding rotation law. Finally, for the differentially rotating fluid described by \eqref{DFluid}, only the frame–dragging function is modified, while all other physical quantities retain the same form as in the rigidly rotating cases. The resulting expression reads
\begin{eqnarray}
\omega_{IV}(x,\chi)&=&\frac{2j}{x^3}+F_{IV}(x,\chi)j^3+\mathcal{O}(j^5),\\
F_{IV}(x,\chi)&=&-\frac{666368}{75075 x^3} - \frac{83296}{15015 x^4} - \frac{24016}{6435 x^5} - \frac{380}{143 x^6} 
+ \frac{506}{195 x^7}\nonumber\\
&+&\left(\frac{4864}{6435 x^5} + \frac{912}{715 x^6} - \frac{1256}{195 x^7}\right)P_2(\chi).
\end{eqnarray}
\item 
{\bf{Case 2}}: $K^{(2)}_2=K^{(2)}_0=N^{(2)}_2=N^{(2)}_0=0$.\\ 
An interesting scenario consists in enforcing $\rho^{(2)}_0=\rho^{(2)}_2=0$. Imposing $\rho^{(2)}_0=0$ yields the first–order ODE
\begin{equation}
\dot{B}^{(2)}_0=\frac{6}{r^4}\left(1-\frac{r_0}{r}\right)+\frac{3r_0}{r^5},
\end{equation}
whose solution with the correct asymptotic decay is
\begin{equation}
B^{(2)}_0(r)=-\frac{2}{r^3}+\frac{3r_0}{4r^4}.
\end{equation}
By contrast, demanding $\rho^{(2)}_2=0$ leads to the second–order, nonhomogeneous equation
\begin{equation}
\left(1-\frac{r_0}{r}\right)\ddot{N}^{(2)}_2(r)+\frac{4}{r}\dot{N}^{(2)}_2(r)-\frac{4}{r^2}N^{(2)}_2(r)=-\frac{3}{r^6}\left(1-\frac{r_0}{r}\right)-\frac{3r_0}{2r^7},
\end{equation}
subject to the throat regularity condition
\begin{equation}\label{rcN22}
\dot{N}^{(2)}_2(r_0)=\frac{N^{(2)}_2(r_0)}{r_0}-\frac{3}{8r_0^5}.  
\end{equation}
The general solution can be written as
\begin{equation}
 N^{(2)}_2(r)=c_1 r+ c_2 h(r)+g(r)   
\end{equation}
where the functions $h(r)$ and $g(r)$ behave near the throat as
\begin{equation}
h(r)=\frac{r_0^4}{(r-r_0)^3}+\mathcal{O}((r-r_0)^{-2}),\quad
g(r)=\frac{1}{r_0}\left(\frac{47}{24}+\ln{r_0}\right)\frac{1}{(r-r_0)^3}+\mathcal{O}((r-r_0)^{-2}).
\end{equation}
Consequently, $N^{(2)}_2(r)$ either diverges at spatial infinity or develops a pole at the throat, violating the regularity condition \eqref{rcN22}. We thus conclude that the constraint $\rho^{(2)}_2=0$ is not physically admissible, and the consistent choice is $N^{(2)}_2(r)=0$. This approach nevertheless shows that slowly rotating wormholes can support a monopolar shape correction $B^{(2)}_0(r)\neq0$. In terms of the dimensionless variables $x=r/r_0$ and $j=J/r_0^2$, it reads
\begin{equation}\label{shapeB}
B(x)=1-\left(\frac{2}{x^3}-\frac{3}{4x^4}\right)j^2+\mathcal{O}(j^4).    
\end{equation}
Rotation, therefore, shifts the effective throat position inward, while the rapid decay of $B^{(2)}_0(x)$ at infinity ensures that the ADM mass remains unchanged. Equation \eqref{shapeB} also quantifies the backreaction on the throat location. Using the general free-throat expression \eqref{deltarchi} and the fact that $b'(r_0)=0$ for the spatial–Schwarzschild seed, one finds $|\delta r|/r_0 \simeq \tfrac{5}{4} j^2$ in this case. Thus the relative shift of the throat is $\lesssim 3\%$ for $j=0.15$ and would reach only $\sim 5\%$ for
$j\simeq0.2$, remaining safely in the perturbative regime. Values $j\gtrsim 0.3$ would correspond to displacements of order $10\%$ or more, at which point higher-order terms in the slow-rotation expansion are expected to become quantitatively important. Finally, it is worth noting that \eqref{shapeB} is independent of the specific choice of the fluid’s angular velocity profile.  In this case as well, rotation induces a nonvanishing comoving density,
\begin{equation}
    \widetilde{\rho}(x)=\frac{3}{4\pi}\left(\frac{1}{x^6}-\frac{1}{2x^7}\right)j^2+\mathcal{O}(j^4),
\end{equation}
even though the static configuration has $\rho = 0$. The rotational correction is positive and reaches its maximum near the throat at $x_m = 7/6$, illustrating how slow rotation generates an effective matter distribution in an otherwise vacuum wormhole background. If we introduce the rescalings $j=J/r_0^2$, and $x=r/r_0$, we find
\begin{eqnarray}
\widehat{\omega}_I(x,\chi)&=&\frac{2j}{x^3}+\widehat{F}_I(x,\chi)j^3+\mathcal{O}(j^5),\\
\widehat{F}_I(x,\chi)&=&-\frac{76416}{25025 x^3} - \frac{9552}{5005 x^4} - \frac{2008}{2145 x^5} - \frac{18}{55 x^6} + \frac{216}{65 x^7}+\left(\frac{4288}{2145 x^5} + \frac{2412}{715 x^6} - \frac{252}{65 x^7}\right)P_2(\chi).
\end{eqnarray}
and the corresponding principal pressures read 
\begin{eqnarray}
\widehat{P}_{r,I}(x,\chi)&=&-\frac{1}{8\pi x^3}+\frac{1}{\pi}\left[\frac{1}{x^6}-\frac{27}{32 x^7}+\left(-\frac{3}{4x^6}+\frac{3}{4x^7}\right)P_2(\chi)\right]j^2+\mathcal{O}(j^4),\\
\widehat{P}_{\chi,I}(x,\chi)&=&\frac{1}{16\pi x^3}+\frac{1}{\pi}\left[-\frac{5}{4x^6}+\frac{63}{64 x^7}+\left(\frac{3}{4x^6}-\frac{3}{4x^7}\right)P_2(\chi)\right]j^2+\mathcal{O}(j^4),\\
\widehat{P}_{\varphi,I}(x,\chi)&=&\frac{1}{16\pi x^3}+\frac{1}{\pi}\left[-\frac{11}{4x^6}+\frac{135}{64 x^7}+\left(\frac{9}{4x^6}-\frac{15}{8x^7}\right)P_2(\chi)\right]j^2+\mathcal{O}(j^4).  
\end{eqnarray}
For Model II, the density and all pressure components remain unchanged relative to Model I. The only modification appears in the frame–dragging function, which now takes the form
\begin{eqnarray}
\widehat{\omega}_{II}(x,\chi)&=&\frac{2j}{x^3}+\widehat{F}_{II}(x,\chi)j^3+\mathcal{O}(j^5),\\
\widehat{F}_{II}(x,\chi)&=&-\frac{1349248}{75075 x^3} - \frac{168656}{15015 x^4} - \frac{5336}{715 x^5} - \frac{3734}{715 x^6} - \frac{34}{65 x^7}+\left(\frac{4288}{2145 x^5} + \frac{2412}{715 x^6} - \frac{252}{65 x^7}\right)P_2(\chi).
\end{eqnarray}
Finally, for Model IV, we obtain
\begin{eqnarray}
\widehat{\omega}_{IV}(x,\chi)&=&\frac{2j}{x^3}+\widehat{F}_{IV}(x,\chi)j^3+\mathcal{O}(j^5),\\
\widehat{F}_{IV}(x,\chi)&=&-\frac{453248}{75075 x^3} - \frac{56656}{15015 x^4} - \frac{16024}{6435 x^5} - \frac{1234}{715 x^6} + \frac{398}{195 x^7}\nonumber\\
&+&\left(\frac{4864}{6435 x^5} + \frac{912}{715 x^6} - \frac{1256}{195 x^7}\right)P_2(\chi).
\end{eqnarray}
\item 
{\bf{Case 3}}: $K^{(2)}_2=K^{(2)}_0=0$ and $N^{(2)}_2=\beta/r^4$, $N^{(2)}_0=\alpha/r^4$.\\ 
Teo \cite{Teo1998PRD} considered an ansatz of the form $N=1+(16J^{2}/r)\cos^2{\theta}$ (see his equation (26)). In geometrized units $G=c=1$, the lapse function  $N$ is dimensionless, while the angular momentum $J$ has dimensions of length squared, i.e. $[J]=L^2$, where $L$ denotes a unit of length. Consequently, the term $J^2/r$ carries dimension $L^3$, and therefore, it is not dimensionally consistent. A dimensionally correct and even in $J$, equatorially symmetric choice is instead
\begin{equation}
N(r,\chi)=1+\frac{1}{r^4}\left[\alpha+\beta P_{2}(\chi)\right]J^2+\mathcal{O}(J^{4}),
\qquad \chi=\cos{\theta},
\end{equation}
or, if one wishes to preserve the zero–tidal property at the monopole level,
\begin{equation}
N(r,\chi)=1+\frac{\beta}{r^{4}} P_{2}(\chi)J^{2}+\mathcal{O}(J^{4}).
\end{equation}
In our notation this corresponds to $N^{(2)}_{0}(r)=0$ and $N^{(2)}_{2}(r)=\beta/r^{4}$. Such a choice respects parity, axis regularity, and asymptotic flatness while preserving all dimensions.
\begin{itemize}
\item
{\bf{Case}} $B^{(2)}_0=0$: In this scenario, we immediately find that the shape function, expressed in the rescaled variables $x=r/r_0$ and $j=J/r_0^2$, takes the form
\begin{equation}
B(x,\chi)=1-\frac{10\beta}{x^3}\left(1-\frac{1}{x}\right)P_2(\chi)j^2+\mathcal{O}(j^4).    
\end{equation}
Solving $B(x,\chi)=x$ yields an additional root at $x_1 = \sqrt[4]{5 j \beta (1-3 \chi^2)}$. However, in the static limit ($j\to 0$), the only throat occurs at $x_0=1$, and since $x_1 \rightarrow 0$ as $j\to 0$, this extra root does not represent a continuous deformation of the original throat. It is therefore a spurious solution introduced by truncating the expansion at $\mathcal{O}(j^4)$. This conclusion is independent of the chosen fluid rotation law. Turning now to the comoving density, we find that
\begin{equation}
\widetilde{\rho}(x,\chi)=-\frac{1}{\pi}\left[-\frac{3}{4x^6}+\frac{3}{8x^7}+\left(\frac{3}{4x^6}-\frac{3+40\beta}{8x^7}\right)P_2(\chi)\right]J^{2}+\mathcal{O}(J^{4}).   
\end{equation}
Notice that in the limit $\beta\to 0$, the above expression correctly reproduces the corresponding energy density given in \eqref{SCHM1C1}. Within the framework of Model I, the frame-dragging function takes the form
\begin{eqnarray}
\widetilde{\omega}_{I}(x,\chi)&=&\frac{2j}{x^3}+\widetilde{F}_{I}(x,\chi)j^3+\mathcal{O}(j^5),\\
\widetilde{F}_{I}(x,\chi)&=&-\frac{256(5\alpha+24\beta+576)}{25025 x^3}-\frac{32(5\alpha+24\beta+576)}{5005 x^4}-\frac{16(9\alpha+68\beta+876)}{6435 x^5}
-\frac{12(\alpha+11\beta+75)}{715 x^6}\nonumber\\
&-&\frac{84(2\alpha-3)}{65 x^7}
+\left[-\frac{36(18\beta+7)}{65 x^7}-\frac{12(31\beta-201)}{715x^6}-\frac{64(31\beta-201)}{6435 x^5}\right]P_2(\chi).
\end{eqnarray}
The corresponding principal pressures are then found to be
\begin{eqnarray}
\widetilde{P}_{r,I}(x,\chi)&=&-\frac{1}{8\pi x^3}+\frac{1}{4\pi}\left[\frac{3-4\alpha}{x^6}-\frac{3-4\alpha}{x^7}+\left(-\frac{2\beta+3}{x^6}-\frac{\beta-3}{x^7}\right)P_2(\chi)\right]j^2+\mathcal{O}(j^4),\\
\widetilde{P}_{\chi,I}(x,\chi)&=&\frac{1}{16\pi x^3}+\frac{1}{4\pi}\left[\frac{8\alpha+2\beta-3}{x^6}-\frac{3(3\alpha - 1)}{x^7}
+\left(\frac{2\beta+3}{x^6}+\frac{7\beta - 6}{2x^7}\right)P_2(\chi)\right]j^2+\mathcal{O}(j^4),\\
\widetilde{P}_{\varphi,I}(x,\chi)&=&\frac{1}{16\pi x^3}+\frac{1}{4\pi}\left\{\frac{8\alpha-2\beta-9}{x^6}-\frac{3(6\alpha-5)}{2x^7}\right.\\
&+&\left.\left[\frac{3(2\beta+3)}{x^6}+\frac{7\beta - 15}{2x^7}\right]P_2(\chi)\right\}j^2+\mathcal{O}(j^4).  
\end{eqnarray}
Notice that the parameters $\alpha$ and $\beta$ quantify the rotational backreaction on the stress–energy tensor: $\alpha$ modulates the overall isotropic (monopolar) correction, while $\beta$ governs the angular dependence through the $P_2(\chi)$ sector, introducing a quadrupolar pressure anisotropy characteristic of the rotating configuration. For Model II, we obtain
\begin{eqnarray}
\widetilde{\omega}_{II}(x,\chi)&=&\frac{2j}{x^3}+\widetilde{F}_{II}(x,\chi)j^3+\mathcal{O}(j^5),\\
\widetilde{F}_{II}(x,\chi)&=&-\frac{256(15\alpha+72\beta+6103)}{75075 x^3}-\frac{32(15\alpha+72\beta+6103)}{15015 x^4}-\frac{16(9\alpha+68\beta+3501)}{6435x^5}\nonumber\\
&-&\frac{4(3\alpha+33\beta+1100)}{715 x^6}-\frac{2(84\alpha-1)}{65 x^7}-\frac{84(2\alpha-3)}{65 x^7}\nonumber\\
&+&\left[-\frac{36(18\beta+7)}{65 x^7}-\frac{12(31\beta-201)}{715x^6}-\frac{64(31\beta-201)}{6435 x^5}\right]P_2(\chi),\label{141}
\end{eqnarray}
while for Model IV, the frame–dragging function reads
\begin{eqnarray}
\widetilde{\omega}_{IV}(x,\chi)&=&\frac{2j}{x^3}+\widetilde{F}_{IV}(x,\chi)j^3+\mathcal{O}(j^5),\\
\widetilde{F}_{IV}(x,\chi)&=&-\frac{256(15\alpha+72\beta+2603)}{75075 x^3}-\frac{32(15\alpha+72\beta+2603)}{15015 x^4}-\frac{16(9\alpha+68\beta+1501)}{6435 x^5}\nonumber\\
&-&\frac{4(3\alpha+33\beta+475)}{715x^6}-\frac{2(252\alpha - 253)}{195x^7}\nonumber\\
&+&\left[-\frac{64(31\beta-76)}{6435 x^5}-\frac{12(31\beta-76)}{715 x^6}-\frac{8(243\beta+157)}{195x^7}\right]P_2(\chi).
\end{eqnarray}
In both cases, the density and principal pressures coincide with those obtained for Model I.
\item
{\bf{Case}} $B^{(2)}_0\neq 0$: In this scenario, $B^{(2)}_2$ is evaluated in the same way as in the previous case. To determine $B^{(2)}_0$, we impose the condition $\rho^{(2)}_0=0$, which yields
\begin{equation}
    B^{(2)}_0(r)=-\frac{2}{r^3}+\frac{3r_0}{4r^4}.
\end{equation}
Introducing the dimensionless variables $x=r/r_0$ and $j=J/r_0^2$, the corresponding shape function becomes
\begin{equation}\label{145}
B(x,\chi)=1+\left[-\frac{2}{x^3}+\frac{3}{4x^4}-\frac{10\beta}{x^3}\left(1-\frac{1}{x}\right)P_2(\chi)\right]j^2+\mathcal{O}(j^4),
\end{equation}
while the comoving energy density is
\begin{equation}
\widetilde{\rho}(x,\chi)=\frac{1}{4\pi}\left(\frac{3}{x^6}-\frac{40\beta+3}{2x^7}\right)P_2(\chi)j^2+\mathcal{O}(j^4).    
\end{equation}
Within the framework of Model I, the frame-dragging function takes the form
\begin{eqnarray}
\widetilde{\omega}^{I}(x,\chi)&=&\frac{2j}{x^3}+\widetilde{F}^{I}(x,\chi)j^3+\mathcal{O}(j^5),\\
\widetilde{F}^{I}(x,\chi)&=&-\frac{128(10\alpha+48\beta+597)}{25025 x^3}-\frac{16(10\alpha+48\beta+597)}{5005x^4}-\frac{8(18\alpha+136\beta+753)}{6435x^5} 
-\frac{6(2\alpha+22\beta+39)}{715x^6}\nonumber\\
&-&\frac{24(7\alpha-9)}{65x^7}
+\left[-\frac{64(31\beta-201)}{6435x^5}-\frac{12(31\beta-201)}{715 x^6}-\frac{36(18\beta+7)}{65 x^7}\right]P_2(\chi).
\end{eqnarray}
The corresponding principal pressures are then found to be
\begin{eqnarray}
\widetilde{P}_r^{I}(x,\chi)&=&-\frac{1}{8\pi x^3}+\frac{1}{\pi}\left[\frac{1-\alpha}{x^6}+\frac{32\alpha-27}{32x^7}
+\left(-\frac{2\beta+3}{4x^6}+\frac{3-\beta}{4x^7}\right)P_2(\chi)\right]j^2+\mathcal{O}(j^4),\\
\widetilde{P}_{\chi}^{I}(x,\chi)&=&\frac{1}{16\pi x^3}+\frac{1}{4\pi}\left[\frac{8\alpha+2\beta-5}{x^6}-\frac{9(16\alpha-7)}{16x^7}
+\left(\frac{2\beta+3}{x^6}+\frac{7\beta - 6}{2x^7}\right)P_2(\chi)\right]j^2+\mathcal{O}(j^4),\\
\widetilde{P}_{\varphi}^{I}(x,\chi)&=&\frac{1}{16\pi x^3}+\frac{1}{4\pi}\left\{\frac{8\alpha-2\beta-11}{x^6}-\frac{9(16\alpha-15)}{16x^7}\right.\\
&+&\left.\left[\frac{3(2\beta+3)}{x^6}+\frac{7\beta-15}{2x^7}\right]P_2(\chi)\right\}j^2+\mathcal{O}(j^4).  
\end{eqnarray}

Finally, for Model II, we obtain
\begin{eqnarray}
\widetilde{\omega}^{II}(x,\chi)&=&\frac{2j}{x^3}+\widetilde{F}^{II}(x,\chi)j^3+\mathcal{O}(j^5),\\
\widetilde{F}^{II}(x,\chi)&=&-\frac{128(30\alpha+144\beta+10541)}{75075 x^3}-\frac{16(30\alpha+144\beta+10541)}{15015 x^4}-\frac{8(18\alpha+136\beta + 6003)}{6435 x^5}\nonumber\\
&-&\frac{2(6\alpha+66\beta+1867)}{715x^6}-\frac{2(84\alpha+17)}{65x^7}\nonumber\\
&+&\left[-\frac{64(31\beta-201)}{6435x^5}-\frac{12(31\beta-201)}{715 x^6}-\frac{36(18\beta+7)}{65 x^7}\right]P_2(\chi),
\end{eqnarray}
while for Model IV, the frame–dragging function reads
\begin{eqnarray}
\widetilde{\omega}^{IV}(x,\chi)&=&\frac{2j}{x^3}+\widetilde{F}^{IV}(x,\chi)j^3+\mathcal{O}(j^5),\\
\widetilde{F}^{IV}(x,\chi)&=&-\frac{128(30\alpha+144\beta+3541)}{75075x^3}-\frac{16(30\alpha+144\beta+3541)}{15015x^4}-\frac{8(18\alpha+136\beta+2003)}{6435x^5}\nonumber\\
&-&\frac{2(6\alpha+66\beta+617)}{715x^6}-\frac{2(252\alpha-199)}{195x^7}\nonumber\\
&+&\left[-\frac{64(31\beta-76)}{6435 x^5}-\frac{12(31\beta-76)}{715x^6}-\frac{8(243\beta+157)}{195x^7}\right]P_2(\chi).
\end{eqnarray}
\end{itemize}
\end{itemize}

\subsection{Ellis-Bronnikov alias Morris-Thorne wormhole: $b(r)=r_0^2/r$ and $\Phi=0$}

We consider the Morris–Thorne (Ellis–Bronnikov) wormhole \cite{Ellis1973JMP, Ellis1974JMP, Ellis1979GRG, Bronnikov1973APP, Morris1988AJP}, characterised by the shape function $b(r)=r_0^2/r$ where $r_0$ denotes the position of the throat, and by a vanishing redshift function $\Phi$. In this geometry, the comoving energy density and principal pressures are
\begin{equation}
\rho(r)=-\frac{r_0^2}{8\pi r^4},\quad
p_r(r)=-\frac{r_0^2}{8\pi r^4},\quad p_t(r)=\frac{r_0^2}{8\pi r^4}.
\end{equation}
This geometry is massless because the ADM mass vanishes. Regarding the NEC one has $\rho+p_r=-r_0^2/(4\pi r^4)<0$ while $\rho+p_t=0$. Independent of the fluid rotation law, the Morris–Thorne background satisfies $\rho-p_r=0$. Consequently, the equation for $\omega_1$ (see \eqref{omega1eq}) reduces to a homogeneous ODE whose coefficients are fixed entirely by the static matter content. In particular, the solution structure and its regularity properties are determined without reference to the specific velocity profile $\Omega_1$.  At this point, it is convenient to introduce the proper radial distance $\ell$ in place of the areal coordinate $r$. Notice that the coordinate $\ell$ remains finite and smooth across the throat, thus providing a regular chart that continuously covers both asymptotic regions without coordinate singularities. In terms of $\ell$, equation \eqref{omega1eq} takes the simpler and singularity-free form
\begin{equation}
\omega_1^{''}(\ell)+\frac{4\ell}{\ell^2+r_0^2}\omega_1^{'}(\ell)=0,    
\end{equation}
which immediately implies $\omega^{''}_1(0)=0$ at the throat. Here, the prime denotes differentiation with respect to $\ell$. Its general solution is
\begin{equation}
\omega_1(\ell)=c_1+c_2\left[\frac{\ell}{2r_0^2(\ell^2+r_0^2)}+\frac{1}{2r_0^3}\arctan{\left(\frac{\ell}{r_0}\right)}\right], 
\end{equation}
with asymptotic expansion
\begin{equation}
  \omega_1(\ell)=c_1+\frac{\pi c_2}{4r_0^3}-\frac{c_2}{3\ell^3}+\mathcal{O}\left(\frac{1}{\ell^5}\right).    
\end{equation}
Requiring that $\omega_1\to 2/\ell^3$ as $\ell\to\infty$  fixes the integration constants to $c_1=3\pi/(2r_0^3)$ and $c_2=-6$. Hence, we end up with the solution
\begin{equation}\label{omega1MT}
  \omega_1(\ell)=\frac{3\pi}{2r_0^3}-\frac{6}{2r_0^2}\left[\frac{\ell}{\ell^2+r_0^2}+\frac{1}{r_0}\arctan{\left(\frac{\ell}{r_0}\right)}\right].
\end{equation}
A few remarks are in order. First, the choice of integration constants can be imposed on only one asymptotic end. To render both ends asymptotically Minkowski, one must introduce two coordinate charts, $(t,\ell_+,\chi,\varphi_+)$ and $(t,\ell_-,\chi,\varphi_-)$ related by a rigid rotation 
\begin{equation}
\varphi_{-}=\varphi_+-\frac{3\pi J}{r_0^3}t.   
\end{equation}
In what follows, we restrict attention to the upper asymptotic region with $0\leq\ell<\infty$. Finally, differentiating \eqref{omega1MT} gives $\omega_1^{''}(\ell)=24\ell/(\ell^2+r_0^2)^3$ which satisfies $\omega^{''}_1(0)=0$ as expected.
\begin{itemize}
\item 
{\bf{Case 1}}: $K^{(2)}_2=K^{(2)}_0=N^{(2)}_2=N^{(2)}_0=B^{(2)}_2=B^{(2)}_0=0$.\\ 
In this scenario \eqref{L30ode} and \eqref{L32ode} become
\begin{eqnarray}
&&{\omega^{(3)}_2}^{''}(\ell)+\frac{4\ell}{\ell^2+r_0^2}{\omega^{(3)}_2}^{'}(\ell)-\frac{10}{\ell^2+r_0^2}\omega^{(3)}_2(\ell)
=-\frac{4}{3}(\ell^2+r_0^2)\left[\Omega_1(\ell)-\omega_1(\ell)\right](\omega_1^{'}(\ell))^2,\label{odeomega32}\\
&&{\omega^{(3)}_0}^{''}(\ell)+\frac{4\ell}{\ell^2+r_0^2}{\omega^{(3)}_0}^{'}(\ell)=\frac{2}{\ell^2+r_0^2}\omega^{(3)}_2(\ell)
+\frac{4}{3}(\ell^2+r_0^2)\left[\Omega_1(\ell)-\omega_1(\ell)\right](\omega_1^{'}(\ell))^2.\label{odeomega30}
\end{eqnarray}
Note that, in the present background, all terms proportional to $p_r-\rho$ vanish. Therefore, the ODEs for $\omega^{(3)}_0$ and $\omega^{(3)}_2$ are independent of $\Omega^{(3)}_0$ and $\Omega^{(3)}_2$, since those quantities appear multiplied by $p_r-\rho$ in \eqref{L30ode} and \eqref{L32ode}. Let us consider the subdominant case where $\Omega_1=0$, \emph{i.e.} $\Omega$ can contribute only through $\mathcal{O}(J^3)$. If one also imposes $\Omega^{(3)}_0=\Omega^{(3)}_2=0$, then to cubic order the frame dragging is entirely determined by the geometry and not by the fluid’s intrinsic rotation. If we introduce the rescaled variables $x=\ell/r_0$ and $j=J/r_0^2$, the general solution to \eqref{odeomega32} reads
    \begin{equation}\label{omega32gen}
    \omega^{(3)}_2(x)=c_1 h_1(x)+c_2 h_2(x)+h_p(x), 
    \end{equation}
    where
    \begin{eqnarray}
    h_1(x)&=&3h_2(x)\arctan{x}+\frac{x(15x^2+13)}{x^2+1},\quad h_2(x)=5x^2+1,\\
    h_p(x)&=&\sum_{n=0}^3\mathfrak{f}_n(x)\arctan^n{x}
    \end{eqnarray}
    with
    \begin{eqnarray}
     \mathfrak{f}_0(x)&=&\frac{9}{4(x^2+1)}\left[2\pi(75x^4+60x^2-1)-\frac{\ell(45x^2+49)}{x^2+1}h_2(x)\right],\\
     \mathfrak{f}_1(x)&=&-\frac{234}{x^2+1}\left[\frac{\pi x}{13}(15x^2+13)+\frac{225}{104}x^4+\frac{75}{52}x^2- \frac{19}{104}\right],\\
     \mathfrak{f}_2(x)&=&-27\left[\pi h_2(x)-\frac{2x}{3}\frac{15x^2+13}{x^2+1}\right],\quad
     \mathfrak{f}_3(x)=18h_2(x).
    \end{eqnarray}
    An asymptotic expansion of \eqref{omega32gen} yields
    \begin{eqnarray}
    \omega^{(3)}_2(x)&=&\mathfrak{A}x^2+\mathfrak{B}+\frac{\mathfrak{C}}{x^5}+\mathcal{O}\left(\frac{1}{x^7}\right),\quad
    \mathfrak{A}=\frac{15\pi}{2}c_1+5c_2-\frac{45\pi^3}{2}+\frac{675\pi}{8},\\
    \mathfrak{B}&=&\frac{3\pi}{2}c_1+c_2+\frac{135\pi}{8}-\frac{9\pi^3}{2},\quad
    \mathfrak{C}=\frac{8}{35}(9\pi^2-90-2c_1 ).
    \end{eqnarray}
    Imposing asymptotic flatness by setting $\mathfrak{A}=0=\mathfrak{B}$ gives the relations
    \begin{equation}
    c_1=c_1,\quad    
    c_2=-\frac{3}{2}\pi c_1+\frac{9\pi}{8}(4\pi^2-15).
    \end{equation}
    Since the regularity condition at the throat, obtained directly from \eqref{odeomega32}
    \begin{equation}\label{regomega32}
    \left.\frac{d^2\omega^{(3)}_2}{dx^2}\right|_{x=0}-10\left.\frac{d\omega^{(3)}_2}{dx}\right|_{x=0}
    =\frac{4}{3}\omega_1(0)\left(\left.\frac{d\omega_1}{dx}\right|_{x=0}\right)^2,    
    \end{equation}
    is automatically satisfied for any value of $c_1$, we may set $c_1=0$ without loss of generality. With this choice, the final expression for $\omega^{(3)}_2(x)$ reads
    \begin{equation}\label{omega32ell}
    \omega^{(3)}_2(x)=\frac{9\pi}{8}(4\pi^2-15)(5x^2+1)+\sum_{n=0}^3\mathfrak{f}_n(x)\arctan^n{x}.
    \end{equation}
    Integrating \eqref{odeomega30} symbolically using \textsc{Maple}, we obtain the general solution
    \begin{equation}\label{sol30}
    \omega^{(3)}_0(x)=d_1 g_1(x)+d_2+g_p(x), 
    \end{equation}
    where
    \begin{equation}\label{gpl}
    g_1(x)=\frac{x}{2(x^2+1)}+\frac{1}{2}\arctan{x},\quad
    g_p(x)=\sum_{n=0}^3\mathfrak{h}_n(x)\arctan^n{x}
    \end{equation}
    with
    \begin{eqnarray}
     \mathfrak{h}_0(x)&=&\frac{9\pi(4\pi^2+45)}{8}x^2-\frac{405}{4}x+\frac{9x(5x^2+1)}{2(x^2+1)^2}
     -\frac{9(x+\pi)}{x^2+1},\\
     \mathfrak{h}_1(x)&=&-\frac{9\left[8\pi x(3x^2+5)+45x^4+38x^2-15\right]}{4(x^2+1)},\\
     \mathfrak{h}_2(x)&=&-9\left[3\pi(x^2+1)-\frac{2x(3x^2+5)}{x^2+1}\right],\quad
     \mathfrak{h}_3(x)=18(x^2+1).\label{h3}
    \end{eqnarray}
    An asymptotic expansion of \eqref{sol30} gives
    \begin{equation}
    \omega^{(3)}_0(x)=\mathfrak{D}+\frac{\mathfrak{E}}{x^3}+\mathcal{O}\left(\frac{1}{x^5}\right),\quad
    \mathfrak{D}=\frac{2\pi d_1+8d_2-36\pi^3+63\pi}{8},\quad
    \mathfrak{E}=-\frac{5d_1-108\pi^2+360}{15}.
    \end{equation}
    Imposing asymptotic flatness, i.e. $\mathfrak{D}=0$, and regularity at the throat, that is
    \begin{equation}\label{regomega30}
    \left.\frac{d^2\omega^{(3)}_0}{dx^2}\right|_{x=0}=2\omega^{(3)}_0(0)
    -\frac{4}{3}\omega_1(0)\left(\left.\frac{d\omega_1}{dx}\right|_{x=0}\right)^2,    
    \end{equation}
    yield $d_1=18$ and $d_2=\pi(36\pi^2-99)/8$. Hence, the final expression for $\omega^{(3)}_0(x)$ reads
    \begin{equation}
        \omega^{(3)}_0(x)=\frac{9\pi}{8}(4\pi^2-11)+9\left[\frac{x}{x^2+1}+\arctan{x}\right]+g_p(x),
    \end{equation}
    where $g_p(x)$ can be retrieved from \eqref{gpl}-\eqref{h3}. The remaining physical quantities are as follows. For the energy density, we obtain
    \begin{eqnarray}
    \rho(x)&=&-\frac{1}{8\pi(x^2+1)^2},\quad\rho^{(2)}_0(x)=-\frac{3}{4\pi(x^2+2)^3},\quad
    \rho^{(2)}_2(x)=\sum_{n=0}^2\mathfrak{c}_n\arctan^n{x},\label{rhocase1}\\
    \mathfrak{c}_0&=&\frac{3}{4\pi(x^2+1)^3}\left[1-\frac{1}{8}(\pi x^2+\pi -2x)^2\right],\quad
    \mathfrak{c}_1=\frac{3\pi(x^2+1)-6x}{8\pi(x^2+1)^2},\quad
    \mathfrak{c}_2=-\frac{3}{8\pi(x^2+1)}.
    \end{eqnarray}
    Moreover, the proper radial pressure can be immediately retrieved from 
    \begin{equation}
    p_r(x)=-\frac{1}{8\pi(x^2+1)^2},\quad
    P^{(2)}_{0r}(x)=-P^{(2)}_{2r}(x)=\frac{3}{4\pi(x^2+1)^3}.
    \end{equation}
    The components of the tangential pressure are
    \begin{equation}
    p_t(x)=\frac{1}{8\pi(x^2+1)^2},\quad
    P^{(2)}_{0\chi}(x)=-P^{(2)}_{0r}(x),\quad
    P^{(2)}_{2\chi}(x)=P^{(2)}_{0r}(x).
    \end{equation}
    Finally, the azimuthal pressure is determined by $P^{(2)}_{0\varphi}(x)=-3P^{(2)}_{0r}(x)$ and $P^{(2)}_{2\varphi}(x)=3P^{(2)}_{0r}(x)$.
\item 
{\bf{Case 2}}: $K^{(2)}_2=K^{(2)}_0=N^{(2)}_2=N^{(2)}_0=B^{(2)}_2=0$.\\ 
Following the same procedure as in the Schwarzschild wormhole and imposing $\rho^{(2)}_0(\ell)= 0$, one can obtain a regular solution for $B^{(2)}_0(\ell)$ at the throat. However, when analysing the differential equation governing $\omega^{(3)}_0(\ell)$, we find that its inhomogeneous part develops a third-order pole at $\ell= 0$. This singular behaviour prevents the regularity condition for $\omega^{(3)}_0(\ell)$ from being satisfied at the throat. Consequently, this case cannot be considered as a physically acceptable configuration for a Morris–Thorne wormhole.
\item 
{\bf{Case 3}}: $K^{(2)}_2=K^{(2)}_0=0$ and $N^{(2)}_2=\beta/r^4$, $N^{(2)}_0=\alpha/r^4$.\\ 
If we further impose $B^{(2)}_0=0$ and $\Omega_1=0$, the first–order frame-dragging function $\omega_1$ coincides with that obtained in Case 1. In this setting, the higher-order corrections $\omega^{(3)}_0$ and $\omega^{(3)}_2$ can be expressed in terms of the rescaled variable $x=\ell/r_0$ as
\begin{equation}
\omega^{(3)}_0(x)=\mathfrak{K}+\sum_{n=0}^3\mathfrak{n}_n(x)\arctan^n{x},\quad
\omega^{(3)}_2(x)=\mathfrak{M}(x)+\sum_{n=0}^3\mathfrak{m}_n(x)\arctan^n{x}
\end{equation}
with
\begin{eqnarray}
\mathfrak{K}&=&-\frac{\pi}{32}\left[(24\pi^2+450)\alpha-(60\pi^2+1797)\beta-144\pi^2+252\right],\\   
\mathfrak{n}_3(x)&=&\frac{3}{2}\left[-2\alpha+(21x^2+5)\beta+12(x^2+1)\right],\\
\mathfrak{n}_2(x)&=&\frac{9}{4}\left[\left(2\pi-\frac{4x}{x^2 + 1}\right)\alpha 
-\left(\pi(21x^2+5)-\frac{2x(21x^2+19}{x^2+1}\right)\beta-12\pi(x^2+1)+\frac{8x(3x^2+5)}{x^2 + 1}\right],\\
\mathfrak{n}_1(x)&=&\frac{1}{2(x^2+1)}\left[\left(18\pi x+\frac{3(75x^4+126x^2+11)}{4(x^2+1)}\right)\alpha\right.\nonumber\\
&&\left.+\left(-9\pi x(21x^2+19)-\frac{3015x^4}{8}-\frac{609x^2}{4}+\frac{1149}{8}\right)\beta\right.\nonumber\\
&&\left.-36\pi x(3x^2+5)-\frac{405x^4}{2}-171 x^2+\frac{135}{2}\right],\\
\mathfrak{n}_0(x)&=&\left[\frac{\pi(9x^2+24)}{2(x^2+1)^2}+\frac{3x(75x^4+192x^2+101)}{8(x^2+1)^3}\right]\alpha\nonumber\\
&&+\left[\frac{9\pi}{32}\left(28\pi^2 x^2+\frac{335x^4+335x^2+72}{x^2+1}\right)-\frac{3x(1005x^4+1412x^2+415)}{16(x^2+1)^2}\right]\beta\nonumber\\
&&+\frac{9\pi}{8}\left(4\pi^2 x^2+\frac{45x^4+45x^2-8}{x^2 + 1}\right) - \frac{x(405x^4+756x^2+423)}{4(x^2+1)^2}.
\end{eqnarray}
and
\begin{eqnarray}
\mathfrak{M}(x)&=&\frac{9\pi}{32}\left[\pi^2(28\beta+16)+215\beta-60\right](5x^2+1),\quad
\mathfrak{m}_3(x)=\left(\frac{63\beta}{2}+18\right)(5x^2+1),\\
\mathfrak{m}_2(x)&=&-\frac{(189\beta+108)}{4}\left[\pi(5x^2+1)-\frac{2x(15x^2+13)}{3(x^2 + 1)}\right],\\
\mathfrak{m}_1(x)&=&-\frac{1}{2(x^2+1)^2}\left[9\pi x(15x^2+13)(x^2+1)(7\beta+4)
+\frac{225}{8}(67\beta+36)x^6+\frac{1}{8}(9513\beta+4716)x^2\right.\nonumber\\
&&\left.+\frac{45}{8}(569\beta+300)x^4-\frac{537\beta}{8}-\frac{171}{2}\right],\\
\mathfrak{m}_0(x)&=&\frac{1}{(x^2+1)^3}\left[675\pi(\beta+2)x^8-\frac{225}{16}(67\beta+36)x^7+\frac{135\pi}{4}(9\beta+28)x^6-\frac{40695\beta+21780}{16}x^5\right.\nonumber\\
&&\left.+9\pi(\beta+97)x^4-\frac{1}{16}(35661\beta+18972)x^3-\frac{\pi}{4}(813\beta-1044)x^2\right.\nonumber\\
&&\left.-\frac{1}{16}(10137\beta+5292)x-\frac{3\pi}{4}(103\beta+6)\right].
\end{eqnarray}
Moreover, we have
\begin{equation}
B^{(2)}_2(x)=-\frac{10\beta x^2}{(x^2+1)^{5/2}}.    
\end{equation}
Regarding the second-order energy-density corrections, $\rho^{(2)}$is given by the second expression in \eqref{rhocase1}, while
\begin{equation}
\rho^{(2)}_2(x)=\widetilde{\rho}^{(2)}_2(x)-\frac{25\beta}{4\pi(x^2+1)^4},    
\end{equation}
where $\widetilde{\rho}^{(2)}_2$ coincides with the third formula in \eqref{rhocase1}. The corresponding second-order corrections to the radial pressure are given by
\begin{equation}
P^{(2)}_{0r}(x)=\frac{(3-4\alpha)x^2+3}{4\pi(x^2+1)^4},\quad
P^{(2)}_{2r}(x)=-\frac{(3+2\beta)x^2+3(1+\beta)}{4\pi(x^2+1)^4}.
\end{equation}
Similarly, the second-order corrections to the pressure along the $\chi$-direction can be written as
\begin{equation}
P^{(2)}_{0\chi}(x)=\frac{(8\alpha+2\beta-3)x^2+2(\beta-\alpha)-3}{4\pi(x^2+1)^4},\quad
P^{(2)}_{2\chi}(x)=\frac{(2\beta+3)x^2+7\beta+3}{4\pi(x^2+1)^4}.
\end{equation}
Finally, the corresponding second-order components of the pressure in the $\varphi$-direction are given by
\begin{equation}
P^{(2)}_{0\varphi}(x)=\frac{(8\alpha-2\beta-9)x^2-2(\beta+\alpha)-9}{4\pi(x^2+1)^4},\quad
P^{(2)}_{2\varphi}(x)=\frac{3(2\beta+3)x^2+11\beta+9}{4\pi(x^2+1)^4}.
\end{equation}
Last but not least, by choosing $B^{(2)}_0\neq 0$ and following the same procedure as in the Schwarzschild wormhole, i.e. imposing $\rho^{(2)}_0(\ell)=0$, a regular solution for $B^{(2)}_0(\ell)$ can indeed be obtained at the throat. However, upon examining the differential equation governing $\omega^{(3)}_0(\ell)$, we find that its inhomogeneous term develops a third-order pole at $\ell=0$. This singularity prevents the regularity condition for $\omega^{(3)}_0(\ell)$ from being fulfilled at the throat. Therefore, this configuration cannot be regarded as a physically admissible case for a Morris–Thorne wormhole.
\end{itemize}

\begin{figure}[!ht]
\centering
    \includegraphics[width=0.3\textwidth]{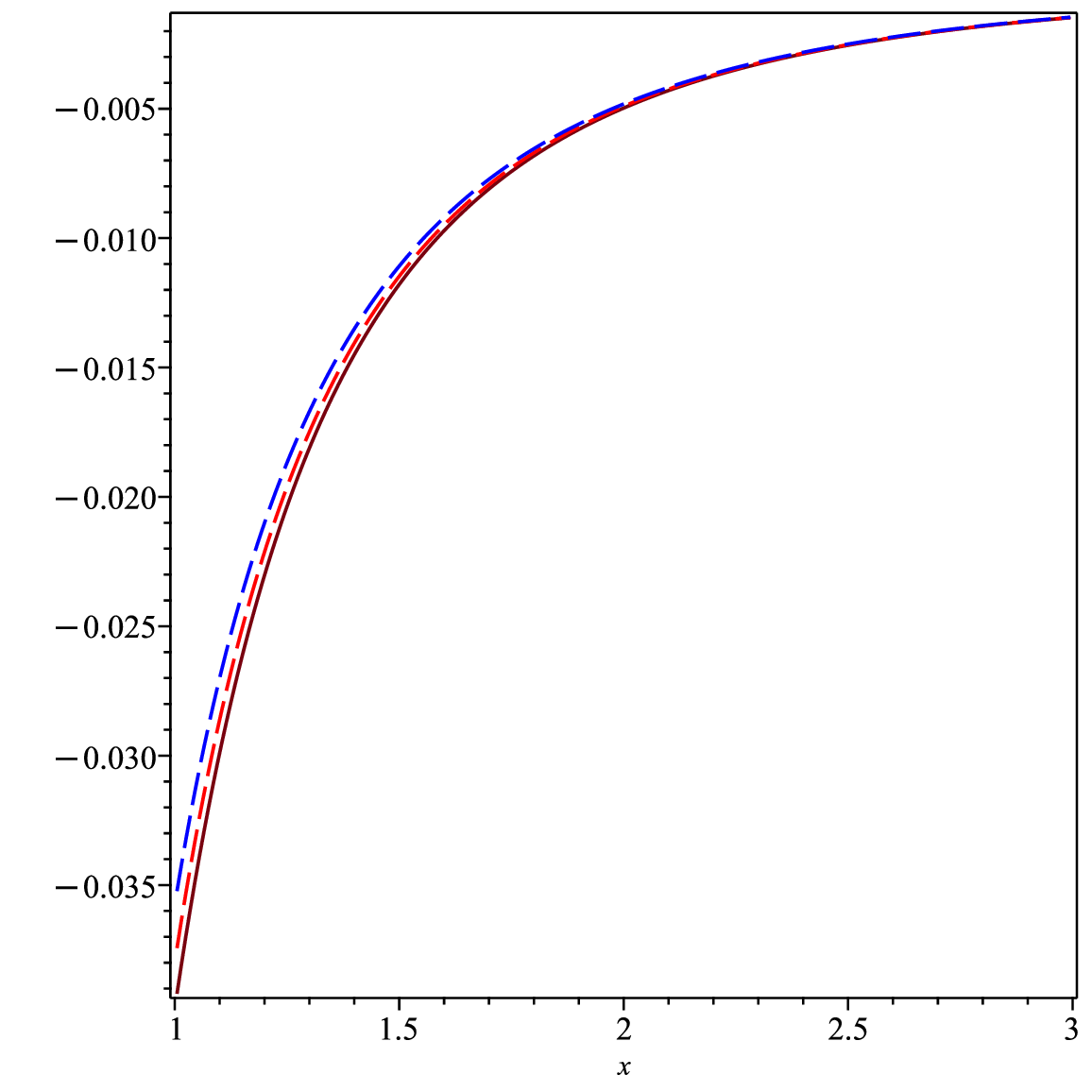}
    \includegraphics[width=0.3\textwidth]{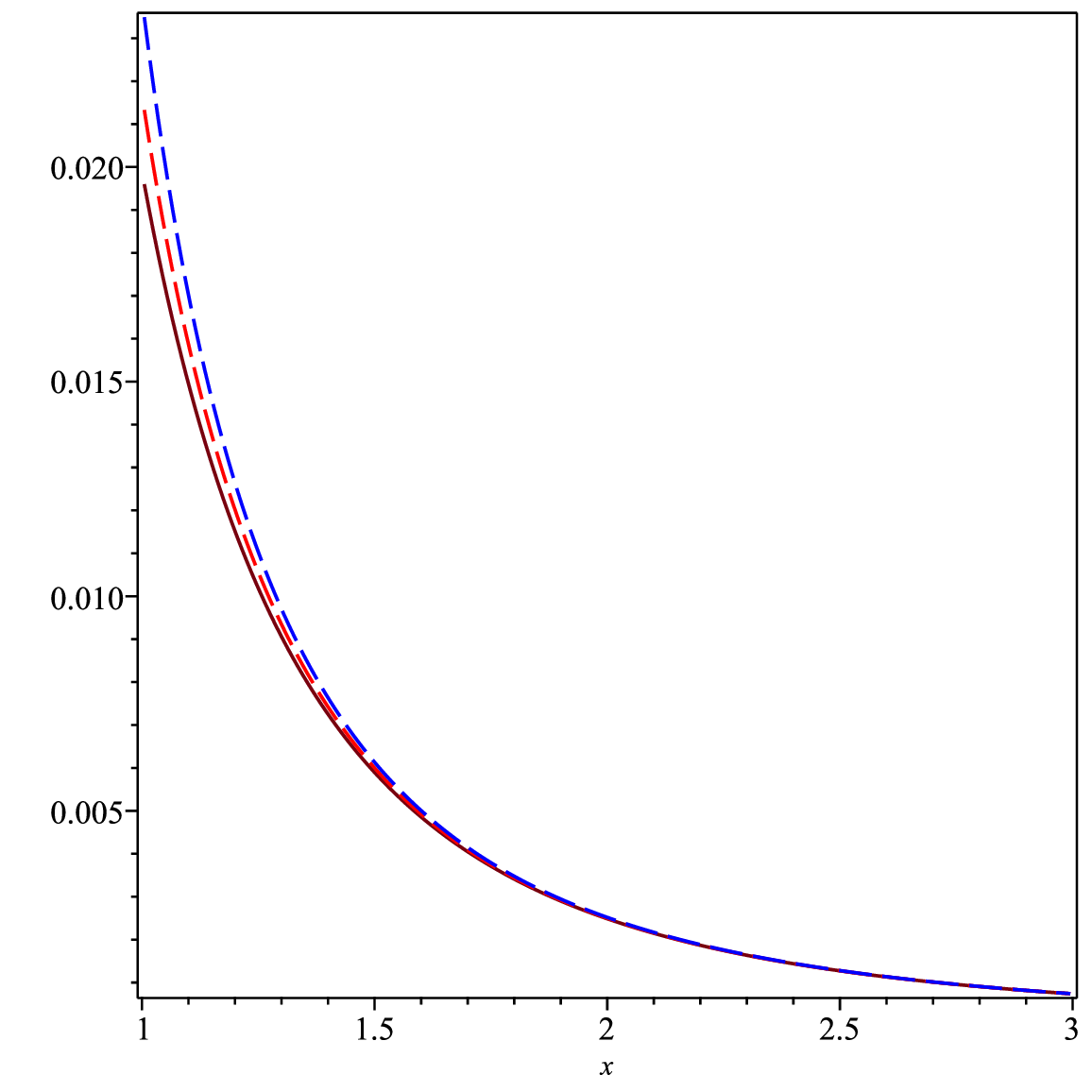}
    \includegraphics[width=0.3\textwidth]{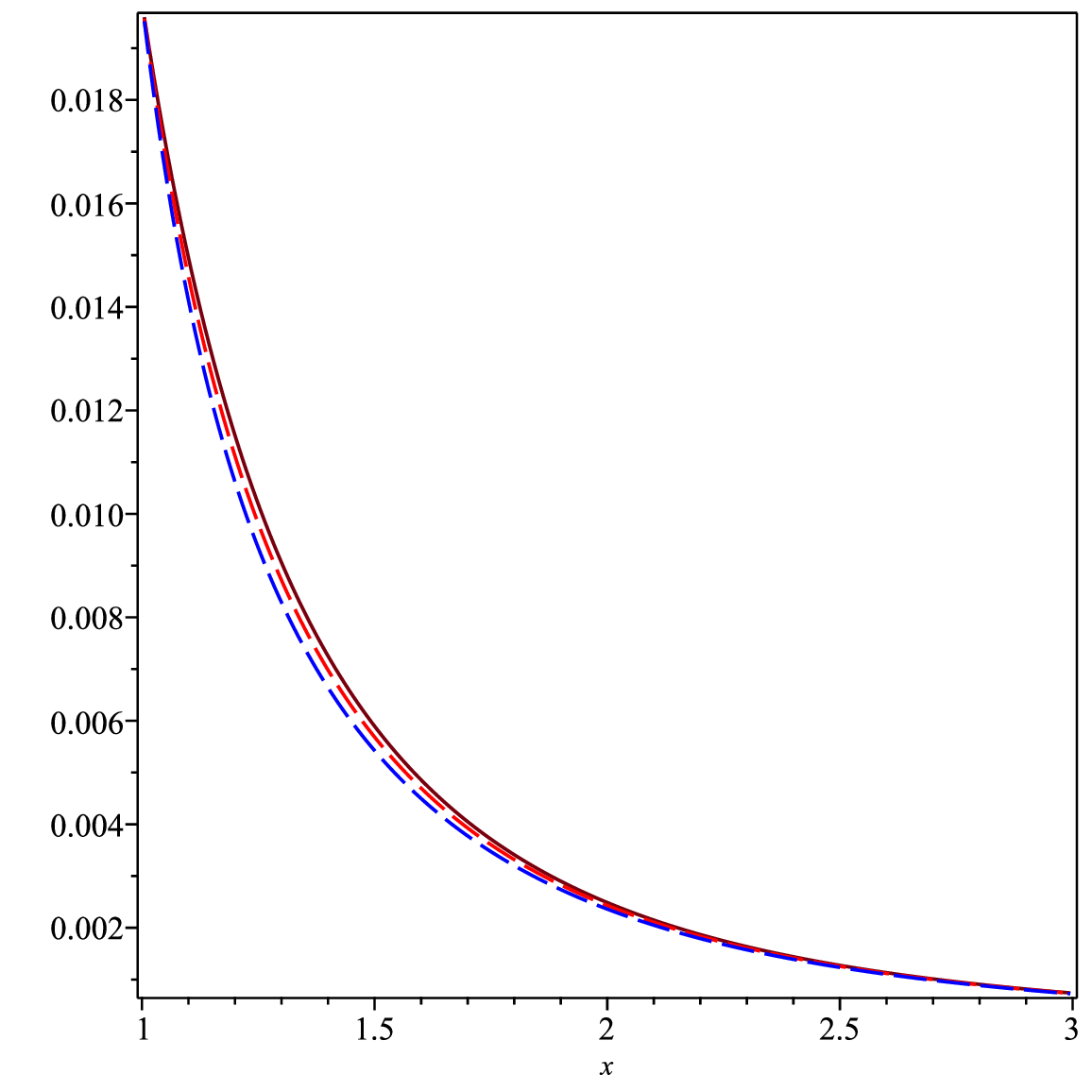}
\caption{\label{NEC_SCHW_MOD1_CASE1} Behaviour of the NEC for Model 1, Case 1 of the Schwarzschild wormhole under slow rotation, for different values of the dimensionless angular momentum parameter $j$. The static configuration ($j=0$) is represented by the black curves, while the rotating cases $j=0.1$ and $j=0.15$ are shown in red and blue, respectively. Dashed curves correspond to quantities evaluated on the equatorial plane of the wormhole. {\bf{Left panel}}: radial NEC component $\widetilde{\rho}+P_r$. {\bf{Central panel}}: tangential NEC in the meridional direction $\widetilde{\rho}+P_\chi$. {\bf{Right panel}}: tangential NEC in the azimuthal direction $\widetilde{\rho}+P_\varphi$.}
\end{figure}

\begin{figure}[!ht]
\centering
    \includegraphics[width=0.3\textwidth]{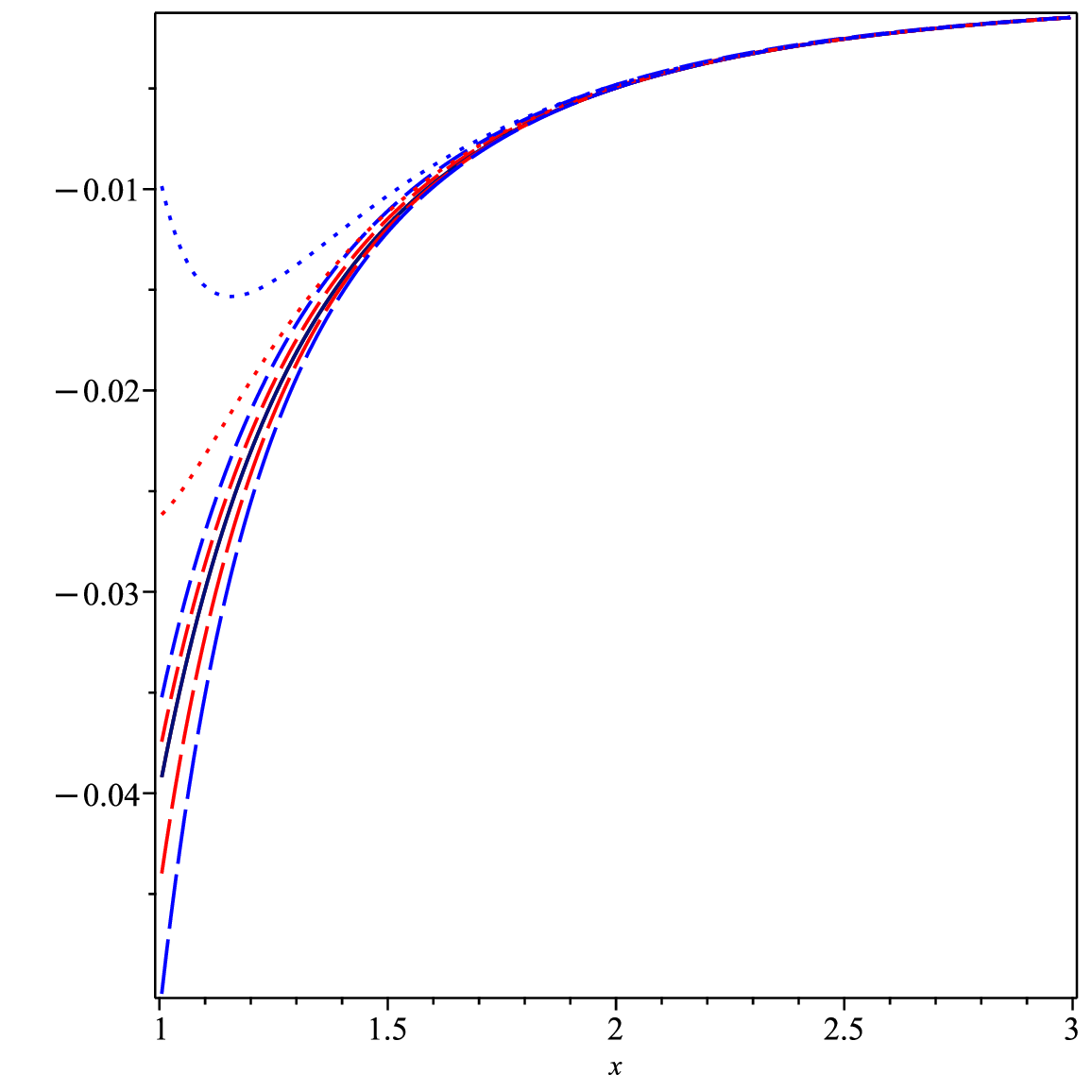}
    \includegraphics[width=0.3\textwidth]{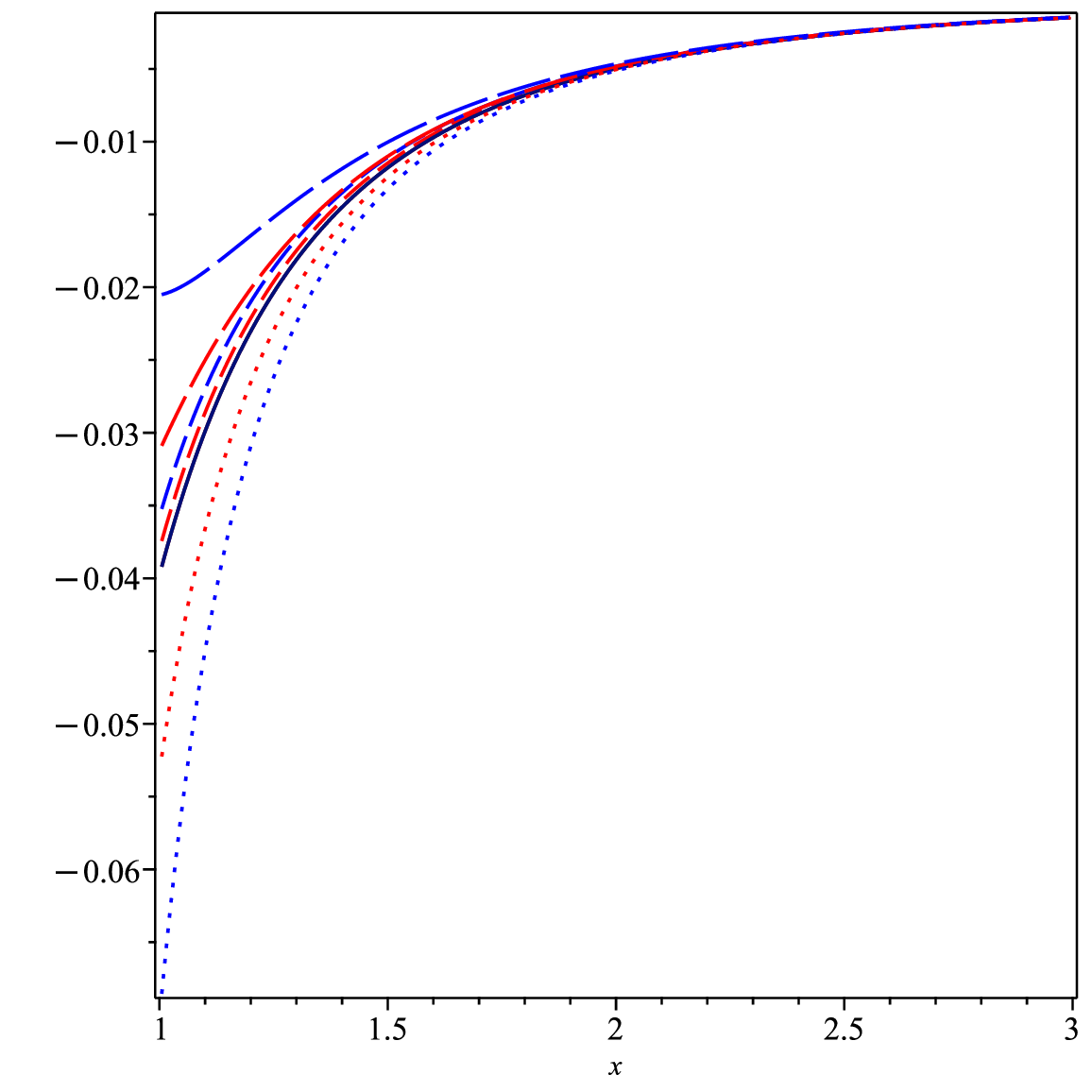}
    \includegraphics[width=0.3\textwidth]{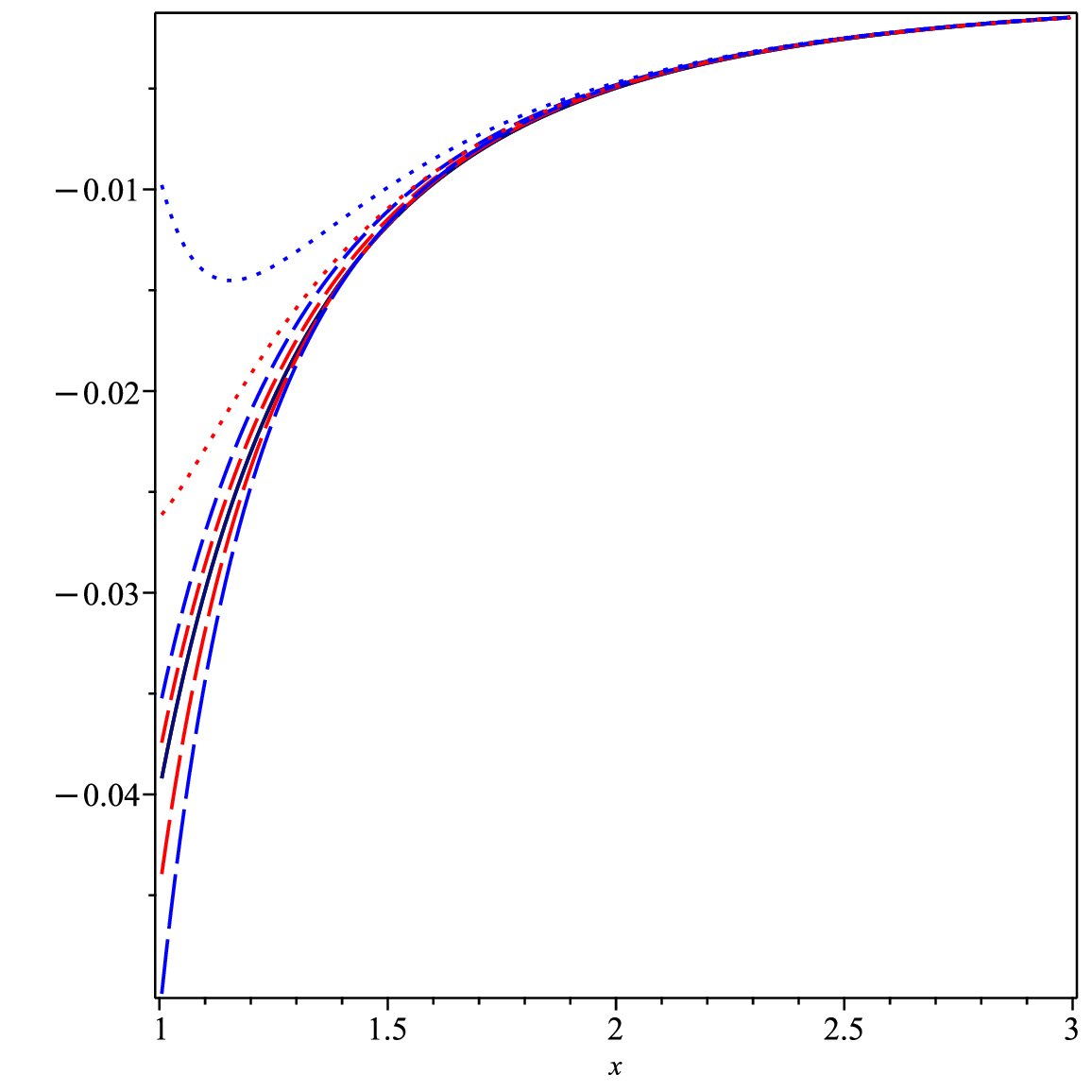}
\caption{\label{NEC_SCHW_MOD1_CASE3} Behaviour of the radial NEC for Model 1, Case 3 of the Schwarzschild wormhole under slow rotation, shown for different values of the dimensionless angular momentum parameter $j$. The static configuration ($j=0$) is represented by the black curves, while the rotating cases ($j=0.1$) and ($j=0.15$) are shown in red and blue, respectively. Dashed curves correspond to quantities evaluated on the equatorial plane of the wormhole for the reference case $(\alpha,\beta)=(0,0)$, whereas long–dashed curves represent configurations with $(\alpha,\beta)\neq(0,0)$. Dotted lines indicate the corresponding profiles along the symmetry axis (poles) of the wormhole. \textbf{Left panel}: comparison between the reference case $(\alpha,\beta)=(0,0)$ and $(\alpha,\beta)=(1,1)$. \textbf{Central panel}: comparison between $(\alpha,\beta)=(0,0)$ and $(\alpha,\beta)=(-1,-1)$. \textbf{Right panel}: comparison between $(\alpha,\beta)=(0,0)$ and $(\alpha,\beta)=(1,-1)$.}
\end{figure}

\begin{figure}[!ht]
\centering
    \includegraphics[width=0.3\textwidth]{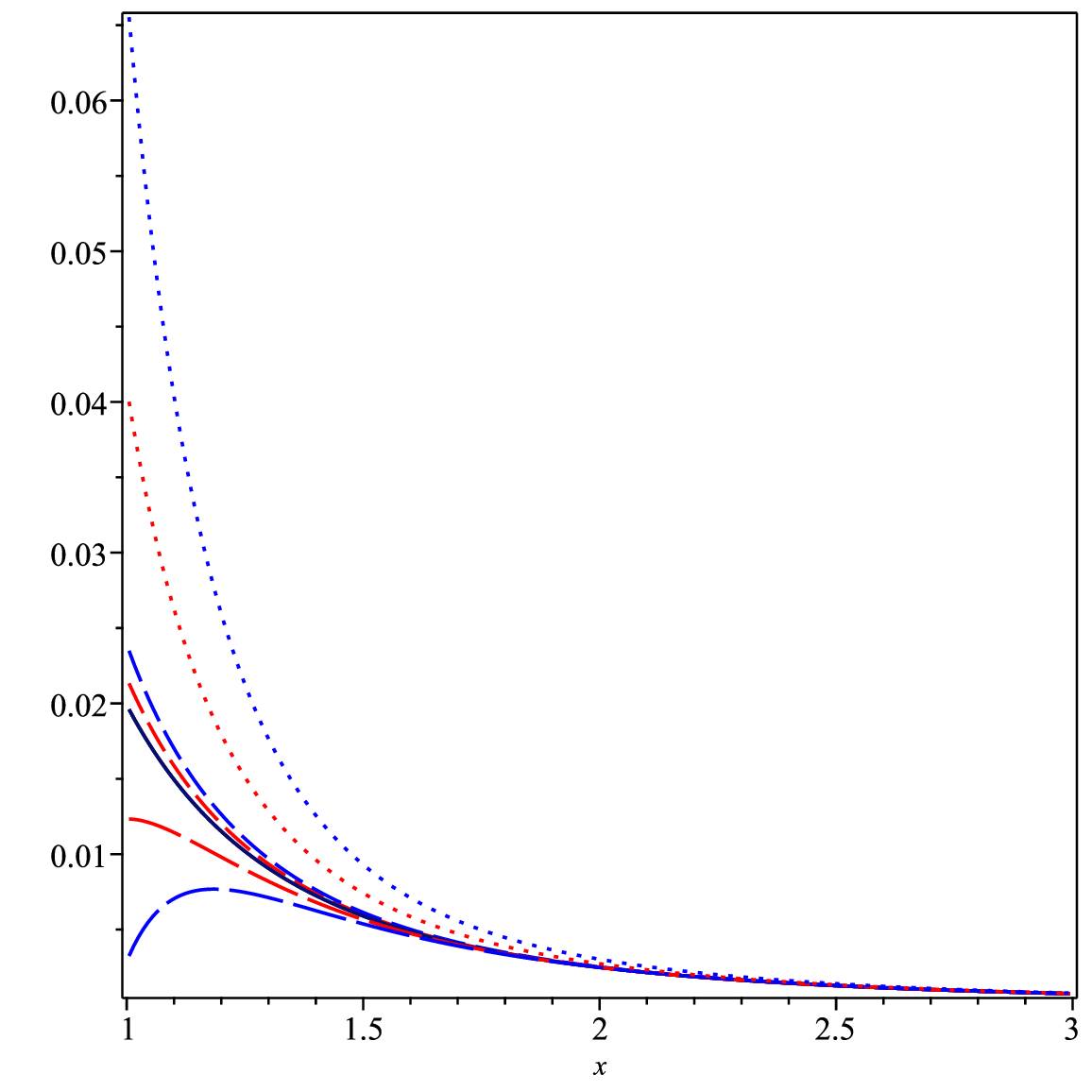}
    \includegraphics[width=0.3\textwidth]{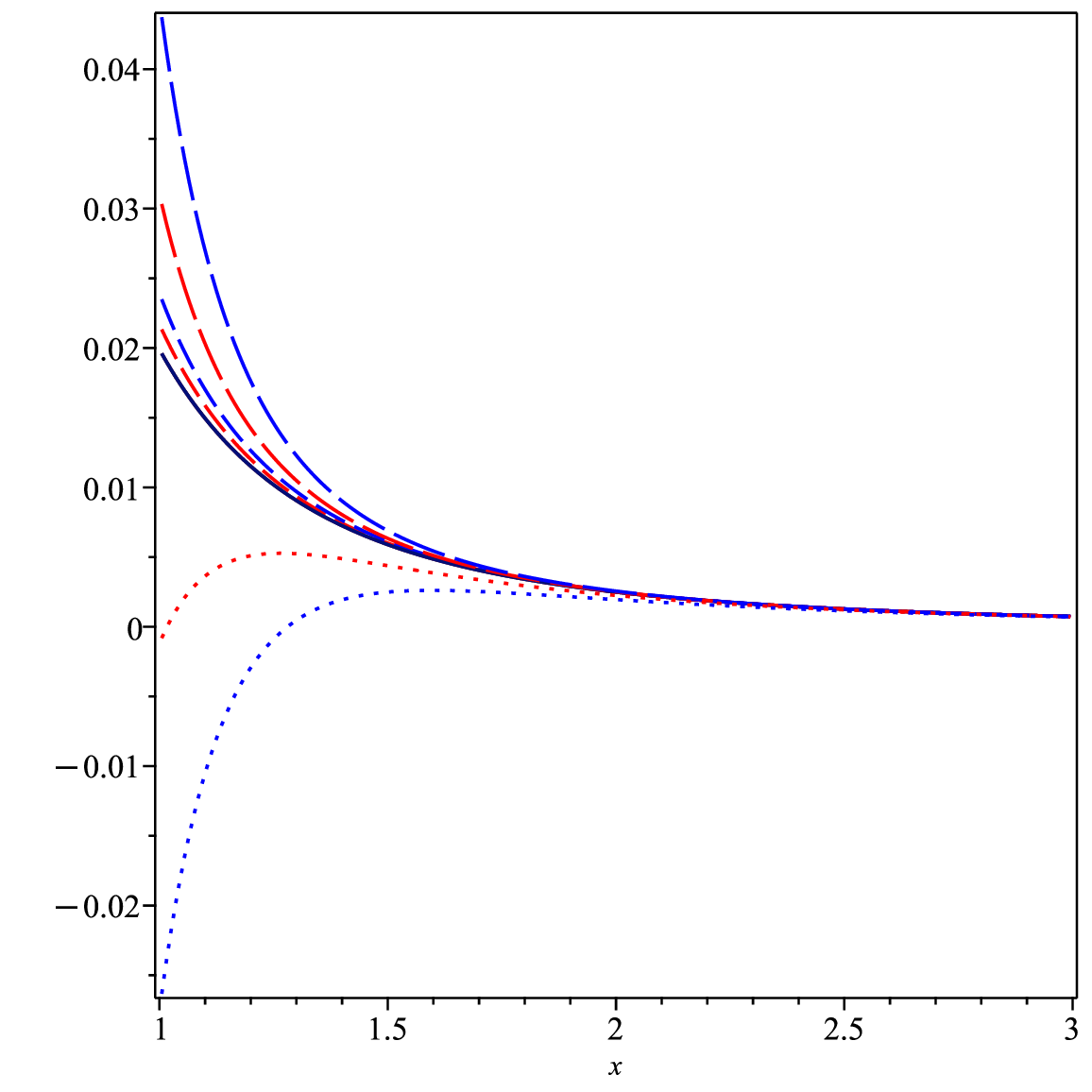}
    \includegraphics[width=0.3\textwidth]{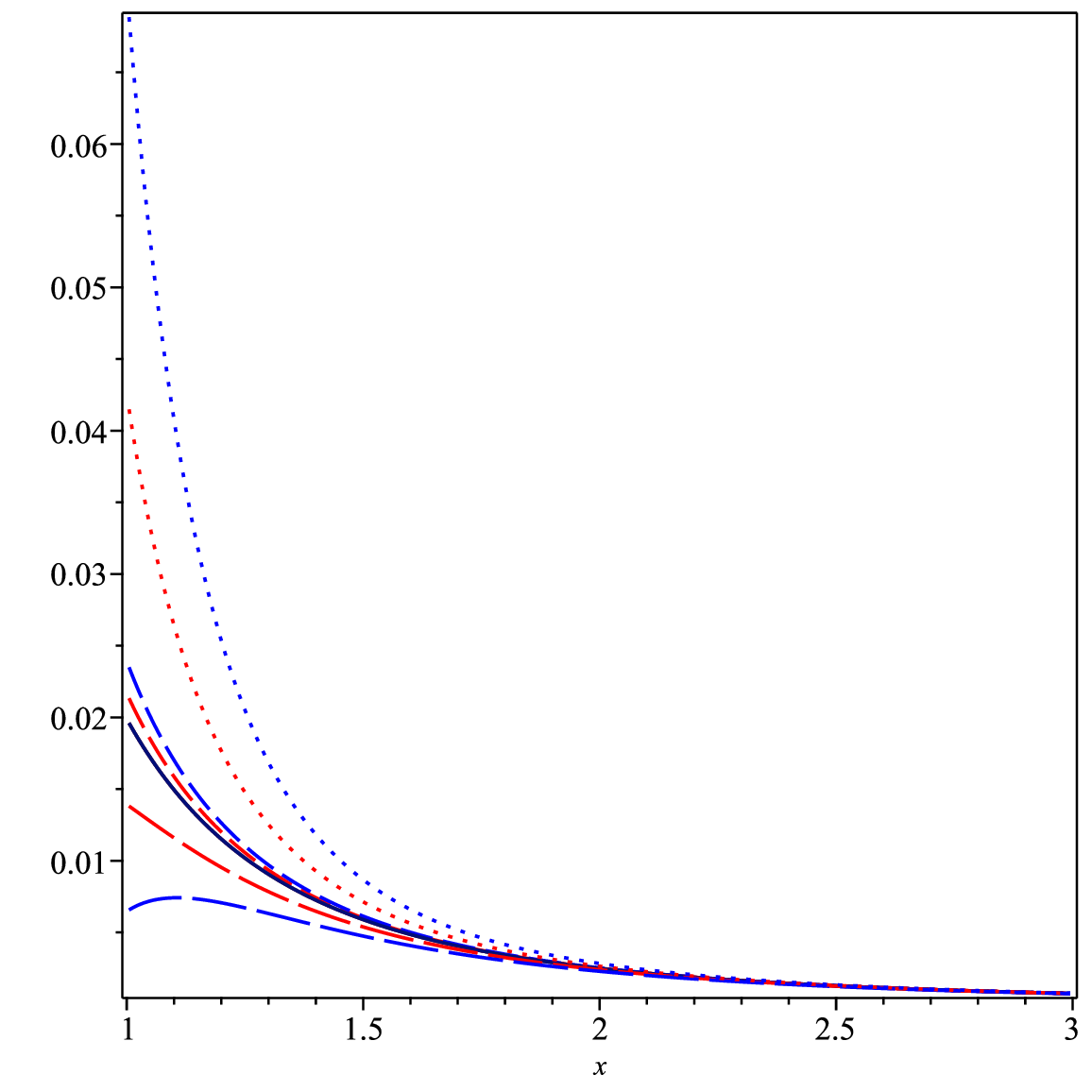}
\caption{\label{NEC_SCHW_MOD1_CASE3_chi} Behaviour of the meridional NEC for Model 1, Case 3 of the Schwarzschild wormhole under slow rotation, shown for different values of the dimensionless angular momentum parameter $j$. The static configuration ($j=0$) is represented by the black curves, while the rotating cases ($j=0.1$) and ($j=0.15$) are shown in red and blue, respectively. Dashed curves correspond to quantities evaluated on the equatorial plane of the wormhole for the reference case $(\alpha,\beta)=(0,0)$, whereas long–dashed curves represent configurations with $(\alpha,\beta)\neq(0,0)$. Dotted lines indicate the corresponding profiles along the symmetry axis (poles) of the wormhole. \textbf{Left panel}: comparison between the reference case $(\alpha,\beta)=(0,0)$ and $(\alpha,\beta)=(1,1)$. \textbf{Central panel}: comparison between $(\alpha,\beta)=(0,0)$ and $(\alpha,\beta)=(-1,-1)$. \textbf{Right panel}: comparison between $(\alpha,\beta)=(0,0)$ and $(\alpha,\beta)=(1,-1)$.}
\end{figure}

\begin{figure}[!ht]
\centering
    \includegraphics[width=0.3\textwidth]{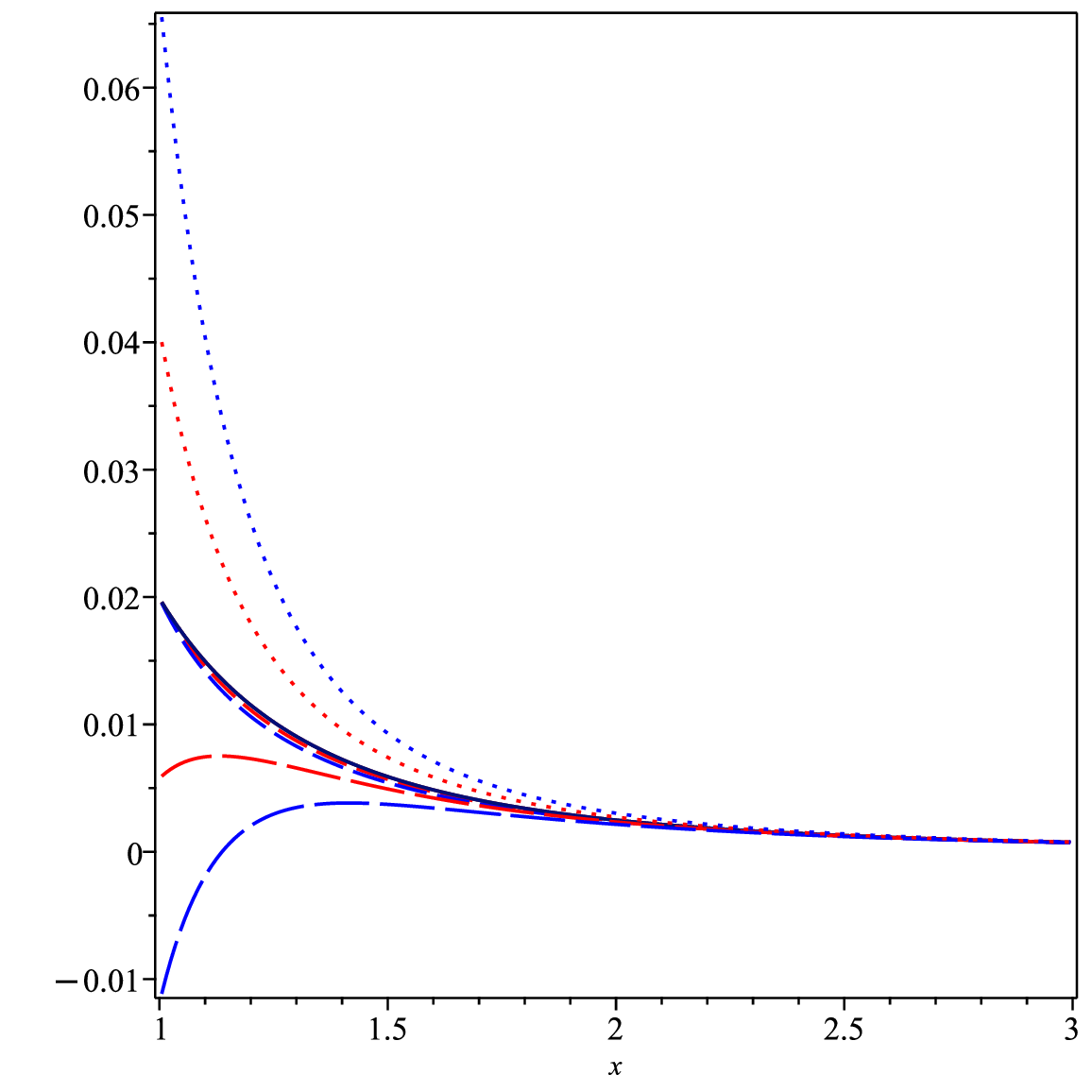}
    \includegraphics[width=0.3\textwidth]{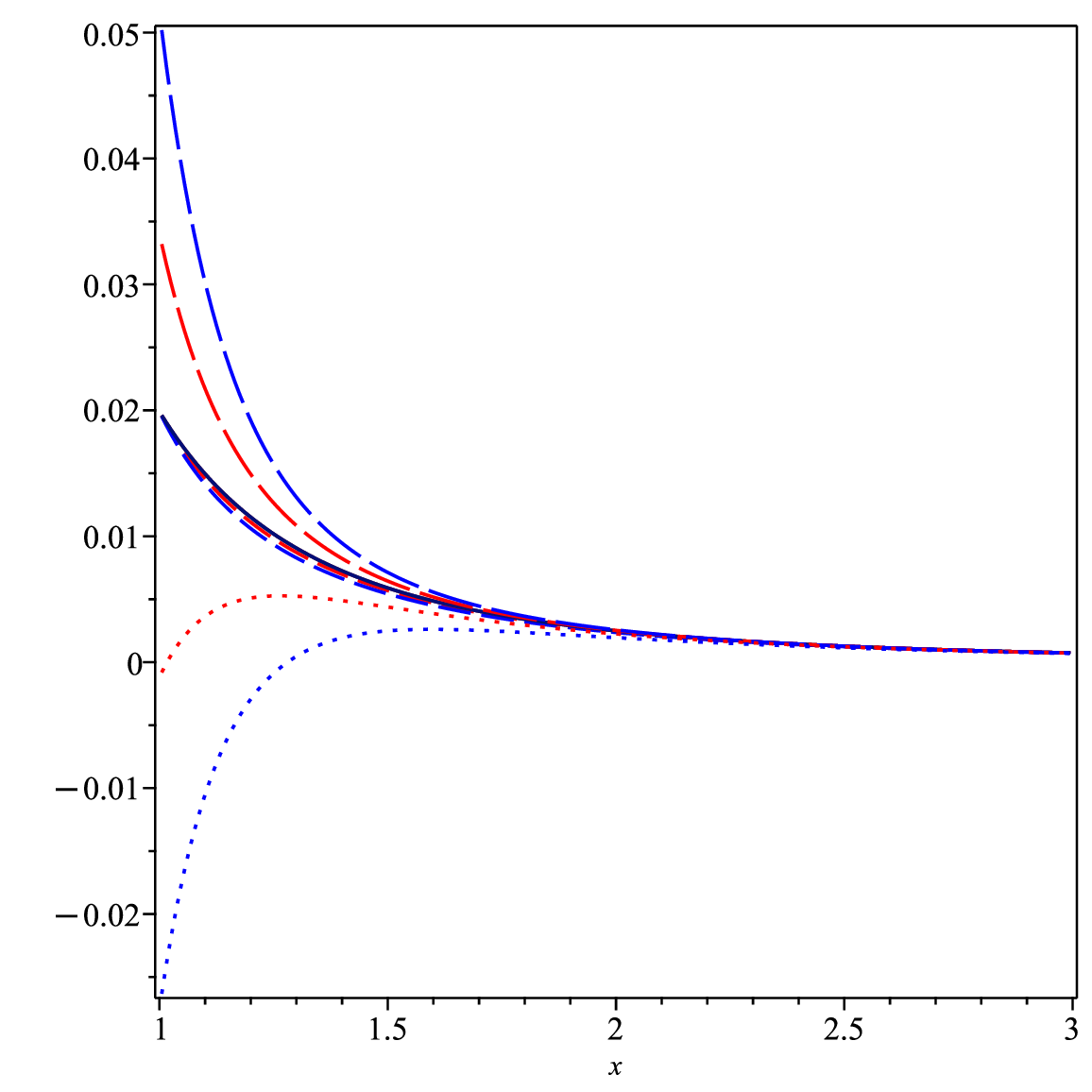}
    \includegraphics[width=0.3\textwidth]{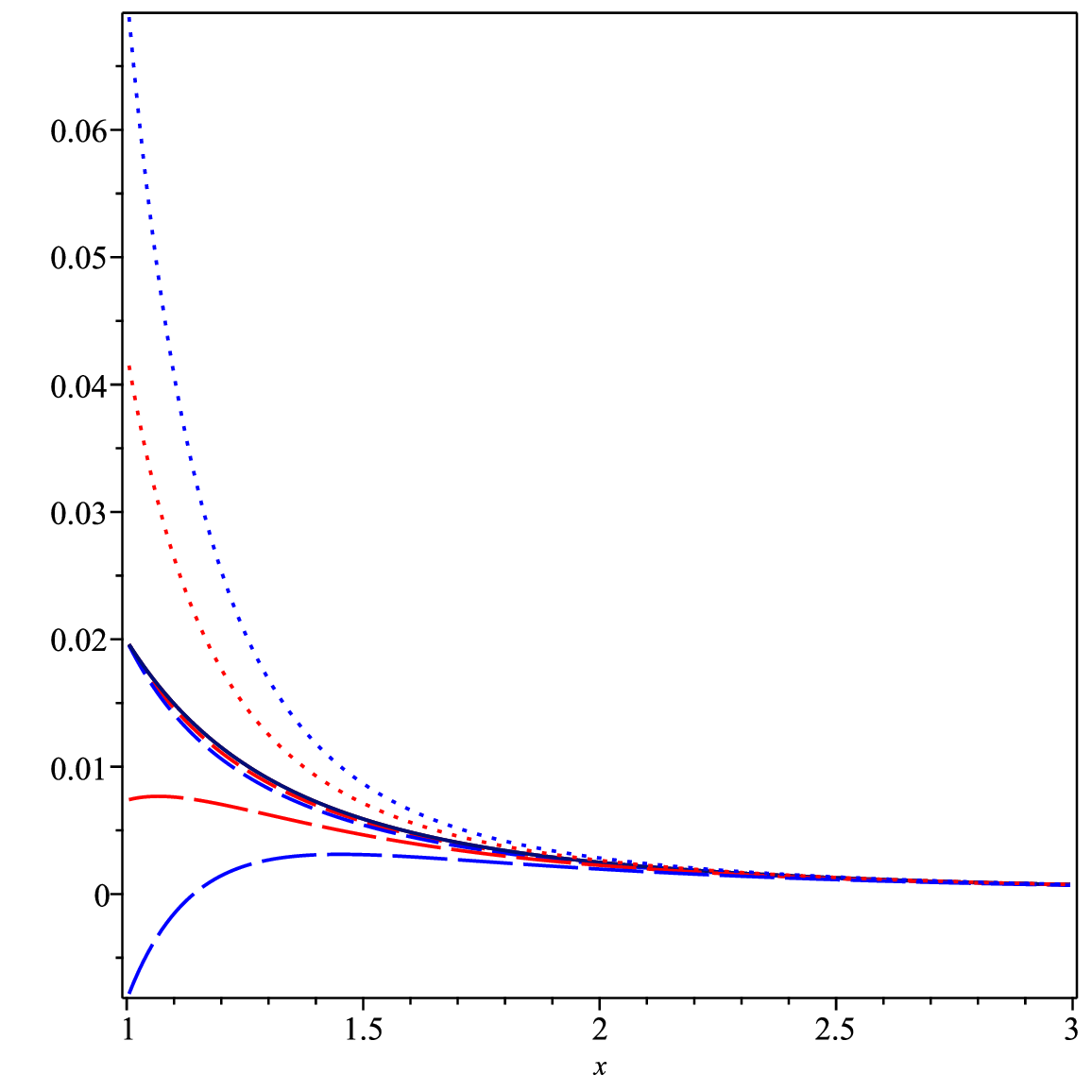}
\caption{\label{NEC_SCHW_MOD1_CASE3_phi} Behaviour of the azimuthal NEC for Model 1, Case 3 of the Schwarzschild wormhole under slow rotation, shown for different values of the dimensionless angular momentum parameter $j$. The static configuration ($j=0$) is represented by the black curves, while the rotating cases ($j=0.1$) and ($j=0.15$) are shown in red and blue, respectively. Dashed curves correspond to quantities evaluated on the equatorial plane of the wormhole for the reference case $(\alpha,\beta)=(0,0)$, whereas long–dashed curves represent configurations with $(\alpha,\beta)\neq(0,0)$. Dotted lines indicate the corresponding profiles along the symmetry axis (poles) of the wormhole. \textbf{Left panel}: comparison between the reference case $(\alpha,\beta)=(0,0)$ and $(\alpha,\beta)=(1,1)$. \textbf{Central panel}: comparison between $(\alpha,\beta)=(0,0)$ and $(\alpha,\beta)=(-1,-1)$. \textbf{Right panel}: comparison between $(\alpha,\beta)=(0,0)$ and $(\alpha,\beta)=(1,-1)$.}
\end{figure}

\begin{figure}[!ht]
\centering
    \includegraphics[width=0.3\textwidth]{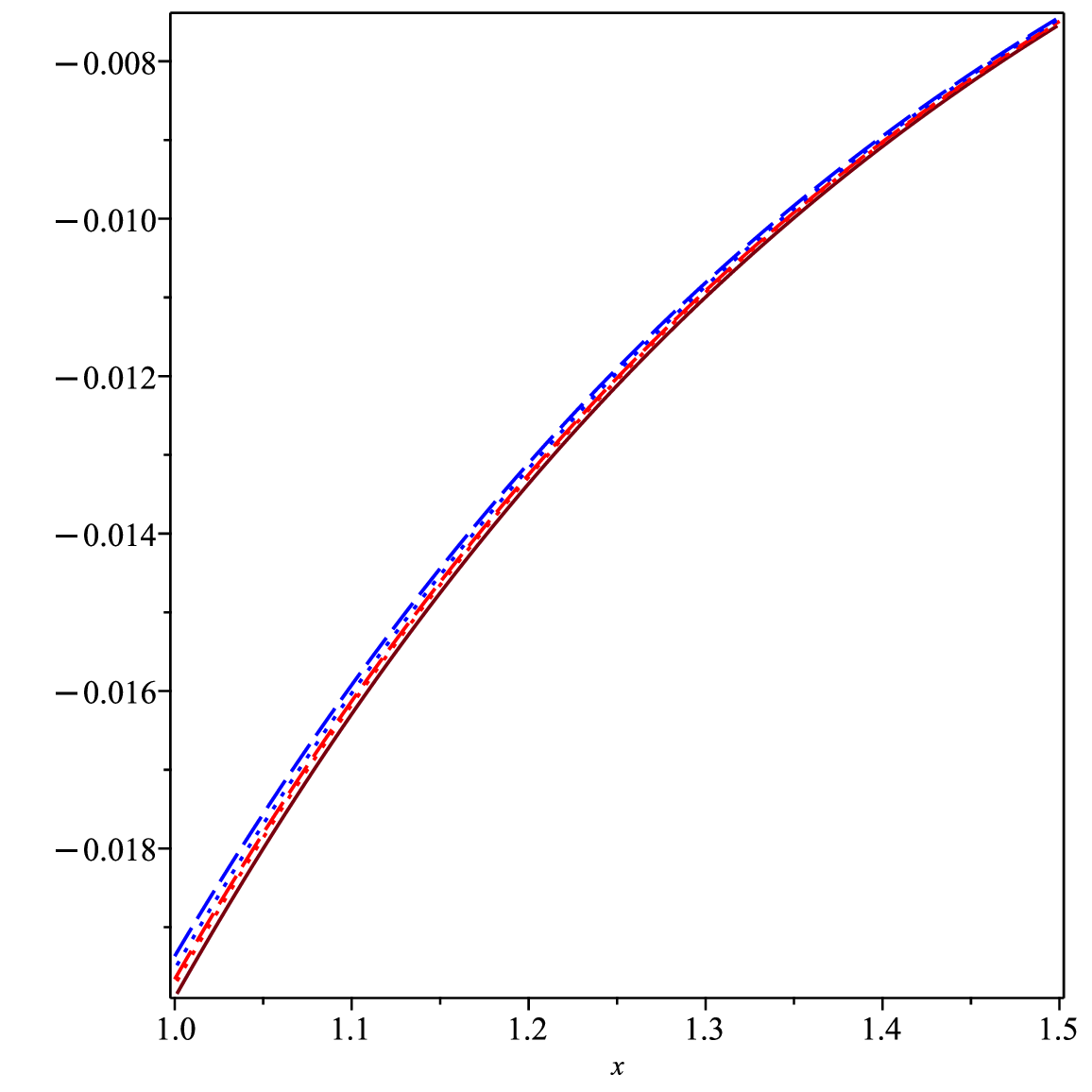}
    \includegraphics[width=0.3\textwidth]{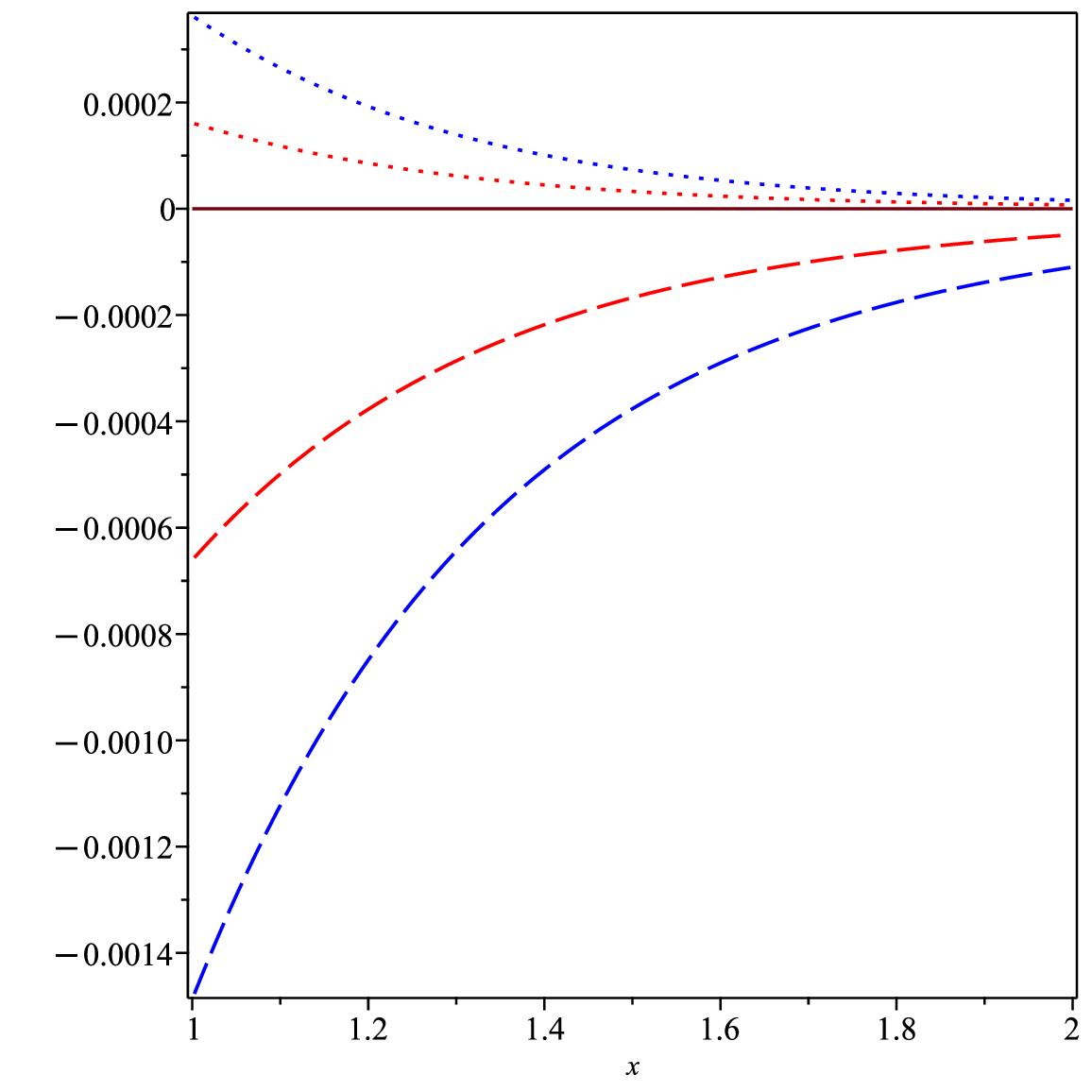}
    \includegraphics[width=0.3\textwidth]{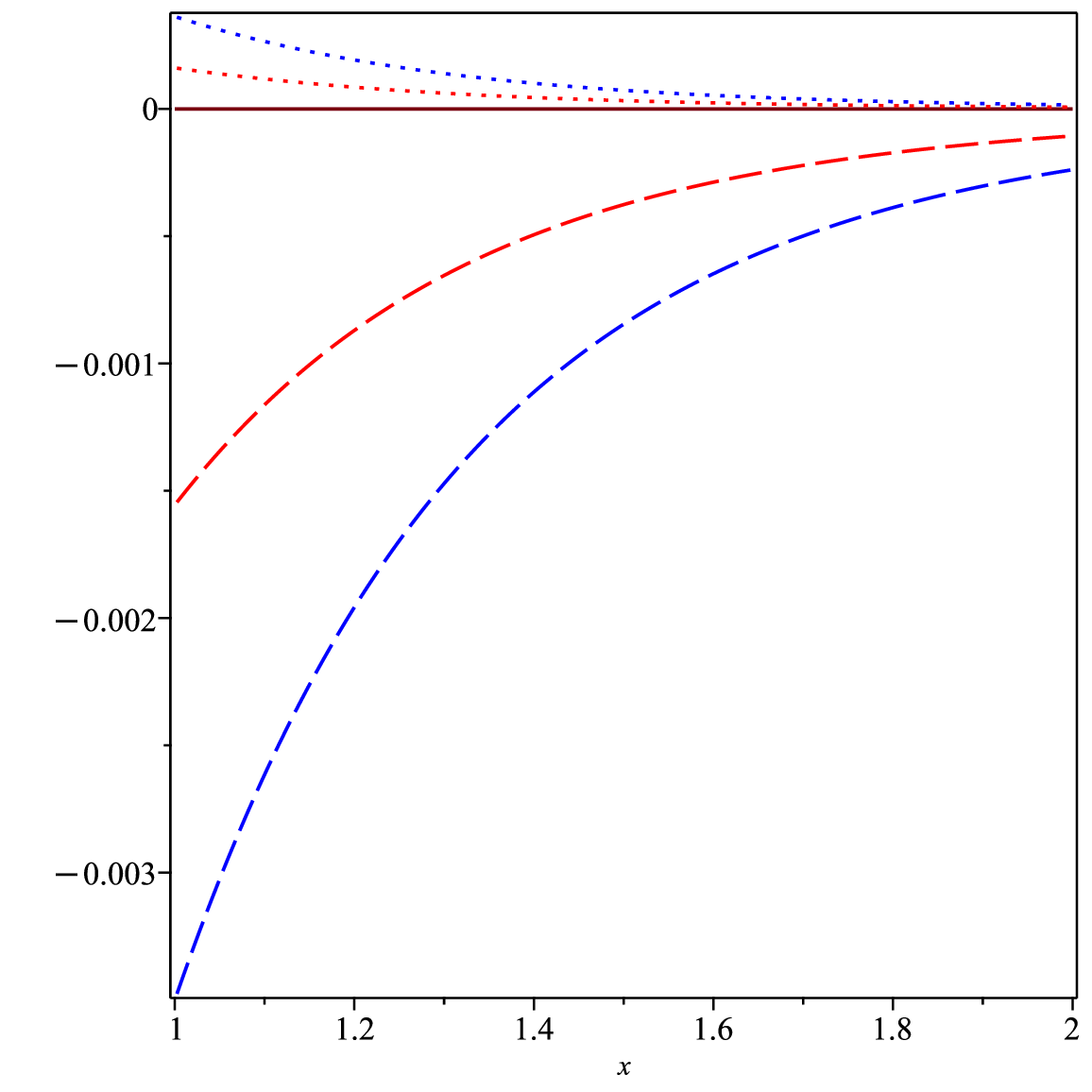}
\caption{\label{NEC_MTW_CASE1} Behaviour of the NEC for Case 1 of the Morris-Thorne wormhole under slow rotation, for different values of the dimensionless angular momentum parameter $j$. Here, $x=\ell/r_0$ denotes the rescaled proper radial distance. The static configuration ($j=0$) is represented by a black curve in the radial case and by a black horizontal line in the remaining cases. The rotating cases $j=0.1$ and $j=0.15$ are shown in red and blue, respectively. Dashed curves correspond to quantities evaluated on the equatorial plane of the wormhole. {\bf{Left panel}}: radial NEC component $\widetilde{\rho}+P_r$. {\bf{Central panel}}: tangential NEC in the meridional direction $\widetilde{\rho}+P_\chi$. {\bf{Right panel}}: tangential NEC in the azimuthal direction $\widetilde{\rho}+P_\varphi$.}
\end{figure}

\begin{figure}[!ht]
\centering
    \includegraphics[width=0.3\textwidth]{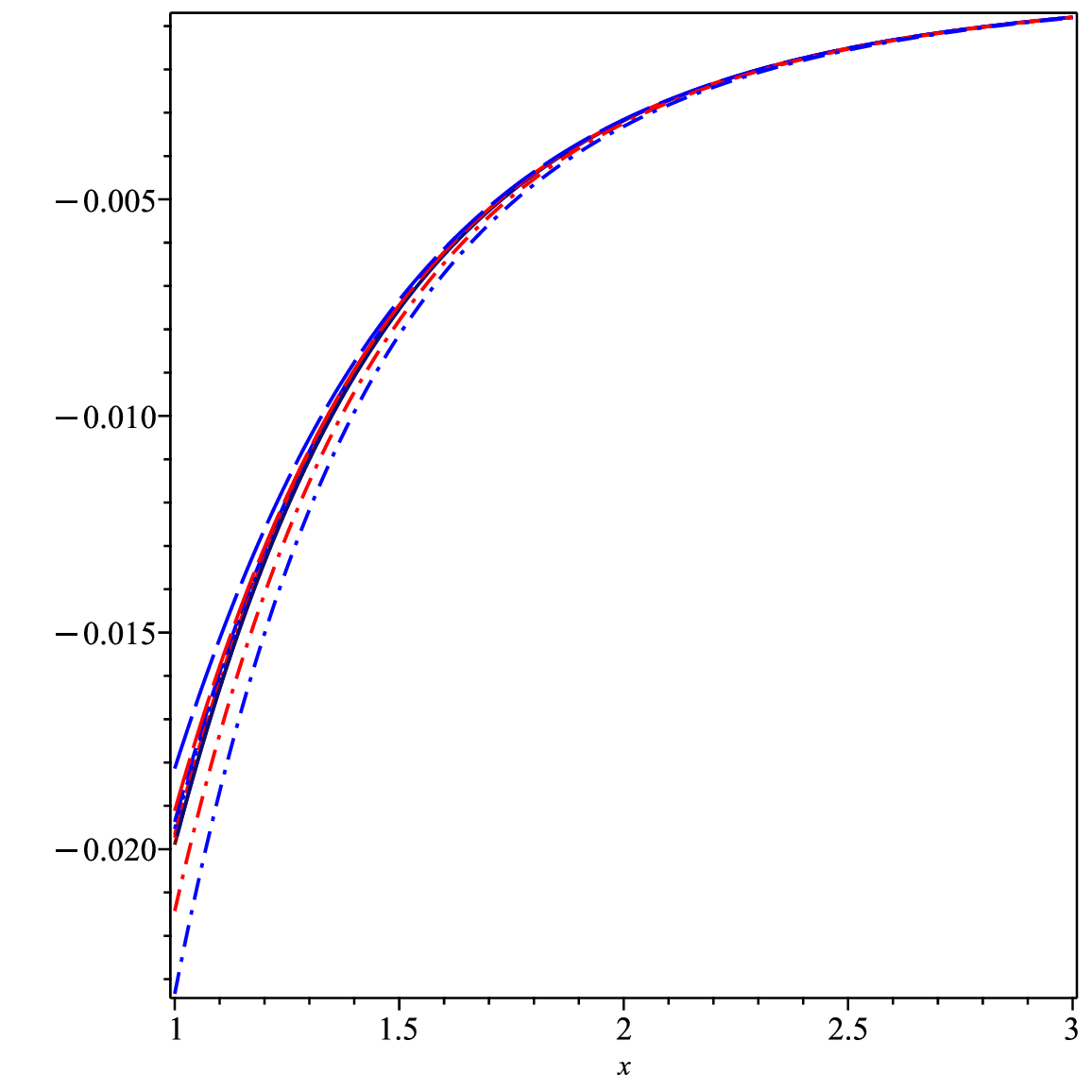}
    \includegraphics[width=0.3\textwidth]{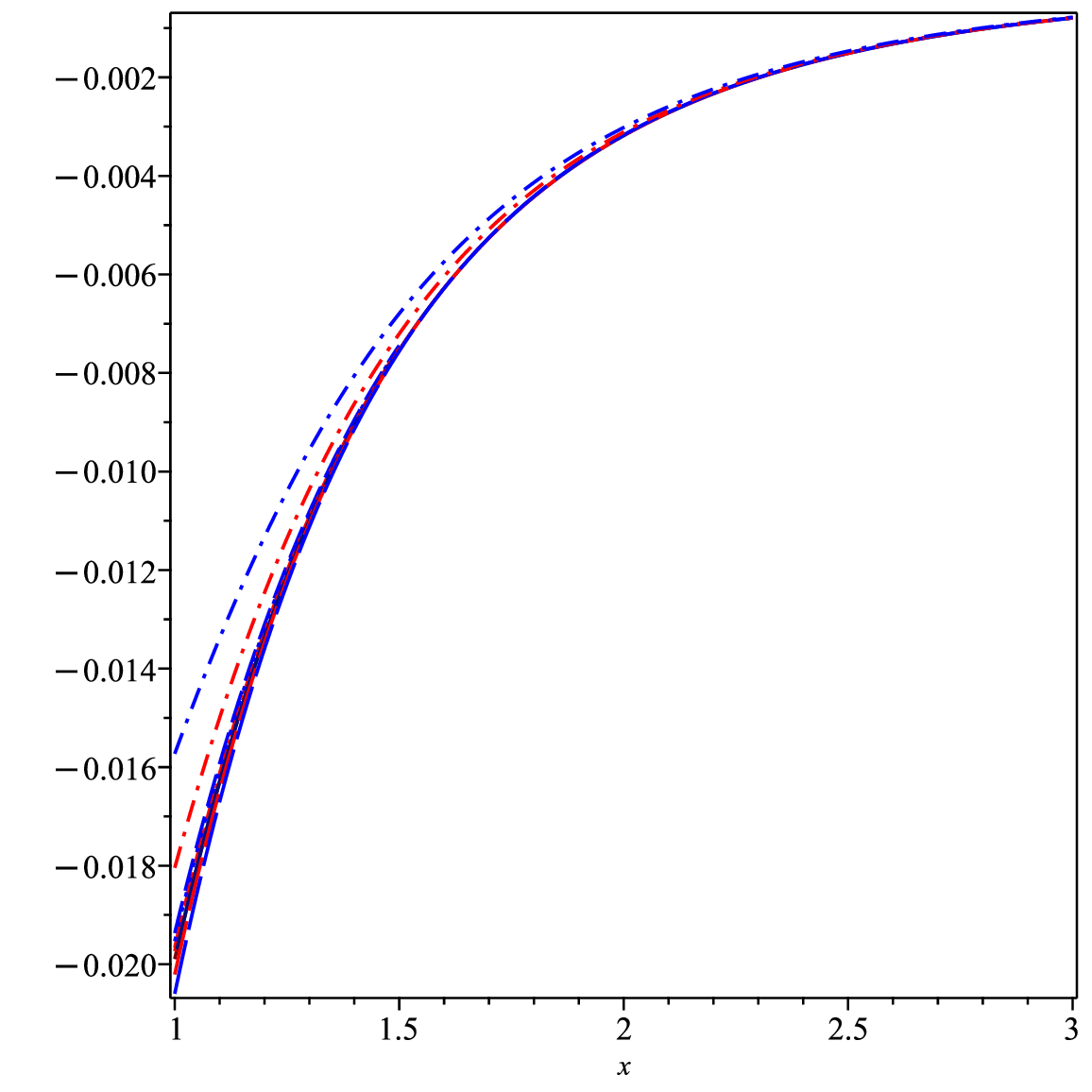}
    \includegraphics[width=0.3\textwidth]{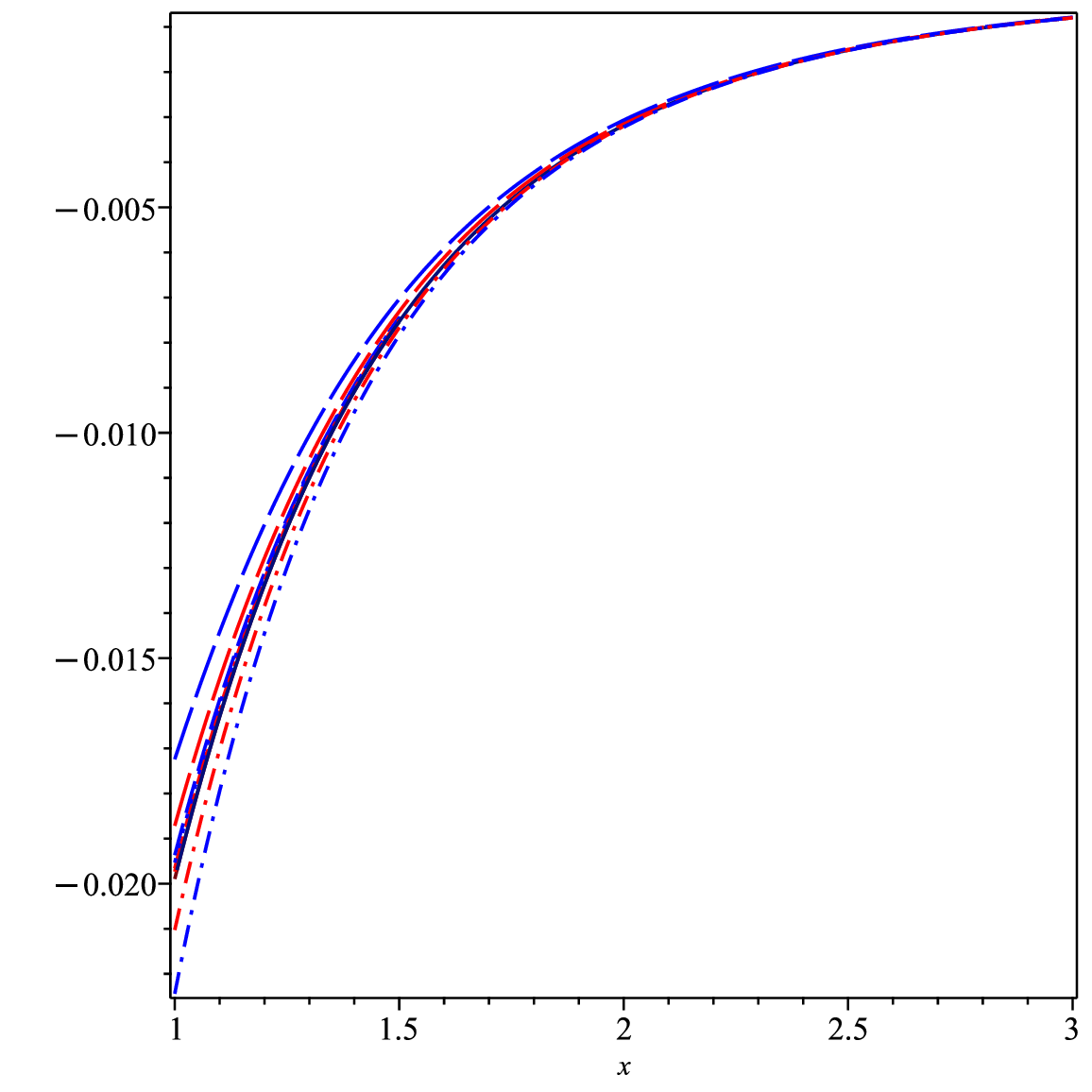}
\caption{\label{NECr-MTW-CASE3}
Behaviour of the radial NEC for Case 3 of the Morris-Thorne wormhole under slow rotation, shown for different values of the dimensionless angular momentum parameter $j$. The static configuration ($j=0$) is represented by the black curves, while the rotating cases ($j=0.1$) and ($j=0.15$) are shown in red and blue, respectively. Dashed and dotted curves denote quantities evaluated on the equatorial plane and along the symmetry axis, respectively, for the reference configuration $(\alpha,\beta)=(0,0)$. In contrast, long–dashed and dash–dotted curves correspond to quantities with $(\alpha,\beta)\neq(0,0)$. \textbf{Left panel}: comparison between the reference case $(\alpha,\beta)=(0,0)$ and $(\alpha,\beta)=(1,1)$. \textbf{Central panel}: comparison between $(\alpha,\beta)=(0,0)$ and $(\alpha,\beta)=(-1,-1)$.
\textbf{Right panel}: comparison between $(\alpha,\beta)=(0,0)$ and $(\alpha,\beta)=(1,-1)$.}
\end{figure}

\begin{figure}[!ht]
\centering
    \includegraphics[width=0.3\textwidth]{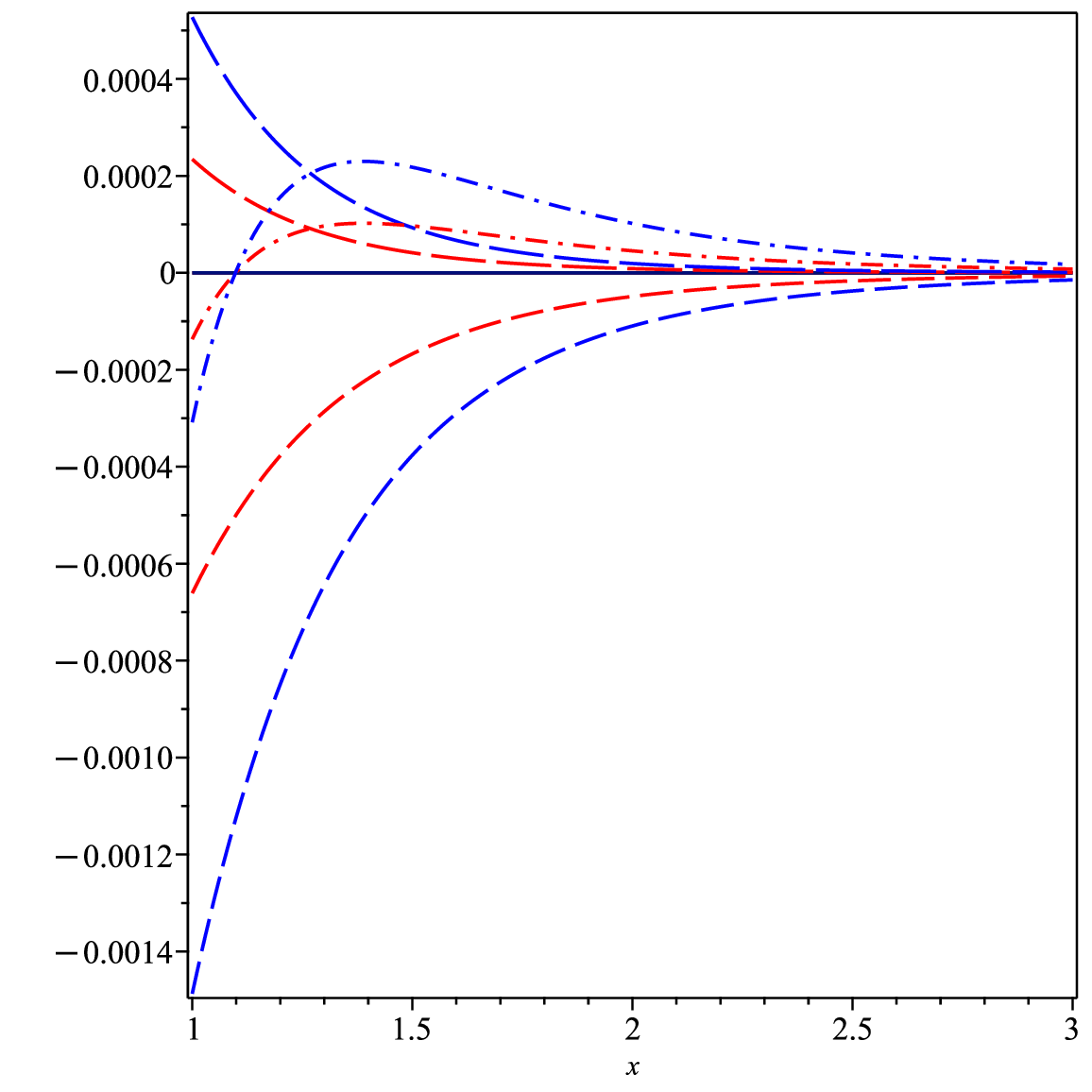}
    \includegraphics[width=0.3\textwidth]{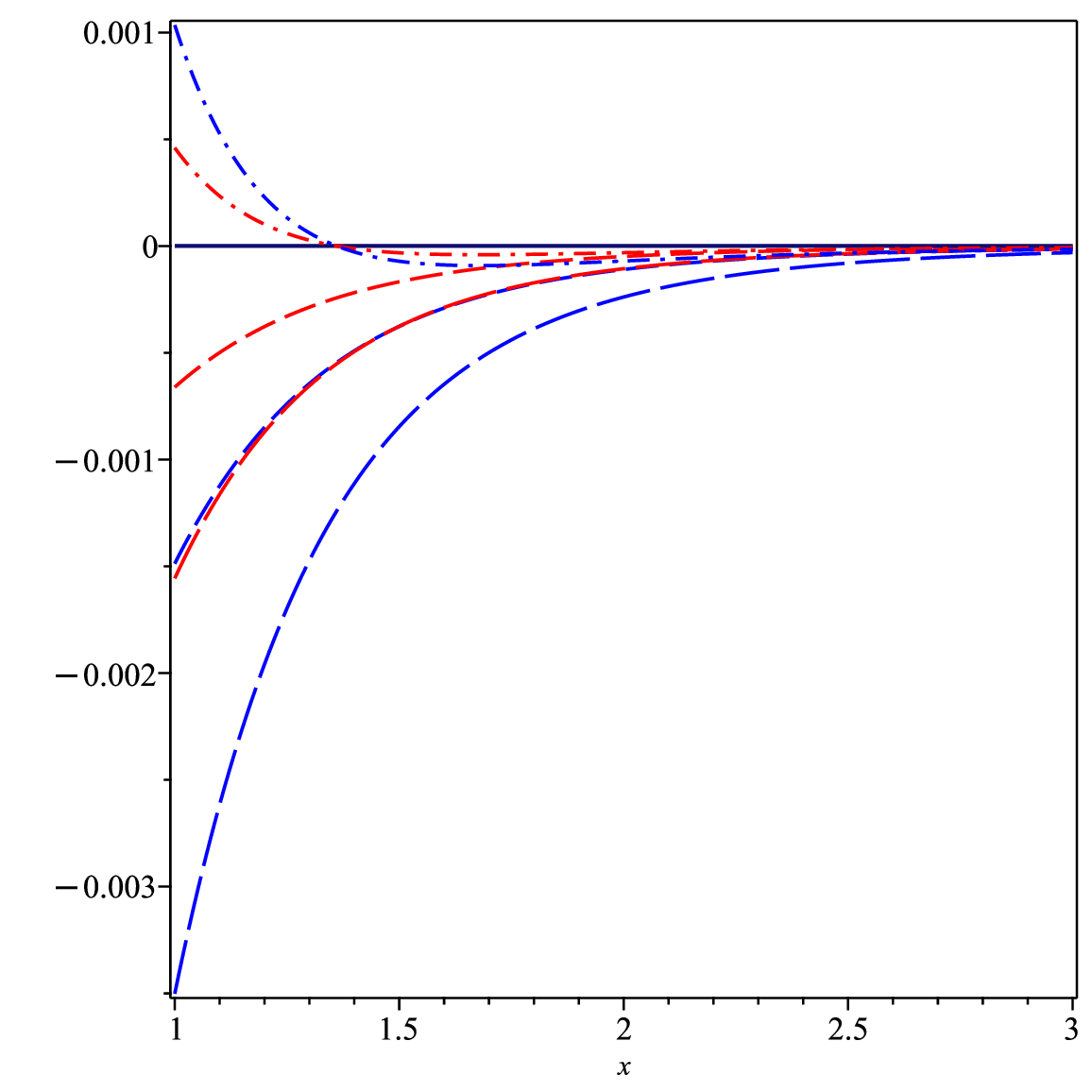}
    \includegraphics[width=0.3\textwidth]{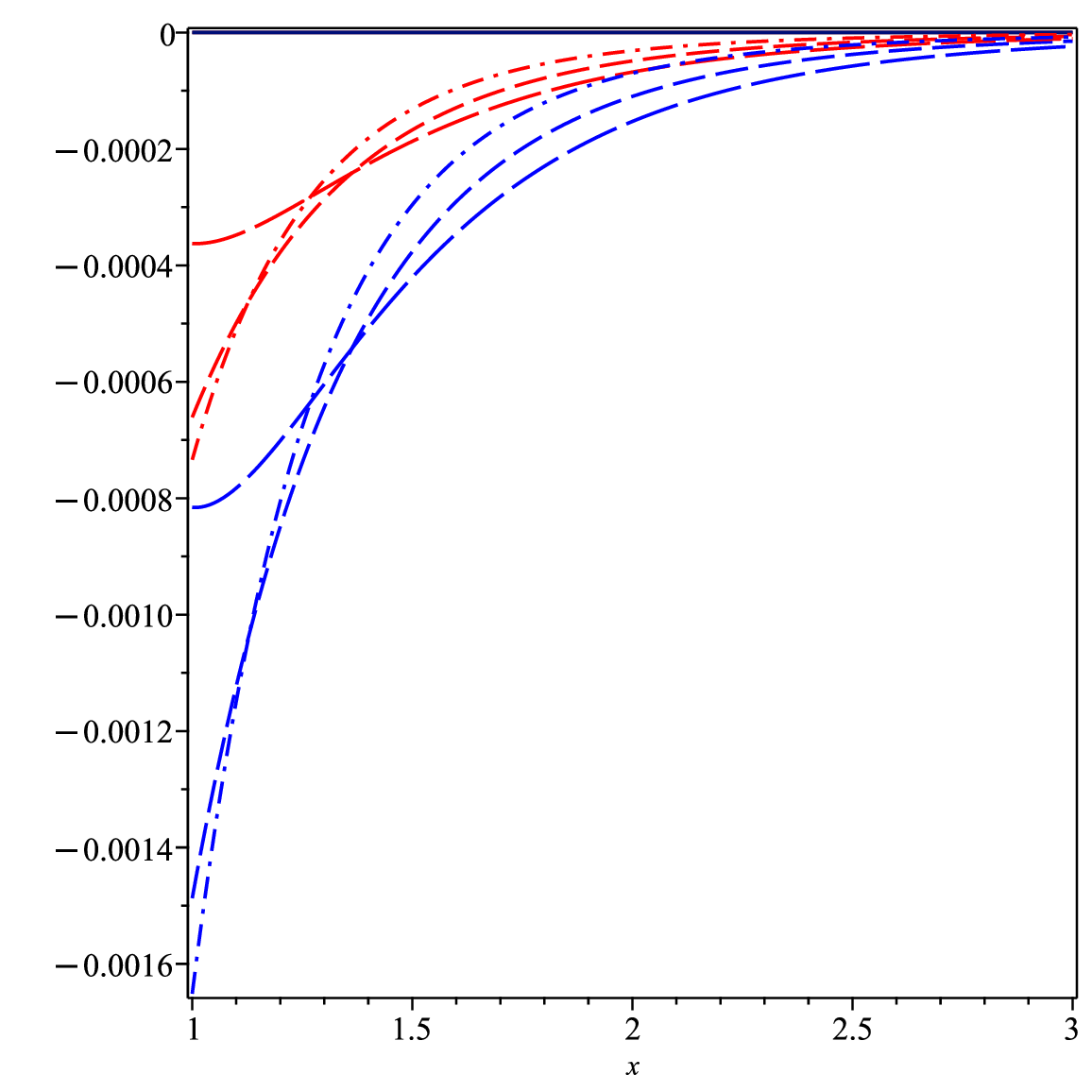}
\caption{\label{NECchi-MTW-CASE3} Behaviour of the meridional NEC for Case 3 of the Morris-Thorne wormhole under slow rotation, shown for different values of the dimensionless angular momentum parameter $j$. The static configuration ($j=0$) is represented by the black curves, while the rotating cases ($j=0.1$) and ($j=0.15$) are shown in red and blue, respectively. Dashed and dotted curves denote quantities evaluated on the equatorial plane and along the symmetry axis, respectively, for the reference configuration $(\alpha,\beta)=(0,0)$. In contrast, long–dashed and dash–dotted curves correspond to quantities with $(\alpha,\beta)\neq(0,0)$. \textbf{Left panel}: comparison between the reference case $(\alpha,\beta)=(0,0)$ and $(\alpha,\beta)=(1,1)$. \textbf{Central panel}: comparison between $(\alpha,\beta)=(0,0)$ and $(\alpha,\beta)=(-1,-1)$. \textbf{Right panel}: comparison between $(\alpha,\beta)=(0,0)$ and $(\alpha,\beta)=(1,-1)$.}
\end{figure}

\begin{figure}[!ht]
\centering
    \includegraphics[width=0.3\textwidth]{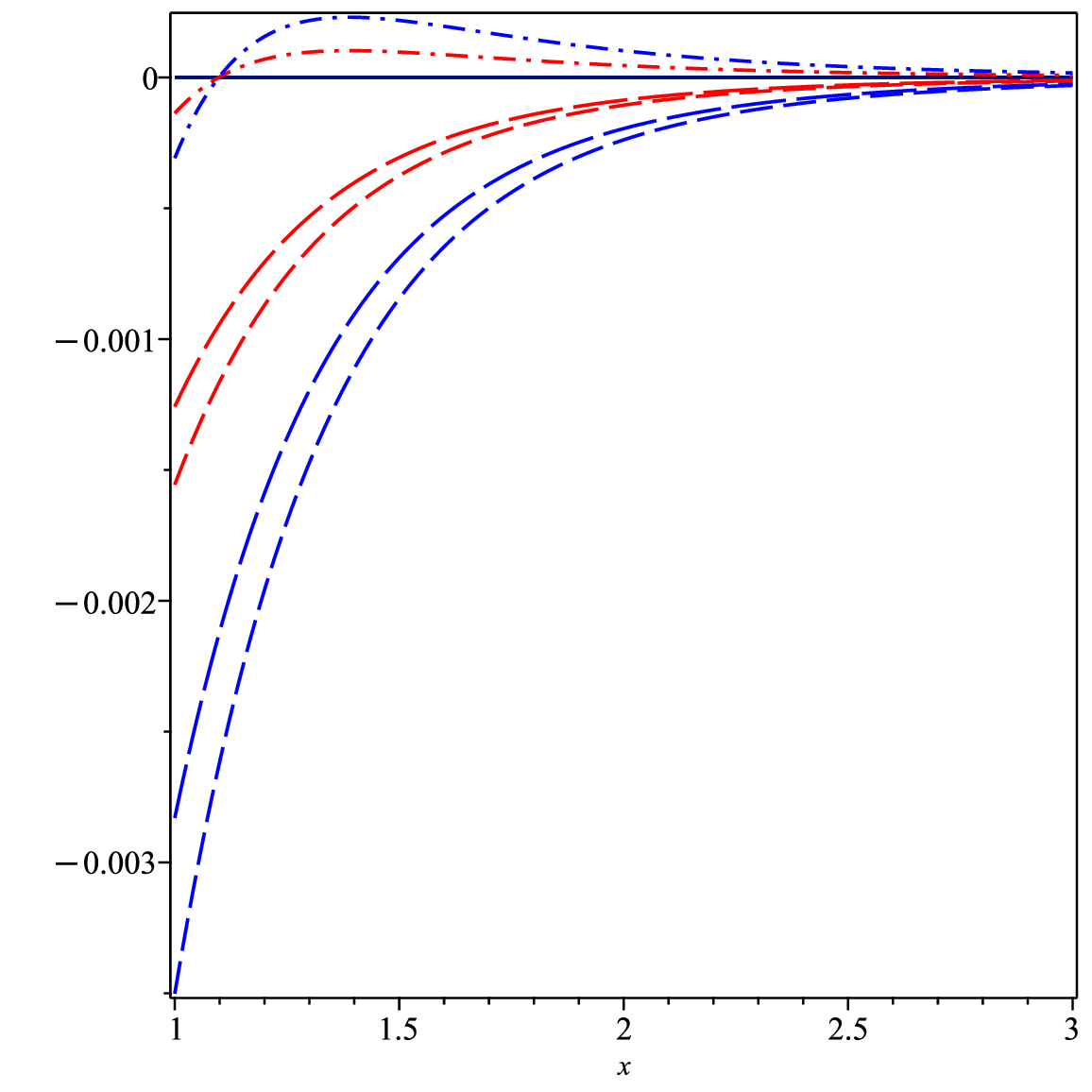}
    \includegraphics[width=0.3\textwidth]{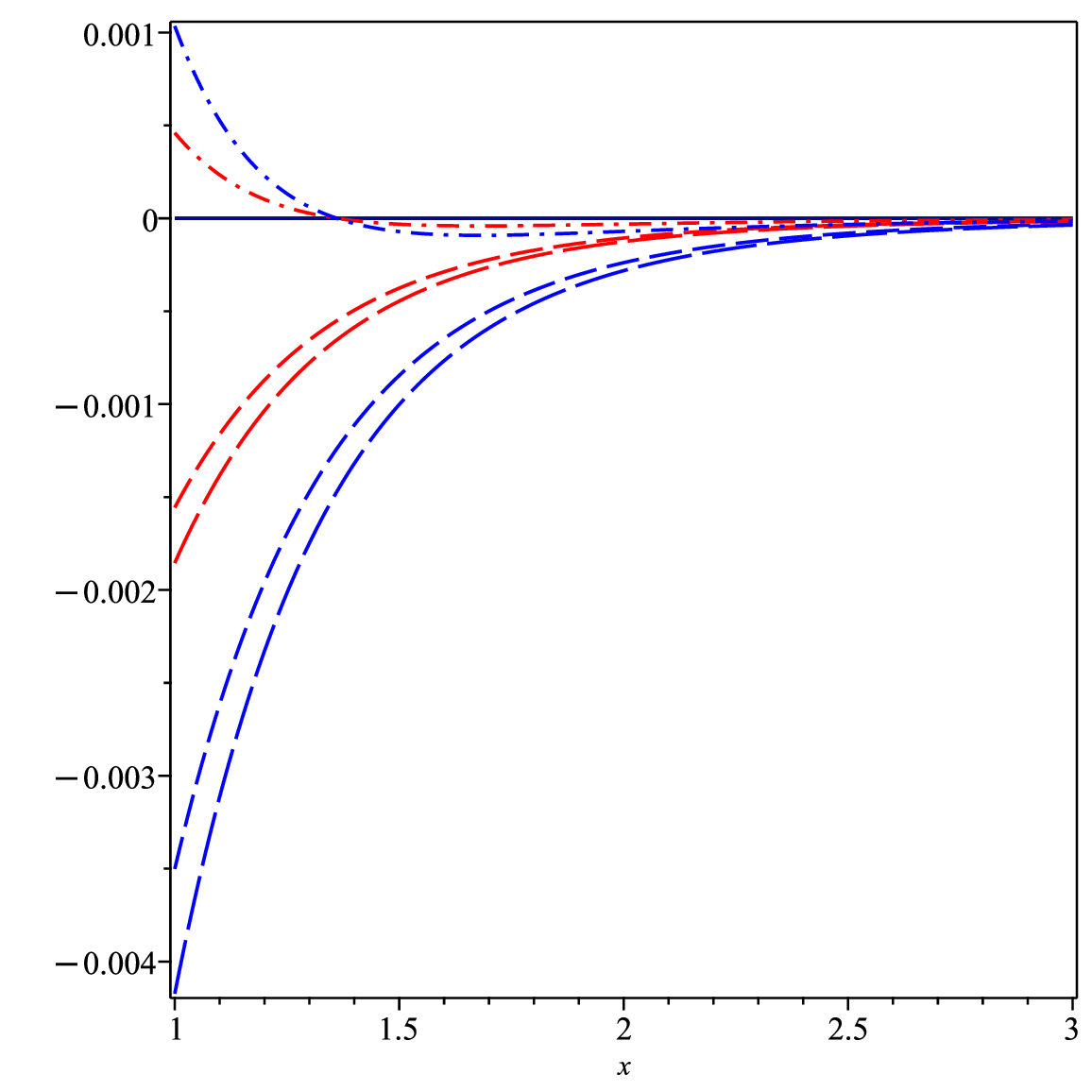}
    \includegraphics[width=0.3\textwidth]{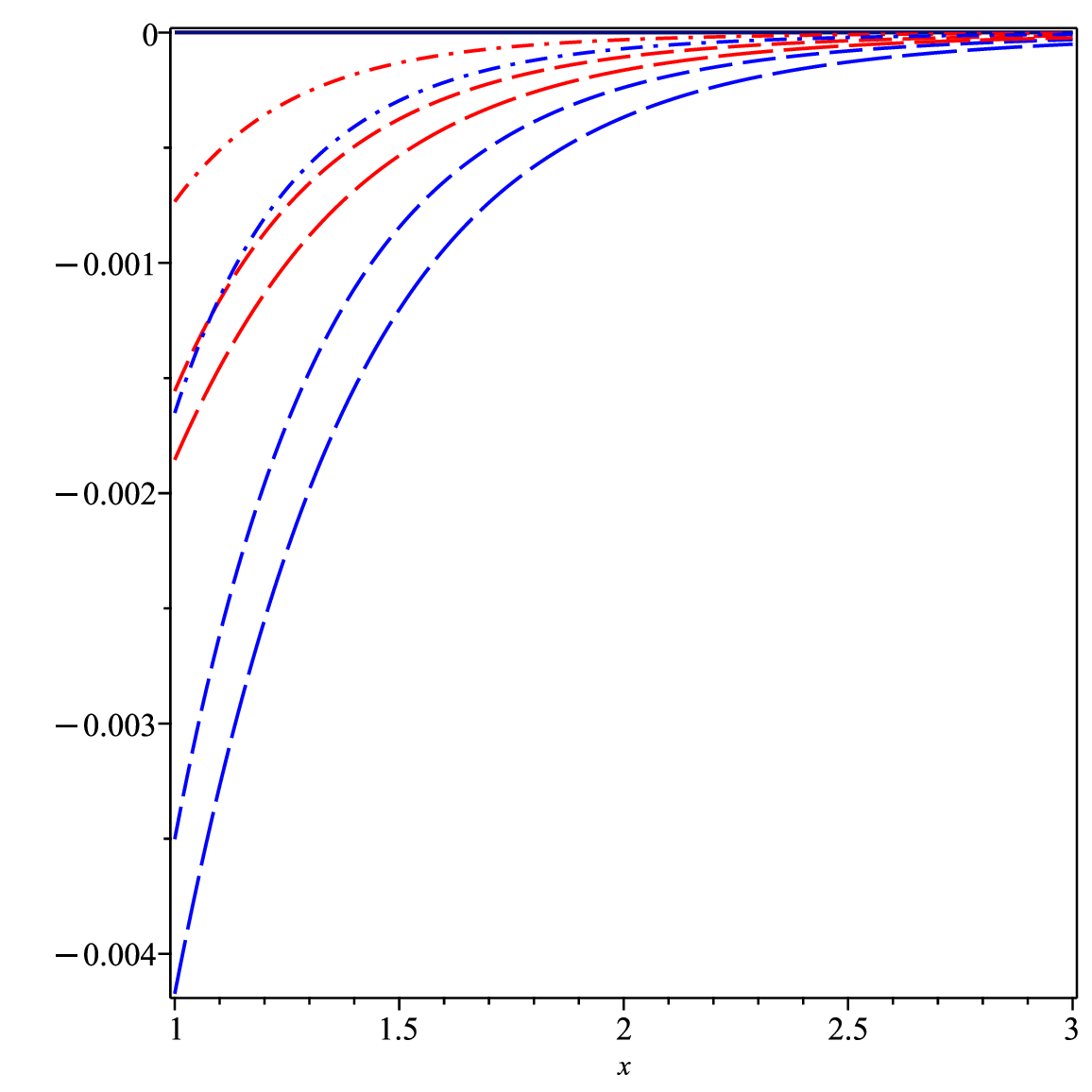}
\caption{\label{NECphi-MTW-CASE3} Behaviour of the azimuthal NEC for Case 3 of the Morris-Thorne wormhole under slow rotation, shown for different values of the dimensionless angular momentum parameter $j$. The static configuration ($j=0$) is represented by the black curves, while the rotating cases ($j=0.1$) and ($j=0.15$) are shown in red and blue, respectively. Dashed and dotted curves denote quantities evaluated on the equatorial plane and along the symmetry axis, respectively, for the reference configuration $(\alpha,\beta)=(0,0)$. In contrast, long–dashed and dash–dotted curves correspond to quantities with $(\alpha,\beta)\neq(0,0)$. \textbf{Left panel}: comparison between the reference case $(\alpha,\beta)=(0,0)$ and $(\alpha,\beta)=(1,1)$. \textbf{Central panel}: comparison between $(\alpha,\beta)=(0,0)$ and $(\alpha,\beta)=(-1,-1)$. \textbf{Right panel}: comparison between $(\alpha,\beta)=(0,0)$ and $(\alpha,\beta)=(1,-1)$.}
\end{figure}

For Model 1, Case 1 of the Schwarzschild wormhole, a direct inspection shows that at the poles all NEC components coincide with their static counterparts, plotted as solid black lines in Fig.~\ref{NEC_SCHW_MOD1_CASE1}. If we introduce a finite rotation parameter $j$, the violation of the radial NEC is slightly improved. On the equatorial plane, increasing $j$ leads to an enhancement of the NEC in the meridional direction, whereas the azimuthal NEC decreases. This indicates that rotation induces an anisotropy in the energy condition.

For Model 1, Case 3, we recall that the configuration $(\alpha,\beta)=(0,0)$ coincides with Model 1, Case 1. From Fig.~\ref{NEC_SCHW_MOD1_CASE3} it is observed that, for $(\alpha,\beta)=(1,1)$, increasing $j$ enhances the violation of the radial NEC on the equatorial plane, while along the symmetry axis the violation becomes weaker and may even display a local minimum. In contrast, for $(\alpha,\beta)=(-1,-1)$, increasing $j$ mitigates the NEC violation on the equatorial plane but amplifies it along the symmetry axis, thus reversing the behaviour found in the previous case. For $(\alpha,\beta)=(1,-1)$, the qualitative behavior closely resembles that of $(\alpha,\beta)=(-1,-1)$.  For this reason, it is not shown in Fig.~\ref{NEC_SCHW_MOD1_CASE3}. Finally, for $(\alpha,\beta)=(-1,1)$, the effect of rotation is opposite, i.e. the NEC violation becomes stronger on the equatorial plane and weaker along the rotational axis, where a local minimum may again appear.

For the NEC in the $\chi$-direction with $(\alpha,\beta)=(1,1)$, Fig.~\ref{NEC_SCHW_MOD1_CASE3_chi} shows that increasing $j$ enhances the NEC along the symmetry axis, while on the equatorial plane it decreases and eventually develops a local maximum. The configuration $(\alpha,\beta)=(-1,-1)$ exhibits a substantially different behavior. In this case, a violation of the NEC can occur along the symmetry axis and becomes more pronounced as $j$ increases. On the equatorial plane, by contrast, the NEC remains positive and grows with the angular momentum. A qualitatively similar trend is found for $(\alpha,\beta)=(1,-1)$. Finally, the case $(\alpha,\beta)=(-1,1)$ displays a behavior analogous to that observed for $(\alpha,\beta)=(1,1)$.

For the azimuthal NEC (see Fig.~\ref{NEC_SCHW_MOD1_CASE3_phi}), when $(\alpha,\beta)=(1,1)$, the NEC exhibits a local positive maximum near the throat but can become violated on the equatorial plane in its vicinity as $j$ increases. Along the symmetry axis, by contrast, the NEC grows with increasing angular momentum. For $(\alpha,\beta)=(-1,-1)$, the situation is reversed, that is, the NEC can be violated along the symmetry axis, while on the equatorial plane it increases with $j$. The configuration $(\alpha,\beta)=(1,-1)$ displays an overall behaviour analogous to the latter case, whereas $(\alpha,\beta)=(-1,1)$ behaves similarly to $(\alpha,\beta)=(1,1)$.

For Case~1 of the slowly rotating Morris–Thorne wormhole, Fig.~\ref{NEC_MTW_CASE1} shows that the violation of the radial NEC is progressively mitigated as $j$ increases, with the most significant improvement occurring on the equatorial plane. The same qualitative behaviour was already observed for the radial NEC in Model~1, Case~1 of the slowly rotating Schwarzschild wormhole (see the left panel of Fig.~\ref{NEC_SCHW_MOD1_CASE1}). Along the symmetry axis, the NEC is also reduced, although to a lesser extent than on the equatorial plane. The NEC in the $\chi$-direction becomes increasingly violated on the equatorial plane as $j$ grows, whereas no violation is detected along the symmetry axis. It is worth noting that for $j=0$ this quantity reduces to $\rho + p_t$, which vanishes identically in the static Morris–Thorne wormhole. This case is represented by the solid black horizontal line in the central and right panels. Finally, the NEC in the azimuthal direction exhibits a behaviour analogous to that of the $\chi$-direction. It is particularly interesting that the introduction of rotation can locally induce violations of the meridional and azimuthal NECs, even though these conditions are exactly saturated in the static configuration.

For the radial NEC in Case~3 of the Morris–Thorne wormhole, we refer to Fig.~\ref{NECr-MTW-CASE3}. When $(\alpha,\beta)=(0,0)$, the radial NEC coincides with that of Case~1 of the same wormhole model. It is included here as a reference to highlight the influence of the parameters $\alpha$ and $\beta$. When $(\alpha,\beta)=(1,1)$, the radial NEC becomes increasingly violated along the symmetry axis as $j$ grows (see the dash–dotted lines in the left panel of Fig.~\ref{NECr-MTW-CASE3}), while on the equatorial plane the violation is reduced with increasing $j$. This overall behaviour contrasts with that observed in Case~1 (see the left panel of Fig.~\ref{NEC_MTW_CASE1}). For $(\alpha,\beta)=(-1,-1)$, the situation is reversed, i.e. the NEC violation is mitigated along the symmetry axis but becomes stronger on the equatorial plane. The configuration $(\alpha,\beta)=(1,-1)$ exhibits a behaviour similar to that of $(\alpha,\beta)=(-1,-1)$, and is therefore not shown. Finally, for $(\alpha,\beta)=(-1,1)$, the violation of the radial NEC decreases on the equatorial plane but intensifies along the symmetry axis.

For the NEC in the $\chi$-direction (see Fig.~\ref{NECchi-MTW-CASE3}), we observe that for $(\alpha,\beta)=(1,1)$ it is satisfied on the equatorial plane, whereas along the symmetry axis the behaviour differs significantly. As $j$ increases, the NEC develops a positive maximum but becomes negative near the throat. The extent of this negative region, however, decreases with increasing $j$. In the case $(\alpha,\beta)=(-1,-1)$, an overall violation of the NEC occurs both on the equatorial plane and along the symmetry axis, with the latter remaining positive only in a narrow region close to the throat. For $(\alpha,\beta)=(1,-1)$, the NEC is satisfied along the symmetry axis but remains violated across the equatorial plane. Finally, for $(\alpha,\beta)=(-1,1)$, the NEC is consistently violated both along the symmetry axis and on the equatorial plane.

For the azimuthal NEC (see Fig.~\ref{NECphi-MTW-CASE3}), we observe that for $(\alpha,\beta)=(1,1)$ the NEC becomes increasingly violated on the equatorial plane as $j$ grows, while it remains positive along the symmetry axis except in the immediate vicinity of the throat. For $(\alpha,\beta)=(-1,-1)$, the behaviour on the equatorial plane follows the same pattern, but along the symmetry axis it is reversed, i.e. the NEC is positive near the throat and becomes violated farther away along the rotational axis. Finally, in the case $(\alpha,\beta)=(-1,1)$, the NEC remains violated both on the equatorial plane and along the symmetry axis for all considered values of $j$.

\begin{figure}[!ht]
\centering
    \includegraphics[width=0.45\textwidth]{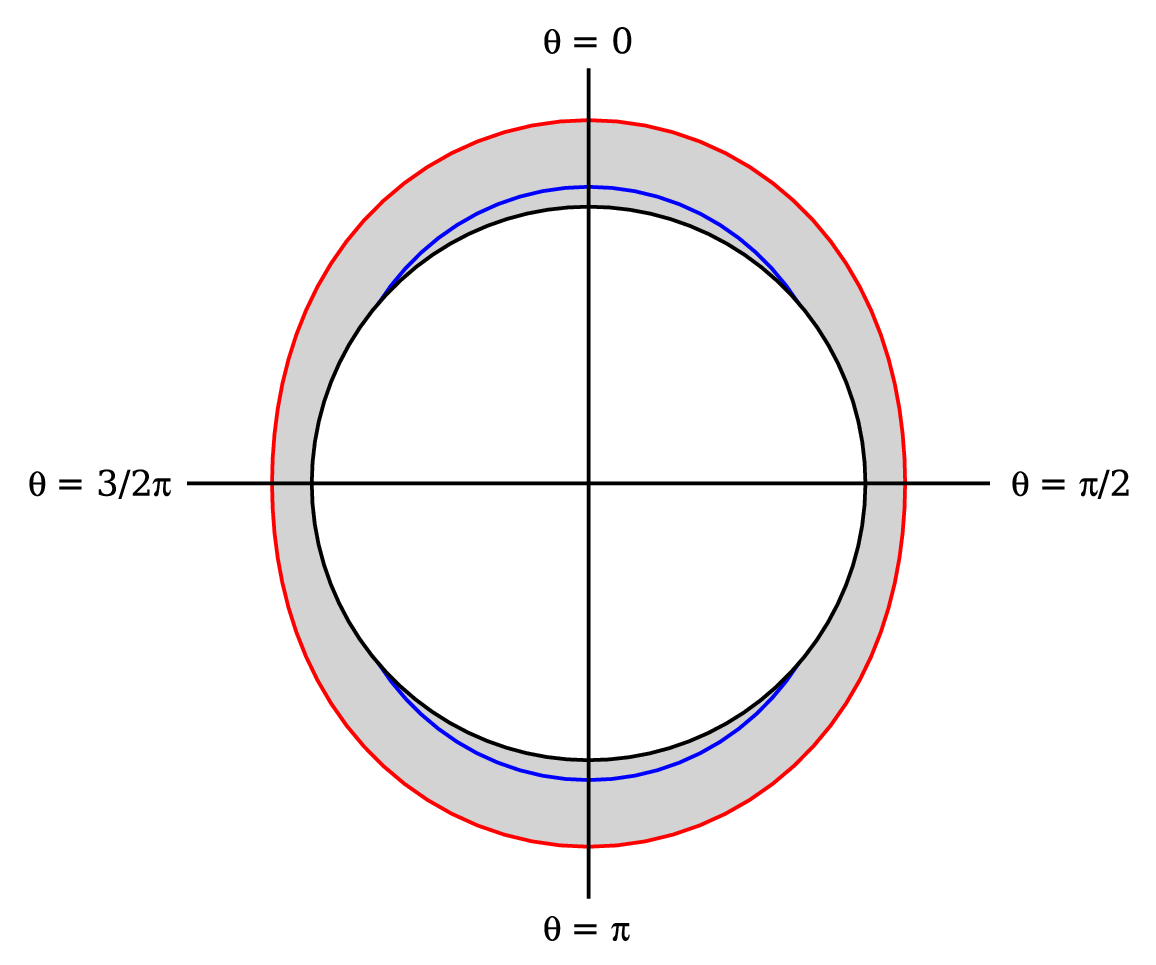}
\caption{\label{e1}
Cross section of the wormhole throat for Case 3 of the Schwarzschild wormhole with $B_0^{(2)} =0$ under slow rotation, with parameters $\alpha=-46$ and $\beta=-20$, for different values of the dimensionless angular momentum parameter $j$. The solid blue and red curves correspond to the boundary of the ergoregion corresponding to $j=0.1$ and $j=0.15$, respectively, while the solid black curve denotes the throat position.}
\end{figure}

\begin{figure}[!ht]
\centering
    \includegraphics[width=0.45\textwidth]{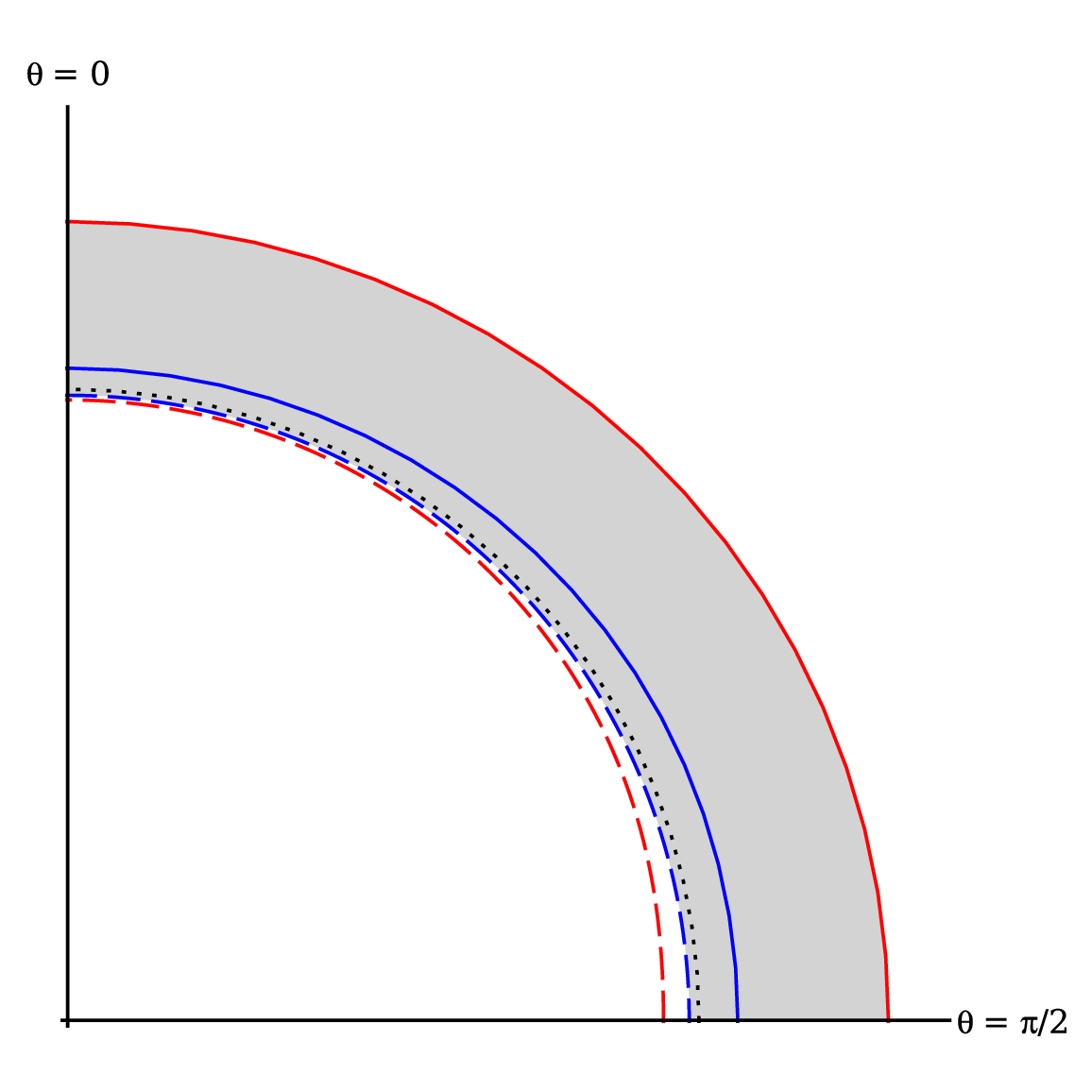}
\caption{\label{e2}
Cross section of the wormhole throat for Case 3 of the Schwarzschild wormhole with $B_0^{(2)} \neq 0$ under slow rotation, with parameters $\alpha=-60$ and $\beta=3$, shown in a quarter view for improved visibility. The solid blue and red curves indicate the boundary of the ergoregion for $j=0.1$ and $j=0.15$, respectively. The corresponding throats are represented by dashed blue and dashed red curves, while the dotted black curve denotes the throat of the static configuration ($j=0$).}
\end{figure}

\begin{figure}[!ht]
\centering
    \includegraphics[width=0.45\textwidth]{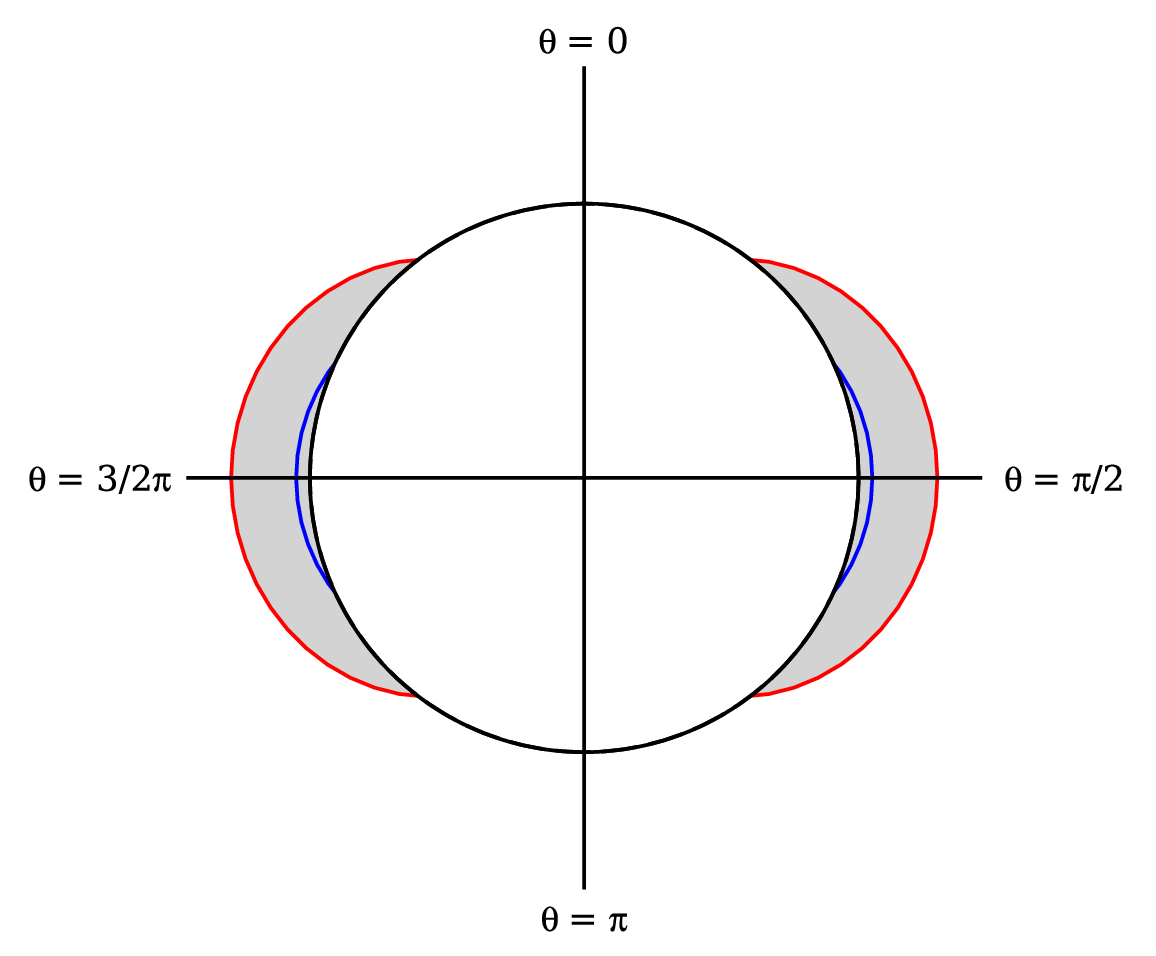}
\caption{\label{e3}
Cross section of the wormhole throat for Case 3 of the Morris–Thorne wormhole under slow rotation, with parameters $\alpha=-40$ and $\beta=40$. The solid blue and red curves correspond to the boundary of the ergoregion corresponding to $j=0.1$ and $j=0.15$, respectively, while the solid black curve denotes the throat. It is observed that, in this configuration, the ergoregion does not surround the throat.}
\end{figure}

Finally, the ergoregion of a rotating wormhole is defined by the stationary limit condition $g_{tt}=0$, which locates the surface beyond which no observer can remain static with respect to infinity. Figures~\ref{e1}–\ref{e3} illustrate how this surface depends on the parameters of the geometric background, and on the spin $j$. In the spatial Schwarzschild case with ($B^{(2)}_0=0$) depicted in Fig.~\ref{e1}, the stationary limit surface forms a closed contour around the throat and expands as $j$ increases. Allowing a monopolar correction ($B^{(2)}_0\neq0$) leads to a visibly asymmetric ergoregion and a rotation‑induced shift of the throat relative to the static throat (see Fig.~\ref{e2}). The ergosurface continues to grow with $j$ but is no longer centred at the throat. For the Morris–Thorne background represented by Fig.~\ref{e3}, the behaviour is qualitatively different. An ergoregion appears for $j>0$ yet remains detached from the throat for the parameter choices shown. Taken together, Figs.~\ref{e1}–\ref{e3} show that the presence and morphology of the ergoregion are sensitive to the background (spatial–Schwarzschild vs. Morris–Thorne), to $B^{(2)}_0$, the parameters $(\alpha,\beta)$, and the rotation parameter $j$.

From a physical standpoint, our NEC diagnostics show that slow rotation does not remove the need for exotic matter at the throat, but it does reshape how the exoticity is distributed. For the spatial–Schwarzschild seed, the static violation of the radial NEC is complemented by a rotation–induced negative comoving density and anisotropic shifts of the principal pressures. The NEC combinations then develop a pronounced angular dependence. The polar regions remain close to the static configuration, while the equatorial plane carries most of the additional NEC violation or mitigation, depending on the chosen lapse deformation. In this sense, spin redistributes the exotic sector between equatorial and polar regions and between radial and tangential null directions, without providing a fully NEC–respecting branch. For the Morris–Thorne seed, the effect of spin is slightly more favourable. The magnitude of the radial NEC violation at the throat decreases with increasing $j$, especially on the equatorial plane, but this is accompanied by the appearance of modest meridional and azimuthal NEC violations in regions where the static geometry saturated the NEC. Overall, the rotational corrections constitute an $\mathcal{O}(J^2)$ reshuffling and partial softening of the exotic stress–energy rather than a mechanism that eliminates it.

\section{Conclusions and outlook}

We have developed a slow‑rotation framework for traversable wormholes in the presence of a co‑rotating anisotropic fluid. Starting from a Teo‑type stationary, axisymmetric ansatz, we expanded the geometry and matter fields in terms of the rotation parameter $J$, imposed an area gauge that preserves the geometric meaning of the areal coordinate order by order in $J$, and developed two complementary treatments of the throat (fixed vs. free). Within this approach, the Einstein equations and the fluid conservation laws close, yielding a linear ODE for the leading frame dragging and coupled cubic corrections equations, together with algebraic relations that simplify the second‑order backreaction. In particular, the $\ell=2$ sector of $G_{r\chi}=0$ fixes $B^{(2)}_{2}$ explicitly, and the conservation law enforces $K^{(2)}_{2}=0$ with the area gauge already implying $K^{(2)}_{0}=0)$. We also quantified the throat displacement in the free-throat scheme and showed that it arises only at order $\mathcal{O}(J^{2})$.

We applied the formalism to two canonical zero‑redshift seeds. For the spatial Schwarzschild wormhole with $b(r)=2M$, we obtained a fully regular solution for the leading frame dragging, $\omega_1(r)=2/r^3$, after calibrating the rotation law to the standard Lense–Thirring tail. We presented closed expressions for the cubic corrections $\omega^{(3)}_{0,2}(r)$ and for all second-order stress–energy tensor components. In the simplest regular case (see Case 1), slow rotation induces a negative comoving density and anisotropic shifts of the principal pressures, consistent with the interpretation that rotation acts as a geometric source of exoticity that partially compensates the tension supporting the throat. Allowing a monopolar backreaction in the shape function (see Case 2) yields an explicit expression $B^{(2)}_{0}(r)\neq 0$ and a positive comoving density at order $\mathcal{O}(J^{2})$ even though the static seed has $\rho=0$, while preserving asymptotic flatness and keeping the ADM mass unchanged. A parity‑even lapse deformation (see Case 3) parameterised by the pair $(\alpha,\beta)$ introduces a well-behaved quadrupolar structure in the slow-rotation backreaction. We also observed an additional root introduced by the truncated shape function and showed that it is a truncation artefact rather than a genuine deformation of the static throat. Across the rotation laws we have considered, Models II and III coincide with Model I through $\mathcal{O}(J^{3})$ up to a monopolar shift in $\omega^{(3)}_{0}$. As a consequence, the comoving energy density, the principal pressures, and all NEC combinations at order $\mathcal{O}(J^2)$ are identical across Models I–III. Only the monopolar part of the cubic frame-dragging correction differs between rotation laws. Regarding the Morris–Thorne wormhole with $b(r)=r_0^2/r$, we derived a closed analytic expression for $\omega_1(\ell)$ in terms of the proper radial distance $\ell$, which is independent of the chosen rotation law at this order due to the equation $\rho-p_r=0$. We solved the cubic equations to obtain $\omega^{(3)}_{0,2}(\ell)$ and gave explicit corrections for the comoving density and the principal pressures. We also showed that introducing a monopolar shape correction $B^{(2)}_{0}\neq 0$ makes the equation for $\omega^{(3)}_0$ singular at the throat, thereby excluding that branch as physically inadmissible in this background. Last but not least, we run energy‑condition diagnostics and study ergoregions when they arise. For the spatial–Schwarzschild seed, rotation redistributes the NEC violation in an anisotropic manner. Moreover, the transformation $(\alpha,\beta)\to -(\alpha,\beta)$ reverses the relative behaviour of the NECs on the equatorial plane and along the symmetry axis across the radial, meridional, and azimuthal components. For the Morris–Thorne seed, rotation tends to mitigate the radial NEC violation, with the greatest improvement occurring on the equatorial plane, while it can also induce local violations in the meridional and azimuthal components despite these being saturated in the static geometry. Regarding ergoregions, the stationary limit surface in the spatial Schwarzschild case encloses the throat and grows with $J$. If we also allow $B^{(2)}_{0}\neq0$, we obtain pronounced asymmetries and a rotation‑induced shift of the throat. In the Morris–Thorne case considered, the ergoregion remains detached from the throat, thus highlighting a qualitative departure from Kerr-like behaviour. Finally, our energy–condition diagnostics confirm that slow rotation leaves the qualitative need for exotic matter intact. In all solutions considered, at least one of the combinations $\widetilde{\rho}+P_{r,\chi,\phi}$ is negative in a neighbourhood of the throat, so a NEC violation is never removed altogether. For the spatial–Schwarzschild seed, rotation chiefly redistributes the NEC–violating sector in an anisotropic fashion, concentrating or relieving it in specific angular regions. Thus, within our slow–rotation framework, the angular momentum of the wormhole acts more as a geometric mechanism that redistributes and mildly softens the exotic stress–energy than as a way of eliminating it. 
We have also verified that, within the family of admissible rotation laws considered here, the matter sector through $\mathcal{O}(J^2)$ is insensitive to the detailed choice of $\Omega(r,\chi)$, so that our conclusions on exoticity and ergoregions reflect generic features of the slow rotation regime rather than artefacts of a particular angular velocity profile. Moreover, all diagnostics presented in Sec. IV have been computed for $0 \leq j \leq 0.15$. For Models I, III, and IV, the explicit coefficients of the slow rotation expansion imply that the cubic corrections to the frame dragging function never exceed about $10\%$ of the linear contribution at $j \simeq 0.1$ and remain at the $\lesssim 20$–$25\%$ level at $j=0.15$. Model II provides the most conservative case. There, the ratio of the cubic to the linear frame dragging term is bounded by $\lesssim 20 j^2$, so that even at $j=0.15$ the cubic contribution stays below $\sim 45\%$ of the linear one. Since the geometry–matter coupling determining the comoving density, principal pressures, and NEC combinations is fixed at $\mathcal{O}(J^2)$, where Models I–III coincid,e and Model IV yields the same stress–energy, the energy–condition and ergoregion analyses remain under quantitative perturbative control throughout the parameter range used in this work.

A natural next step is to relax the zero-redshift assumption and allow for nontrivial lapse functions. This would allow examination of how redshift or blueshift gradients interact with the area gauge and with the fixed- versus free-throat prescriptions, thereby broadening the class of anisotropic fluids that can be treated within the same framework. It would also be worthwhile to extend the slow-rotation expansion to the following order in $J$, both to sharpen curvature and energy-condition diagnostics near the throat and to assess the robustness of the present truncation. A further direction is to move beyond the imposed angular-velocity profiles and derive rotation laws from explicit choices of anisotropic matter. The present construction should thus be viewed as an effective slow-rotation framework. Once a static wormhole seed and an admissible rotation law $\Omega=\Omega(r,\chi)$ are specified, the Einstein equations and the conservation law uniquely determine a regular, co-rotating anisotropic fluid that sources the Teo-type geometry. We do not, at this stage, commit to a specific underlying microphysical model for the exotic sector. Instead, we provide explicit, analytically tractable spacetimes that can serve as benchmarks for more detailed matter models (e.g., scalar, vector, or multi-form fields) and for phenomenological applications. Deriving rotation laws from explicit anisotropic matter models in this way would show how the properties of the supporting fluid set the radial behaviour of the angular velocity and reduce the residual freedom in the slow-rotation expansion.

On the applications side, the explicit metrics obtained here are directly suitable for analysing photon trajectories, lensing and shadow formation, wave propagation, and quasinormal spectra in the slow rotation regime. It would also be helpful to compare the present slow-rotation solutions with existing stationary wormhole models constructed by other methods, to gauge the range of validity of the expansion and identify quantities that remain stable under changes in gauge or throat prescription. At a qualitative level, several features uncovered in Sec. IV are expected to leave characteristic imprints on observables. The exact slow-rotation frame-dragging profiles and their cubic corrections determine the Lense--Thirring precession and enter directly into the locations and stabilities of timelike and null circular orbits, and thus into lensing patterns and shadow morphology. The quadrupolar throat deformations and anisotropic shifts of the principal pressures suggest small but systematic changes in photon-ring radii, image distortion, and time-delay structures in multi-image configurations. The topology of the ergoregion is likewise phenomenologically relevant. For the spatial–Schwarzschild seed, the stationary limit surface always encloses the throat and expands with $J$, whereas for the Morris–Thorne seed, the ergoregion remains detached from the throat. This qualitative difference should translate into distinct regimes for superradiant scattering and energy extraction, as well as for the structure of magnetospheres anchored to the wormhole throat. Finally, the rotation-induced redistribution and partial mitigation of the NEC violation modify the effective scattering potentials experienced by perturbations and are therefore expected to affect quasinormal spectra and possible echo signals. Quantifying these effects in detail will require a dedicated analysis of geodesics (including lensing and shadow formation) and wave propagation on the backgrounds constructed here, which we leave for future work.

\appendix
\section{Covariant components of the anisotropic energy–momentum tensor}\label{AppendixA}

The following covariant components of the anisotropic energy–momentum tensor are obtained from \eqref{AEMT} and have been verified in \texttt{Maple}. Specifically, we find
\begin{eqnarray}
T_{tt}&=&\frac{\widetilde{\rho}N^4 + r^2(1-\chi^2 )K^2
\left\{r^2(1-\chi^2)\omega^2 K^2(\Omega-\omega)^2\widetilde{\rho}
+ N^2\left[2\omega(\Omega-\omega)\widetilde{\rho} + \Omega^2 P_\varphi\right]\right\}}
{N^2 - r^2(1-\chi^2)K^2(\Omega - \omega)^2},\\  
T_{t\varphi}&=&-\frac{r^2(1-\chi^2)K^2\left\{r^2(1-\chi^2)K^2\widetilde{\rho}\omega(\Omega-\omega)^2-N^2\left[\widetilde{\rho}\omega-\Omega(\widetilde{\rho}+P_\varphi)\right]\right\}}
{N^2 - r^2(1-\chi^2)K^2(\Omega - \omega)^2},\\
T_{\varphi\varphi}&=&\frac{r^2(1-\chi^2)K^2\left[N^2 P_\varphi+r^2(1-\chi^2)K^2\widetilde{\rho}(\Omega-\omega)^2\right]}
{N^2 - r^2(1-\chi^2)K^2(\Omega - \omega)^2},\\
T_{rr}&=&\frac{P_r}{1-\frac{B}{r}},\quad
T_{\chi\chi}=\frac{r^2 K^2 P_\chi}{1-\chi^2}.
\end{eqnarray}

\bibliography{QNMS}

\end{document}